\documentclass[twocolumn]{aastex631}
\turnoffeditone
\turnoffedittwo
\usepackage[all]{nowidow}
\usepackage{stix}
\usepackage{xspace}
\usepackage{hyperref}
\usepackage{graphicx}
\usepackage{xcolor}
\usepackage{enumitem}
\usepackage{amsmath}
\let\tablenum\relax
\usepackage[separate-uncertainty=true]{siunitx}
\usepackage{CJKutf8}

\newcommand{\mjup}{\ensuremath{M_\mathrm{Jup}}\xspace}
\newcommand{\teff}{\ensuremath{T_{\mathrm{eff}}}\xspace}
\newcommand{\logg}{\ensuremath{\log(g)}\xspace}

\graphicspath{{./}{figures/}}

\newcommand{\revise}[1]{\emph{\color{blue}#1}}
\renewcommand{\revise}[1]{{#1}}

\begin{document}
\begin{CJK*}{UTF8}{gbsn}
\title{HST/WFC3 Complete Phase-resolved Spectroscopy of White Dwarf-Brown Dwarf Binaries WD~0137 and EPIC~2122}

\correspondingauthor{Yifan Zhou}
\email{yifan.zhou@utexas.edu}

\author[0000-0003-2969-6040]{Yifan Zhou}
\altaffiliation{51 Pegasi b Fellow}
\affiliation{Department of Astronomy, The University of Texas at Austin, 2515 Speedway, Austin, TX 78712, USA}

\author[0000-0003-3714-5855]{D\'aniel Apai}
\affiliation{Department of Astronomy and Steward Observatory, The University of Arizona, 933 N. Cherry Ave., Tucson, AZ 85721, USA}
\affiliation{Lunar and Planetary Laboratory, The University of Arizona, 1640 E. University Blvd, Tucson, AZ 85721, USA}

\author[0000-0003-2278-6932]{Xianyu Tan}
\affiliation{Atmospheric, Oceanic and Planetary Physics, Department of Physics, University of Oxford, OX1 3PU, UK}

\author[0000-0003-3667-8633]{Joshua D. Lothringer}
\affiliation{Department of Physics and Astronomy, Johns Hopkins University, 3400 N. Charles St., Baltimore, MD 21210, USA}

\author[0000-0003-1487-6452]{Ben W. P. Lew}
  \affiliation{Lunar and Planetary Laboratory, The University of Arizona, 1640 E. University Blvd, Tucson, AZ 85721, USA}
   \affiliation{Bay Area Environmental Research Institute and NASA Ames Research Center, Moffett Field, CA 94035, USA}

\author[0000-0003-2478-0120]{Sarah L. Casewell}
\altaffiliation{STFC Ernest Rutherford Fellow}
\affiliation{School of Physics and Astronomy, University of Leicester, University Road, Leicester, LE1 7RH, UK}

\author[0000-0001-9521-6258]{Vivien Parmentier}
\affiliation{Atmospheric, Oceanic and Planetary Physics, Department of Physics, University of Oxford, OX1 3PU, UK}

\author[0000-0002-5251-2943]{Mark S. Marley}
\affiliation{Lunar and Planetary Laboratory, The University of Arizona, 1640 E. University Blvd, Tucson, AZ 85721, USA}

  \author[0000-0002-8808-4282]{Siyi Xu (许\CJKfamily{bsmi}偲\CJKfamily{gbsn}艺)}
  \affiliation{Gemini Observatory/NSF’s NOIRLab, 670 N. A’ohoku Place, Hilo, HI 96720, USA}

\author[0000-0002-4321-4581]{L. C. Mayorga}
\affiliation{The Johns Hopkins University Applied Physics Laboratory, 11100 Johns Hopkins Rd, Laurel, MD 20723, USA}

\begin{abstract}
Brown dwarfs in close-in orbits around white dwarfs offer an excellent opportunity to investigate properties of fast-rotating, tidally-locked, and highly-irradiated atmospheres. We present Hubble Space Telescope Wide Field Camera 3 G141 phase-resolved observations of two brown dwarf-white dwarf binaries: WD~0137-349 and EPIC~212235321. Their 1.1 to 1.7 \micron{} phase curves demonstrate rotational modulations with semi-amplitudes of $5.27\pm0.02$\% and $29.1\pm0.1$\%; both can be well fit by multi-order Fourier series models. The high-order Fourier components have the same phase as the first order and are likely caused by hot spots located at the substellar points, suggesting inefficient day/night heat transfer. Both brown dwarfs' phase-resolved spectra can be accurately represented by linear combinations of their respective day- and night-side spectra.  Fitting the irradiated brown dwarf model grids to the day-side spectra require a filling factor of ${\sim}50\%$, further supporting a hot spot dominating the day-side emission. The night-side spectrum of WD~0137-349B is reasonably well fit by non-irradiated substellar models and the one of EPIC~212235321B can be approximated by a Planck function. We find strong spectral variations in the brown dwarfs' day/night flux and brightness temperature contrasts, highlighting the limitations of band-integrated measurements in probing heat transfer in irradiated objects.  On the color-magnitude diagram, WD~0137-349B evolves along a cloudless model track connecting the early-L and mid-T spectral types, suggesting that clouds and disequilibrium chemistry have a negligible effect on this object. A full interpretation of these high-quality phase-resolved spectra calls for new models that couple atmospheric circulation and radiative transfer under high-irradiation conditions.
\end{abstract}

\section{Introduction}

\begin{figure*}[t]
  \centering
  \includegraphics[height=0.38\textwidth]{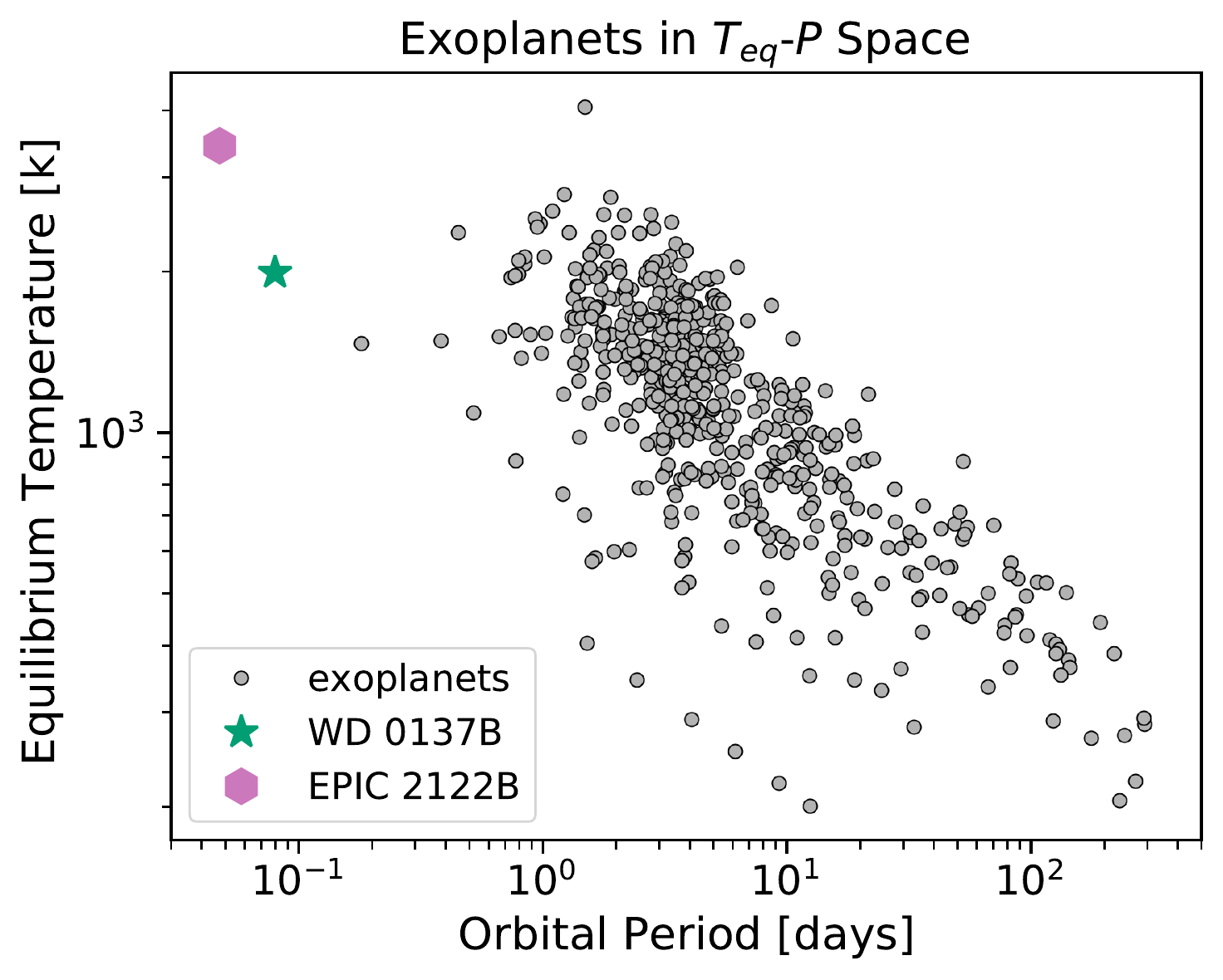}
  \includegraphics[height=0.38\textwidth]{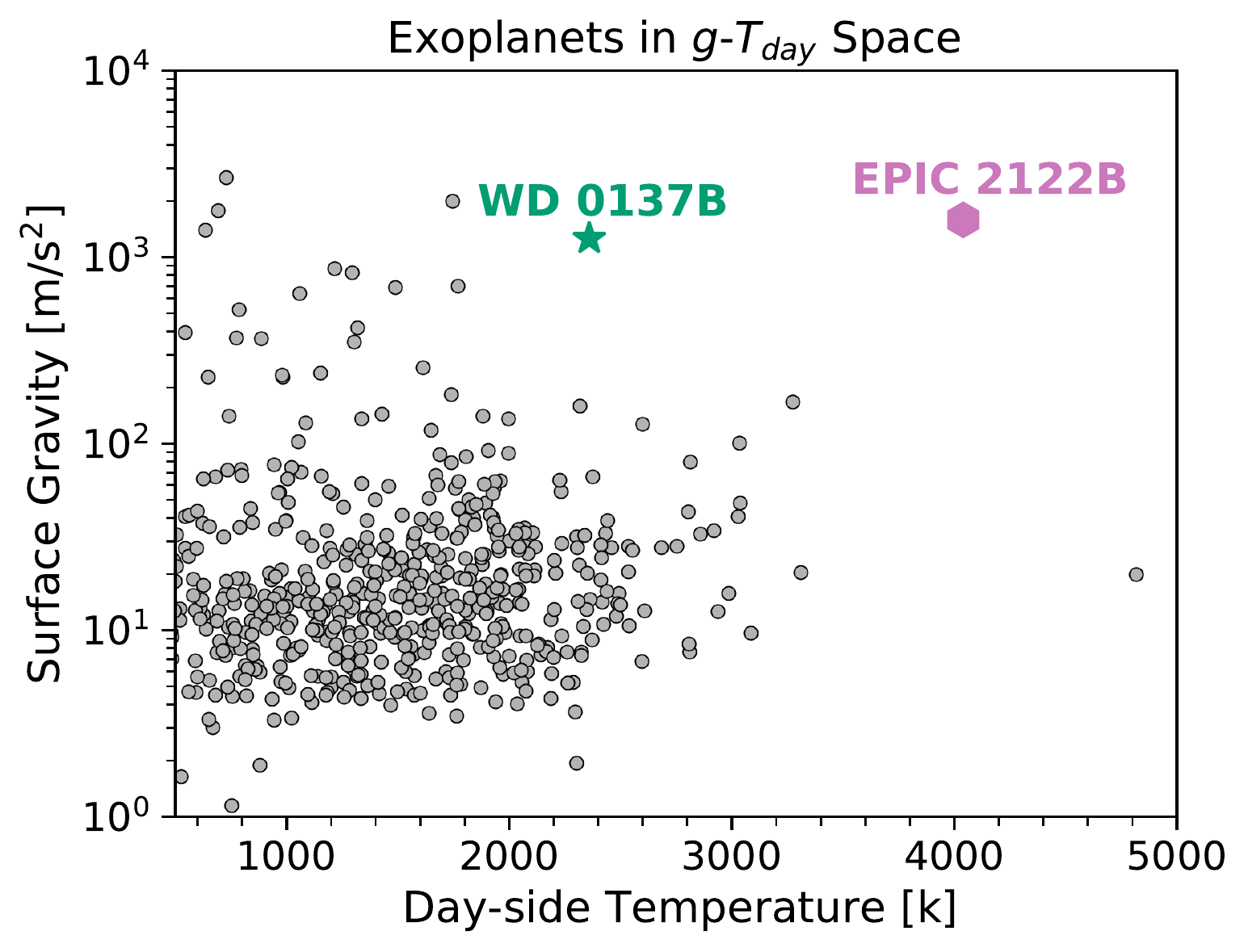}
  \caption{Comparisons of temperatures, orbital periods, and surface gravities between  WD~0137B, EPIC 2122B, and transiting exoplanets. \emph{Left:} The two irradiated brown dwarfs have shorter orbital periods and higher equilibrium temperatures than exoplanets. \emph{Right:} Compared to exoplanets, the two brown dwarfs have higher surface gravity and higher day-side temperatures. We assume day-side only heat redistribution in the derivation of  $T_{\mathrm{day}}$. The exoplanet data are a subset of \href{https://exoplanetarchive.ipac.caltech.edu}{the NASA Exoplanet Archive}. The selection criteria are ``detection method = transit'' and $M_{\mathrm{p}}<13\mjup$.}
  \label{fig:exopop}
\end{figure*}

In planetary atmospheres,  the thermal, compositional, and cloud distributions vary in three dimensions, and these structures are further shaped by atmospheric circulation (e.g., \citealt[][]{Showman2002,Showman2008,Parmentier2013a}, also see a recent review by \citealt{Showman2020}). One of the ultimate objectives for characterizing planetary atmospheres is to precisely map the three dimensional structures and decipher the relationship between the formation of heterogeneous atmospheric structures and external factors such as irradiation strength and spectra, as well as intrinsic factors such as the planet size, effective temperature, surface gravity, and rotation rate. Hot Jupiters have been a focus of this investigation of planetary atmospheres \citep[e.g.,][]{Knutson2012, Stevenson2014, Zellem2014,Parmentier2017,Arcangeli2019}. They occupy a unique parameter space with respect to irradiation, effective temperatures, and day/night contrasts. Because hot Jupiters are typically tidally locked, a permanent and steep longitudinal temperature gradient is established in their atmosphere, which forces heat to transport globally. The self-luminous brown dwarfs and planetary-mass companions also offer critical insights into planetary atmospheric dynamics \citep[e.g.,][]{Showman2012,Apai2017,Tan2020b,Showman2020}. With thermal structures determined by internal heat flux and rotation rates that are similar to solar system gas giants, these objects may display a circulation pattern that are more representative to exoplanets than the extreme cases presented by hot Jupiters. From an observational perspective, compared to other types of close-in or transiting planets, both hot Jupiters and directly imaged brown dwarfs are more favorable for high-precision observations. Their time-resolved spectroscopic observations directly probe the heterogeneous structures \citep[e.g.,][]{Stevenson2014,Apai2013,Kreidberg2018,Arcangeli2019}, delivering the most comprehensive information about the atmospheric dynamics of exoplanets.

A rare group of brown dwarfs that closely orbit their white dwarf hosts offer remarkable opportunities to test substellar atmospheric circulation models beyond the parameter space that has been explored in hot Jupiter and self-luminous brown dwarfs \citep[e.g.,][]{Maxted2006,Casewell2018}. These systems have experienced common envelope evolution, the period when the progenitor of the white dwarf expands and engulfs the brown dwarf. At the same time, the brown dwarf also truncates the red giant's evolution by accelerating expelling the stellar envelope. These interactions exert a drag force onto the brown dwarf, exhaust its kinetic energy, bring it down to a lower orbit. Eventually, as the host evolves into a white dwarf, the binary ends up with a tight orbit (on the order of $10^{-2}$~AU; e.g., \citealt{Maxted2006}). The strong tidal force quickly locks the brown dwarf into synchronous rotation with periods as short as ${\sim}1.1$~\si{\hour} \citep[][]{Casewell2018}. Despite the fact that these brown dwarfs are more massive and formed differently than giant planets, their atmospheres resemble  planetary ones in their compositions, equation of state, and effective temperatures. Therefore, they are interesting targets for detailed investigations of highly irradiated substellar atmospheres.

As a result of their evolution, these post-common-envelope brown dwarfs have three properties that help uniquely constrain substellar atmospheric models: First, due to their close orbits, these brown dwarfs receive approximately the same intensity of irradiation as hot and even ultra-hot Jupiters. Second, white dwarf hosts have much bluer spectra than main sequence stars that host hot Jupiters. As a result, the peak of the irradiating spectrum is in the blue optical or even ultraviolet (UV) wavelengths.  Atoms in the upper atmospheres can more easily absorb the short-wavelength irradiation, which results in temperature inversions as well as ionized hydrogen atoms.  Third, the rotation rates of the brown dwarfs are more than one order of magnitude faster than those of typical tidally locked hot Jupiters. The rapid rotation introduces strong Coriolis force, which can shape the equatorial jets and reduce day/night heat transfer efficiency \citep{Tan2020a,Lee2020}. From an observational point of view, because of the moderate companion-to-host flux contrasts (${\sim}10$ in the infrared, compared to $>10^{2}$ that is typical to hot Jupiters), the phase curves of these brown dwarfs have superior signal-to-noise ratios (SNRs)  than those of hot Jupiters, enabling more strict tests of atmospheric theories. 


The atmospheric properties of these highly irradiated brown dwarfs have been investigated through state-of-the art models. \citet{Tan2020a} investigated the impact of fast rotation rates of these brown dwarfs by comparing GCM simulations of multiple rotation periods. \citet{Lee2020} combined GCM and radiative transfer models for the brown dwarf WD~0137-349B and compared its predicted and observed thermal phase curves. Both studies found that the fast rotation rates significantly decrease the meridional width of the equatorial jets and suppress the day-to-night heat transfer. The day-side hot spot is located near the substellar point, unlike the typically eastward-shifted hot spots in hot Jupiter atmospheres. These works predicted that these white dwarf-irradiated brown dwarfs have greater day/night temperature contrasts than  hot Jupiters.

\citet{Lothringer2020} calculated the spectra from the irradiated hemispheres of two brown dwarfs, WD~0137-349B and EPIC~212235321B. They created models based on the PHOENIX code and carefully incorporated the effect of strong UV irradiation in deriving the emission spectra of the brown dwarfs. They found that the irradiation efficiently heats the middle and upper atmospheres of the brown dwarf, which in turn breaks molecules such as H$_{2}$O and H$_{2}$ through thermal dissociation. Despite the fact that the cooler brown dwarf (WD~0137-349B) of the two has an equilibrium temperature more similar to typical hot Jupiters, its day-side spectrum resembles ultra-hot Jupiters. Both brown dwarfs are predicted to have weak or no molecular absorptions on their day-sides. These model predictions can be tested by phase-resolved spectra.

We present Hubble Space Telescope (\textsl{HST}) Wide Field Camera 3 (WFC3) observations of two WD-BD binary systems WD~0137-349 (hereafter WD~0137 \footnote{A note on our nomenclature: we refer the white dwarf host as WD~0137A, the brown dwarf companion as WD~0137B, and the binary system as WD~0137. The same scheme goes for EPIC~2122A, EPIC~2122B, and EPIC~2122.}) and EPIC~212235321 (hereafter EPIC~2122). Both systems are detached binaries with no on-going mass exchange or accretion, and hence their phase curves only trace the atmospheres of individual components. The irradiated brown dwarfs in these two binaries are in an unexplored parameter space with respect to their irradiation, temperature, surface gravity, and rotation rates (Figure~\ref{fig:exopop}). Brown dwarf WD~0137B is a $M=55 M_{\mathrm{Jup}}$ companion to its $M=0.39\,M_{\mathrm{\odot}}$, $\teff=16,500$~\si{\kelvin} white dwarf host at a separation of 0.65 $R_{\odot}$ (0.003 AU) \citep{Maxted2006}. Because of the tiny orbital semi-major axis, the brown dwarf is tidally locked in a short orbital/rotational period of 116~\si{min} and has a radiative equilibrium temperature of \SI{1990}{\kelvin}. Constant intense irradiation received by the brown dwarf's day-side hemisphere establish a radical temperature difference between its day- and night-side. This day/night contrast is best presented by the large amplitude rotational modulations in its optical ($V/R/I$ bands), near-infrared ($J/H/K$ bands), and  mid-infrared (Spitzer/IRAC Channels 1 to 4) light curves, from which \citet{Casewell2015} found that WD~0137B's day-side brightness temperatures decreased from \SIrange{2400}{1100}{\kelvin} from $J$-band to Spitzer Channel 4 (\SI{8}{\micron}) and night-side brightness temperatures decrease from \SIrange{2100}{400}{\kelvin} within the same wavelength ranges. The day/night brightness temperature differences range from \SIrange{400}{800}{\kelvin}.

The ${\sim}58\,M_{\mathrm{Jup}}$ brown dwarf EPIC~2122B \citep{Casewell2018} is in an even tighter orbit (0.44 $R_{\odot}$ (0.002 AU) semi-major axis) around its $M=0.47\,M_{\odot}$, $\teff=24,900$~\si{\kelvin} white dwarf host. EPIC~2122B's has the shortest orbital/rotational period among non-interacting companions: \SI{68.21}{\minute}. Its equilibrium temperature reaches \SI{3435}{\kelvin}, making it the second hottest irradiated substellar object after KELT-9b \citep{Gaudi2017} and a distinctive target for atmospheric studies. Its extreme day/night contrast has been demonstrated by the enormous 17\% semi-amplitude in its $i$-band light curve \citep{Casewell2018}.

The spectroscopic time series of these two WD-BD systems deliver rich information about the brown dwarfs' fundamental atmospheric properties including circulation patterns and day/night temperature contrasts. In this paper, we present a detailed analysis of the complete phase-resolved spectra of these two binaries. In Section~\ref{sec:obs}, we describe the HST observations and time-resolved spectroscopic data reduction procedures; in Section~\ref{sec:LC}, we analyze and interpret the broadband phase curves; in Section~\ref{sec:spec}, we present the phase-resolved spectra of the two irradiated brown dwarfs and compare them with various models; in Section~\ref{sec:discussion}, we discuss the observations of WD~0137B and EPIC~2122B in the context of irradiated planetary atmospheres and brown dwarf evolution; finally in Section~\ref{sec:conclusions}, we conclude and summarize our results.

\section{Observations and Data Reduction}
\label{sec:obs}

\begin{figure*}[!t]
  \centering
  \includegraphics[width=.83\textwidth]{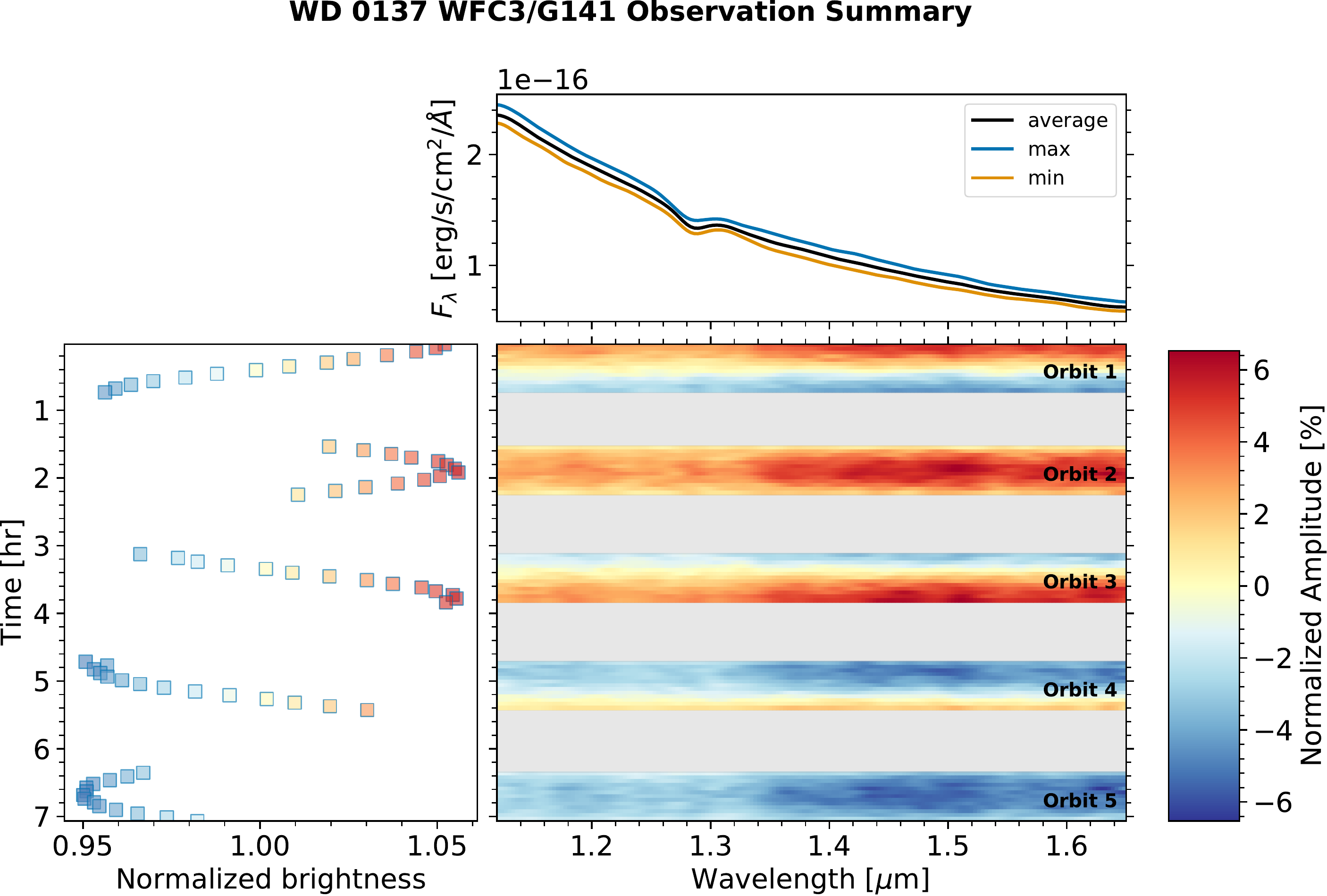}\\
  \vspace{.4em}
  \includegraphics[width=.83\textwidth]{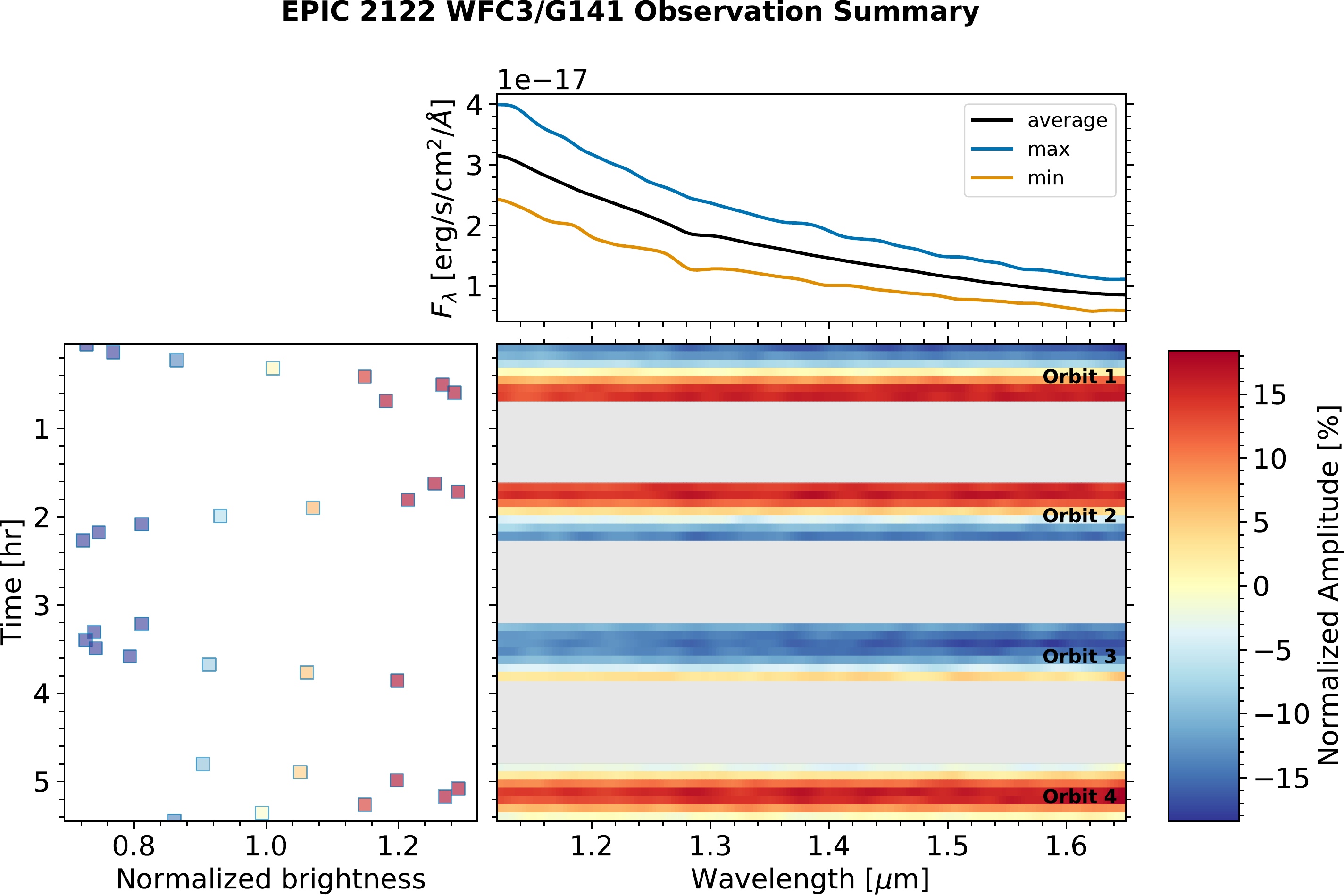}
  \caption{Summary of the time-resolved WFC3/G141 spectroscopic observational results for WD~0137 (upper) and EPIC~2122 (lower). In both figures, the top sub-plots show three representative spectra: the brightest (blue), the faintest (orange), and the average (black). The left sub-plots show the normalized \SIrange{1.12}{1.65}{\micro\meter} integrated phase curves. Our observations captured 4.1 and 5.6 periods of WD~0137 and EPIC~2122, respectively. The color of the points also represent the normalized broadband brightness. The lower right sub-plots show the time-resolved spectra. The color map demonstrates the flux density relative to the mean value of the same wavelength --- deep red/blue colors corresponding to high modulation amplitudes in that wavelength.}
  \label{fig:spectralsequence}
\end{figure*}

Observations of the WD-BD binary systems WD~0137 and EPIC~2122  are part of the Hubble Space Telescope (HST) program GO-15947\footnote{Details of the observation information can be found at \url{https://www.stsci.edu/cgi-bin/get-proposal-info?observatory=HST&id=15947}} (PI: Apai), carried out by the Wide Field Camera 3 infrared Channel (WFC3/IR). WD~0137 was monitored over five consecutive HST orbits from UTC 2020-06-20 18:46:25 to 2020-06-21 01:50:02. Each orbit began with one 29.6~s direct-imaging exposure in the F132N filter. We used this image to identify the target's position on the detector and establish the wavelength solution \citep{Pirzkal2016}. Fourteen 179~s G141 spectroscopic exposures were taken immediately after. In total, seventy spectra were collected, forming a 480 min long spectral time series that covered 4.1 periods of the binary.

EPIC~2122 was observed in a similar set-up over four consecutive HST orbits from UTC 2020-05-06 18:09:08 to 2020-05-06 23:35:11. Because EPIC~2122 is fainter than WD~0137, the direct-imaging frames were taken in the medium band F127M filter to secure sufficient photons for wavelength calibration. In total, thirty-two 313~s G141 exposures were obtained, constituting a  384~min time series that covered 5.6 orbital periods of EPIC~2122.

We downloaded the \texttt{CalWFC3} pipeline product \texttt{flt} files and the reduced the data using a pipeline based on the WFC3/IR spectroscopic software aXe \citep{Kummel2009}. This pipeline has been widely used in time-resolved observations of brown dwarfs \citep[e.g.,][]{Buenzli2012, Apai2013}. A four pixels radius window was adopted to extract the spectra and calculate the associated uncertainties. The G141 spectral resolution was $R\sim130$ at \SI{1.4}{\micro\meter}, corresponding to a velocity resolution of $\Delta v=\SI{2.3e3}{\kilo\meter\per\second}$. Because this resolution is significantly coarser than the maximum radial velocities (RV) of both systems (\SI{210}{\kilo\meter\per\second} for WD~0137B, \citealt{Longstaff2017}, and \SI{308}{\kilo\meter\per\second} for EPIC~2122B, \citealt{Casewell2018}), the RV variations can be safely neglected in wavelength calibration. We then conducted aperture and `ramp effect'' corrections. The aperture correction coefficients were derived by linearly interpolating the look-up table in \citet{Kuntschner2009}.  We modeled and corrected for the detector charge-trapping related ramp effect systematics using the physically motivated RECTE model \citep{Zhou2017}, which calculated the ramp profile based on two free parameters representing numbers of two populations of trapped charges at the beginning to the observations. We simultaneously fit the RECTE model and a sine wave to the \SIrange{1.12}{1.65}{\micro\meter} broadband phase curves, and then divided the optimal  ramp model from each spectrum.  Figure \ref{fig:spectralsequence} shows summaries of the data reduction results. Both targets demonstrate high-amplitude and wavelength-dependent periodic modulations. 

\section{Light Curve Analyses}
\label{sec:LC}
\subsection{The Broadband Light Curves}

\begin{figure*}[!t]
  \centering
  \includegraphics[height=0.35\textwidth]{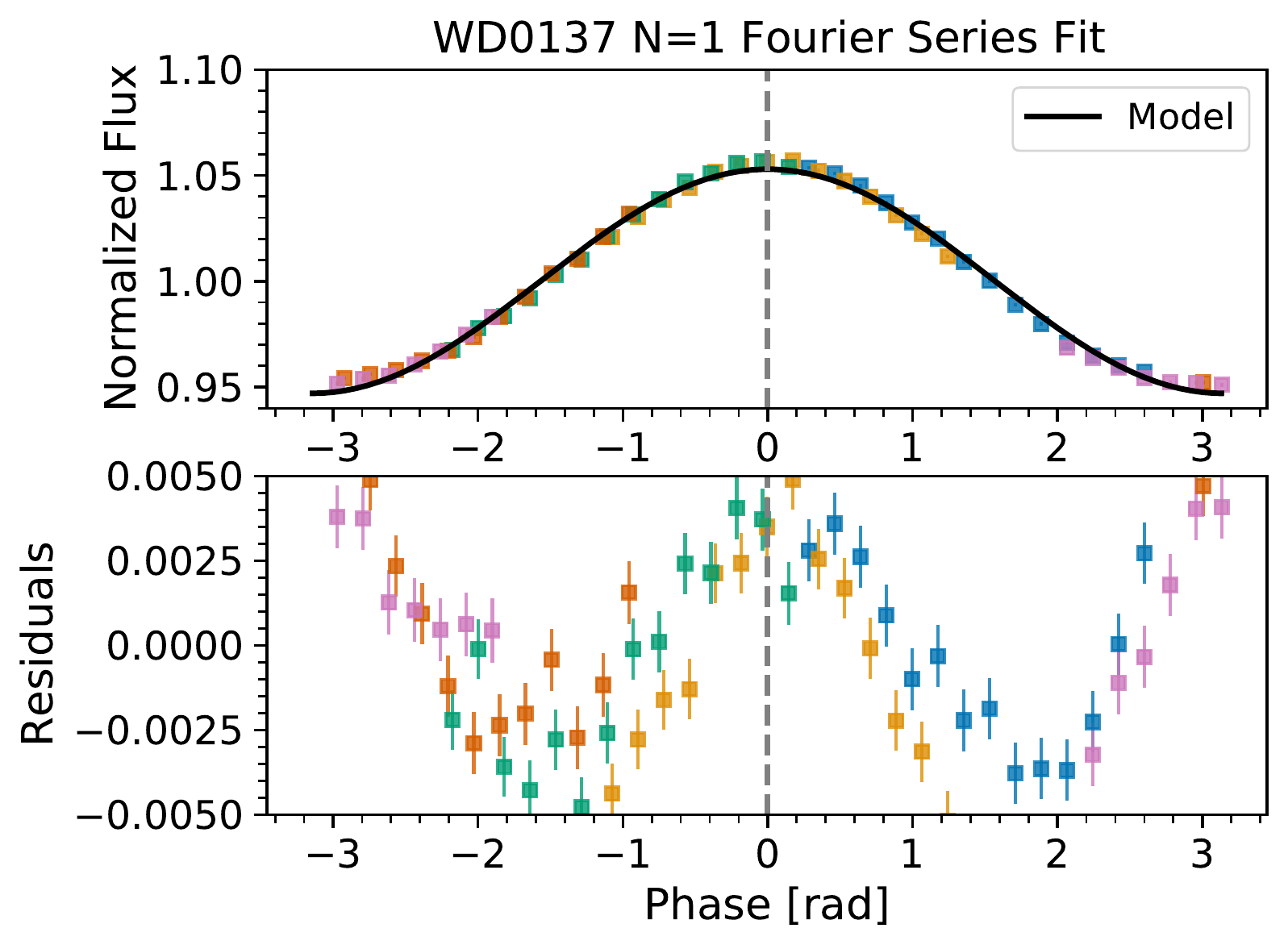}
  \includegraphics[height=0.35\textwidth]{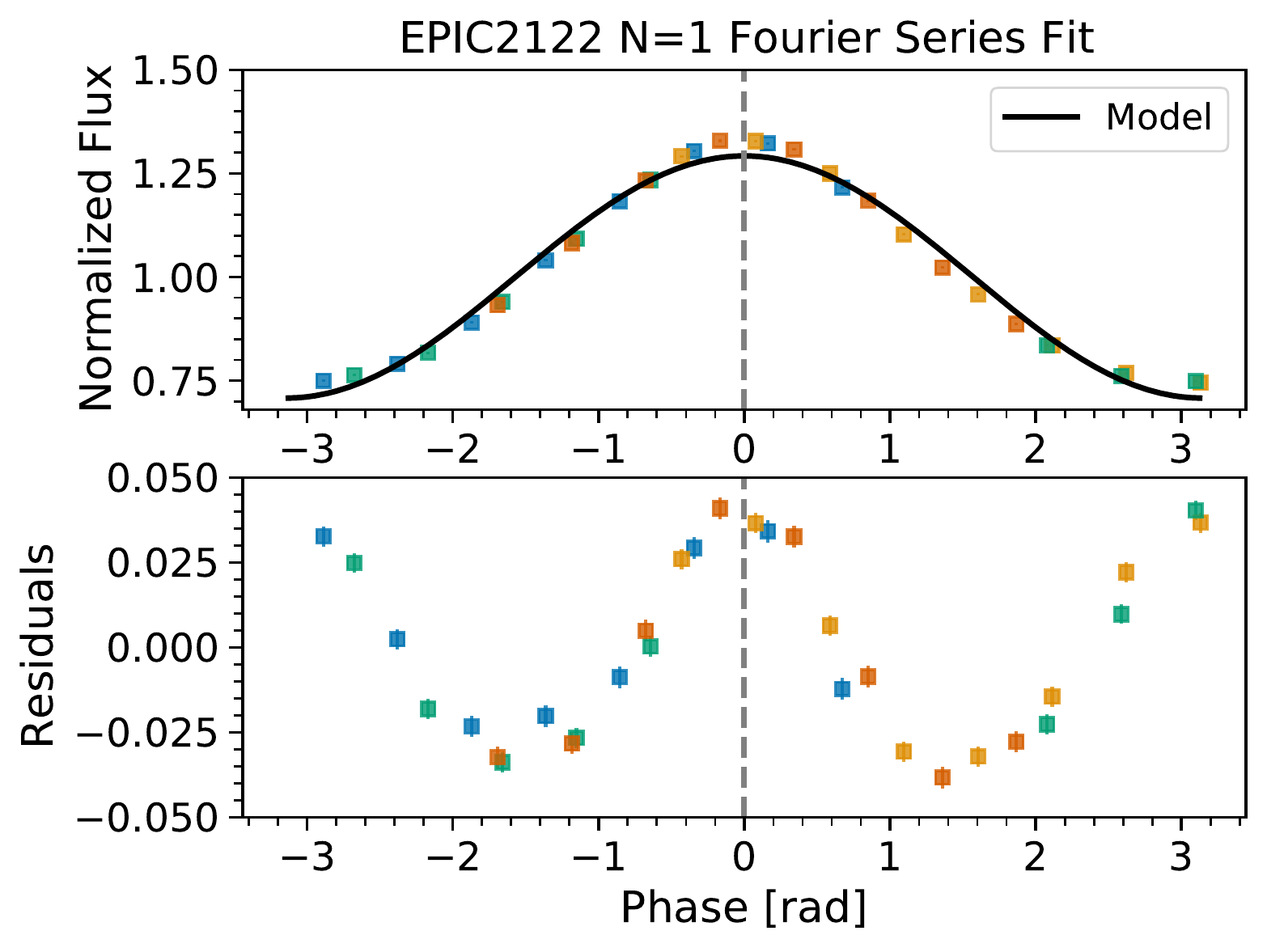}
  \includegraphics[height=0.35\textwidth]{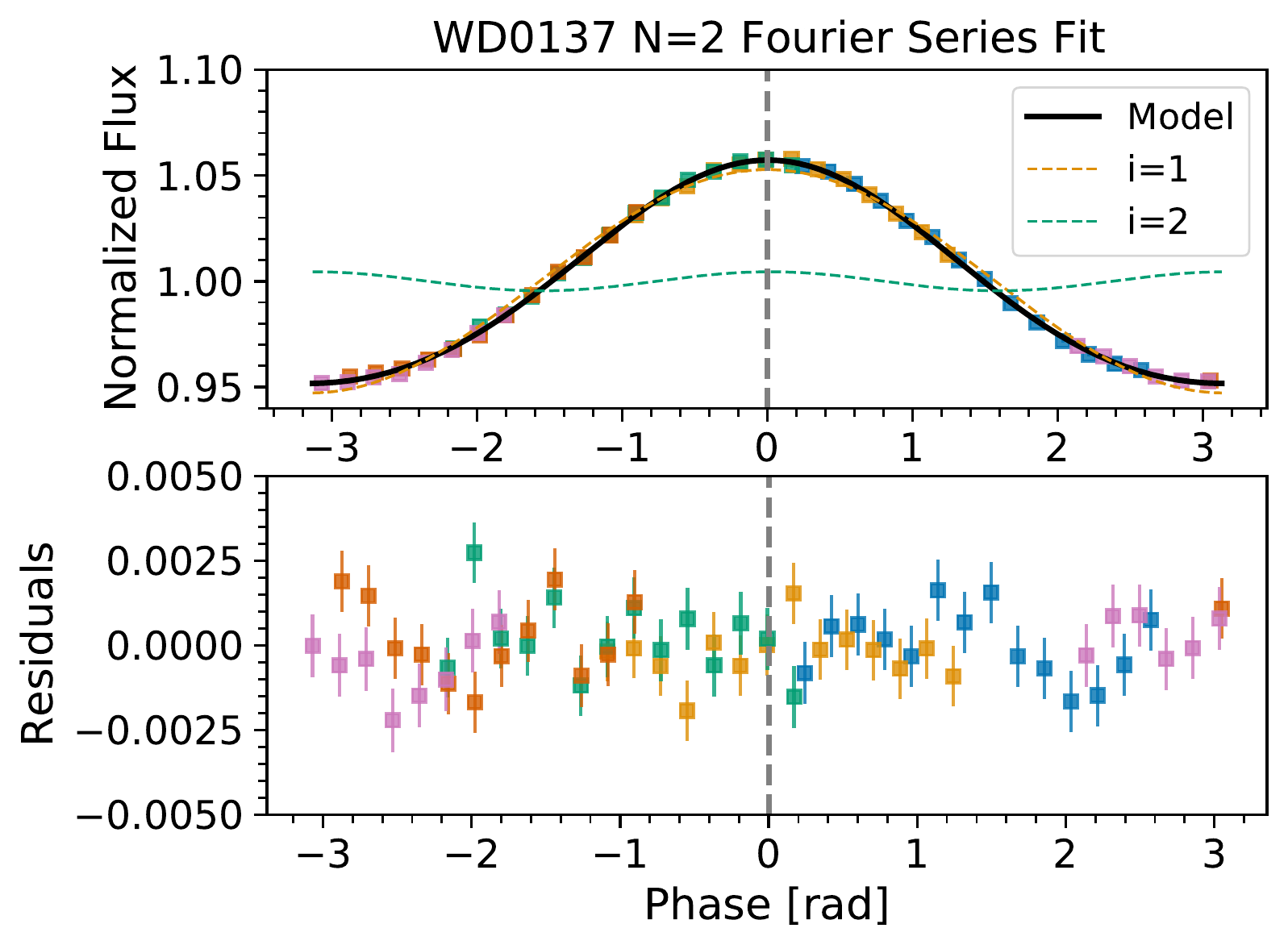}
  \includegraphics[height=0.35\textwidth]{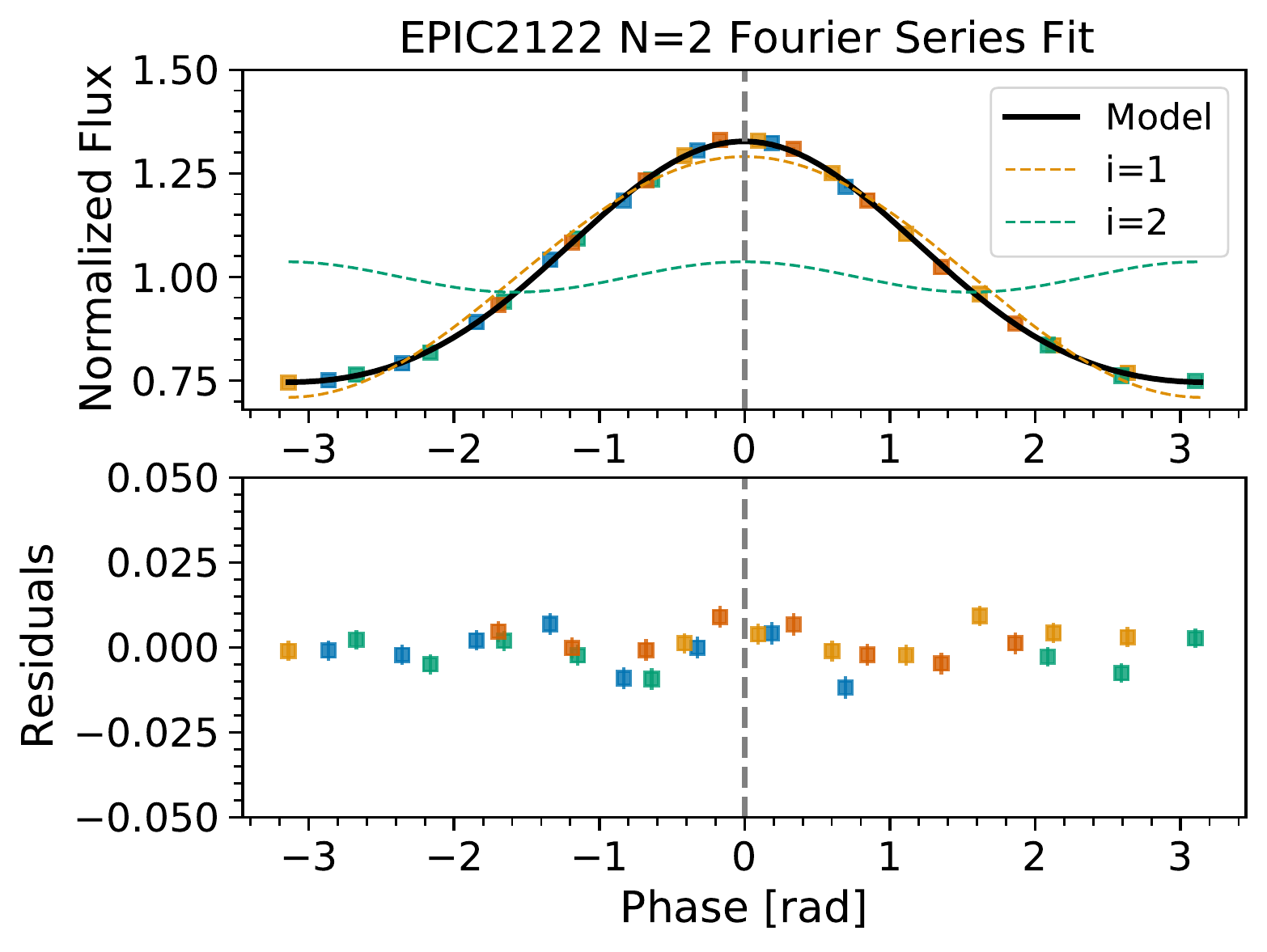}
  \includegraphics[height=0.35\textwidth]{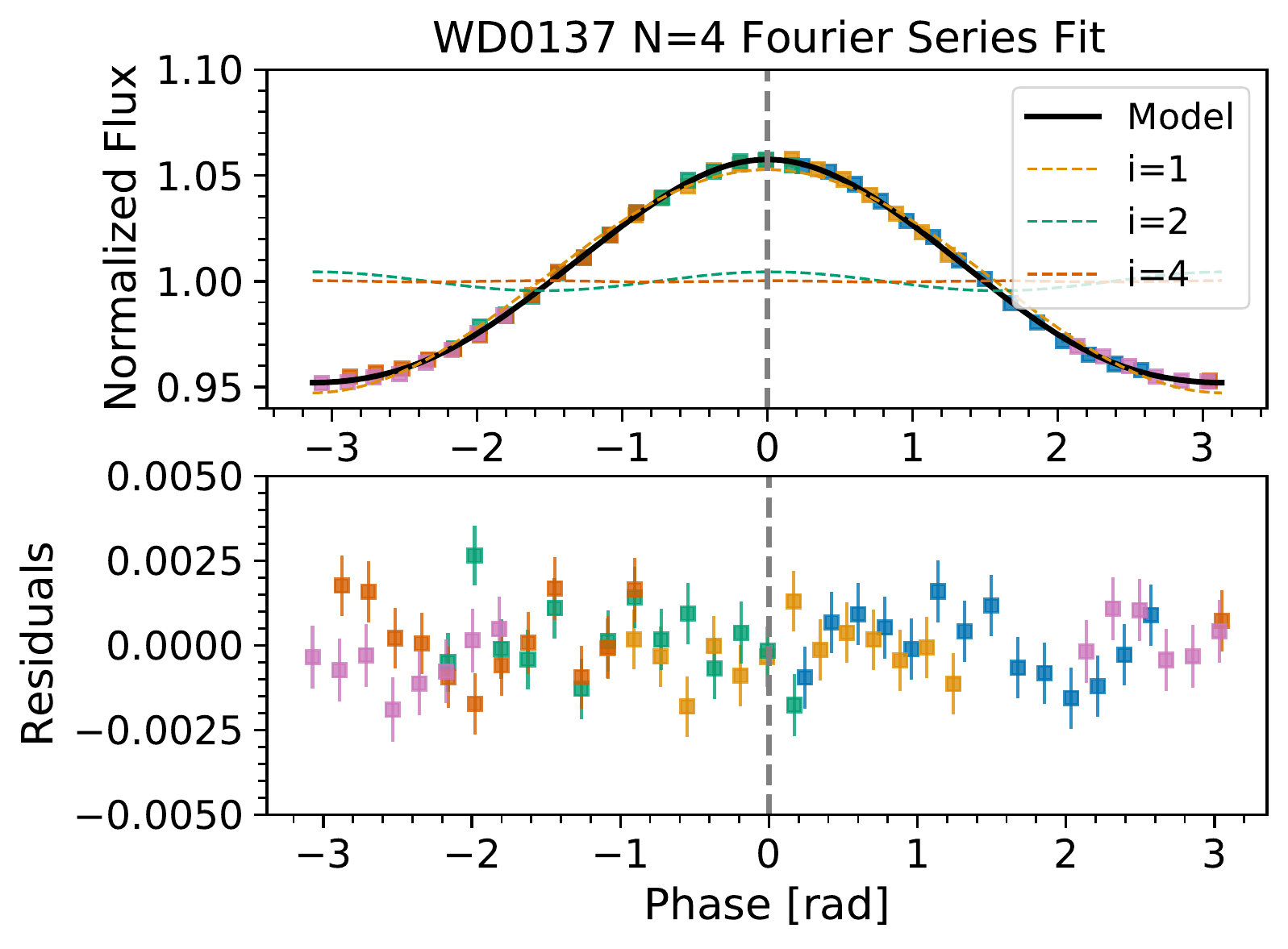}
  \includegraphics[height=0.35\textwidth]{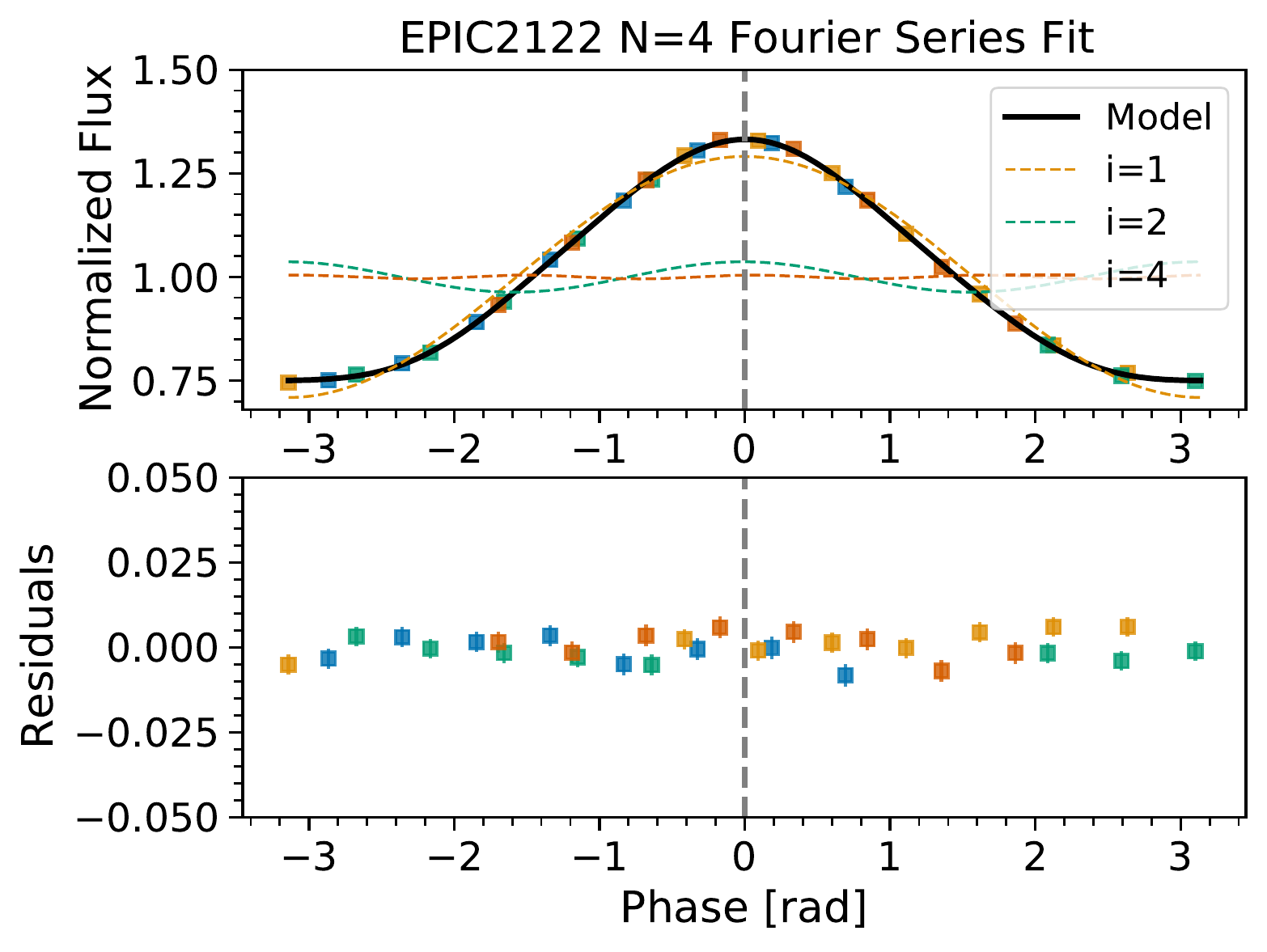}
  \includegraphics[width=0.65\textwidth]{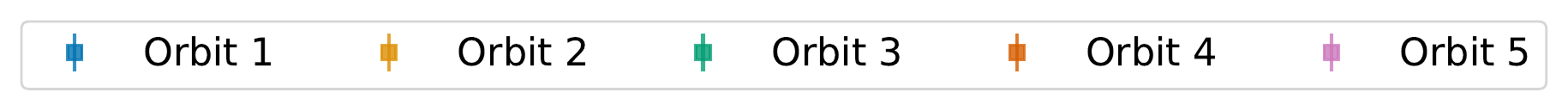}
  \caption{Fitting Fourier series models of various maximum orders (upper: $N=1$, middle: $N=2$, lower $N=4$) to the light curves of WD0137 (left) and EPIC~2122 (right). The light curves are phase-folded to their best-fitting periods (1.917 hr for WD~0137 and 1.13368 hr for EPIC~2122) and shifted so that the $k=1$ wave peak is at zero phase. Squares in different colors show broadband photometry in different HST orbits. The black solid lines show the best-fitting Fourier series models and the color-dashed lines show the model components. The $N=1$ model is insufficient for either target and results in similar residual patterns that peak at $\pm\pi$ and 0 phases. The $N=2$ and $N=4$ models have much better fits in both cases. Based on BIC, $N=2$ model is the most favored one for WD~0137 and $N=4$ is the best one for EPIC~2122.}
  \label{fig:lcfit}
\end{figure*}

We first determine the system period and modulation amplitude by fitting Fourier series models to the normalized broadband (\SIrange{1.12}{1.65}{\micro\meter}) light curves.
The Fourier series is defined by Equation \ref{eq:1}
\begin{equation}
  \label{eq:1}
  \mathcal{F} = 1 + \sum_{k} a_{k}\sin(2\pi t / (P/k)) + b_{k}\cos(2\pi t/ (P/k)),
\end{equation}
in which $P$ is the period, $a_{k}$ and $b_{k}$ are the $k$th order coefficients of the $\sin$ and $\cos$ terms. They are related to the semi-amplitude (amp) and the phase offset ($\phi$) by
\begin{align}
  \label{eq:3}
  &\mathrm{amp}_{k} = \sqrt{a_{k}^{2} + b_{k}^{2}}\\
  &\phi_{k}=\arctan(b_{k}/a_{k}).
\end{align}
In a hemispherically integrated light curve, amplitudes quickly diminish with an increasing $k$. Meanwhile, odd order (except $k=1$) signals are not present in the light curve because of symmetry  \citep{Cowan2008}. Therefore, we start the fit with the simplest model --- only including the base order ($k=1$) --- and gradually increase the model complexity by adding even order ($i=2k, k\ge1$) terms. In each iteration, the best-fitting $P$, $a_{k}$, and $b_{k}$ are determined by a least-$\chi^{2}$ criterion and the Bayesian Information Criterion (BIC) value is recorded. We use $\Delta\mathrm{BIC}_{N}=\mathrm{BIC}_{N}-\mathrm{BIC}_{1}$, the BIC difference between fitting the $k_{\mathrm{max}}=N$th order and the base order Fourier series, to select the best model: the one with the least $\Delta\mathrm{BIC}$.

The best-fitting values and the BICs are listed in Table~\ref{tab:lcfit} and the comparisons between the observed and best-fitting phase-folded light curves are presented in Figure~\ref{fig:lcfit}. The WD~0137 light curve favors the $N=2$ Fourier series  model, which has a best-fitting  period of $P=1.917\pm0.001$~\si{\hour}. This value is in the middle of  published RV measurements  ($1.927\pm0.005$~\si{\hour} in \citealt{Maxted2006}, \ $P=1.9063205(7)$~\si{\hour} in \citealt{Casewell2015}, and $P=1.906318536(2)$~\si{\hour} in \citealt{Longstaff2017}). The discrepancy between ours and the most precise \citet{Longstaff2017} period is 40 seconds, more than 10 times of the joint uncertainties. It may be attributed to the fact that the RV and phase curve observations trace different periodic motions (orbital motion of the system vs. atmospheric rotation). If WD~0137B's atmosphere contains a retrograde jet, the period measured by the phase curve can be slightly prolonged. The semi-amplitudes of the $k=1$ and $k=2$ waves are $5.28\pm0.02$\si{\percent} and $0.45\pm0.02$\si{\percent}, respectively. The two waves are almost in-phase with only a small offset of $0.12\pm0.04$\,\si{rad}. Based on the most recent and precise ephemeris \citep{Longstaff2017}, the phase difference between the light curve peak and substellar point is $\Delta \phi=0.01\pm0.03$\,\si{rad}, i.e., no significant phase shift is detected.

The light curve of EPIC~2122 favors a more complex $N=4$ Fourier series composed of $k$\,=\,1, 2, and 4, i.e., of three waves. The best-fitting period is $1.1368\pm00004$~\si{\hour}, consistent within $1\sigma$ with the much more precise K2 light curve period ($P=1.136869(1)$~\si{\hour}; \citealt{Casewell2018}). The semi-amplitudes of the $k=1, 2$, and 4 waves are $29.1\pm0.1$\si{\percent}, $3.7\pm0.1$\si{\percent}, and $0.5\pm0.1$\si{\percent}. Similar to what is observed in WD~0137, the waves of all three orders are nearly in-phase. The phase offsets of the $k=2$ and 4 waves relative to the $k=1$ wave are $0.07\pm0.03$~\si{rad} and $-0.1\pm0.2$~\si{rad}, respectively.

\begin{deluxetable*}{ccccccccc}
  \tablecolumns{8}
  \tablewidth{0pt}
  \tablecaption{Broadband Light Curve Fitting Results \label{tab:lcfit}}
  \tablehead{
    \colhead{$N$} & \colhead{$P$} & \colhead{$\mathrm{amp}_{1}$}  & \colhead{$\mathrm{amp}_{2}$} & \colhead{$\mathrm{amp}_{4}$} & \colhead{$\Delta\phi_{2}$\tablenotemark{a}}  & \colhead{$\Delta\phi_{4}$\tablenotemark{a}} & \colhead{$\Delta$BIC } & \colhead{Favored}\\
    \colhead{} & \colhead{[h]} & \colhead{[\%]}  & \colhead{[\%]} & \colhead{[\%]} & \colhead{[rad]} & \colhead{[rad]} & \colhead{} & \colhead{}}
  \startdata
  \multicolumn{9}{c}{WD~0137}\\
  1 & 1.928$\pm$0.003& 5.30$\pm$0.06 & - &- & - & - &0 & N\\
  2 & 1.917$\pm$0.001 & 5.28$\pm$0.02 & 0.45$\pm$0.02 &- & 0.12$\pm$0.04 & - &-140.1 & Y\\
  4 & 1.917$\pm$0.001 & 5.27$\pm$0.02 & 0.45$\pm$0.02 &0.04$\pm$0.02 & 0.14$\pm$0.04 & 0.7$\pm$0.5 &-135.8&N \\\hline
  \multicolumn{9}{c}{EPIC~2122}\\
  1 & 1.135$\pm$0.003 & 29.2$\pm$0.7 & - &- & - & - &0&N\\
  2 & 1.1368$\pm$0.0005 & 29.0$\pm$0.1 & 3.7$\pm$0.1 &- & 0.06$\pm$0.04 & - &-100.4&N\\
  4 & 1.1368$\pm$0.0004 & 29.1$\pm$0.1 & 3.7$\pm$0.1 &0.5$\pm$0.1 & 0.07$\pm$0.03 & -0.1$\pm$0.2 &-111.6&Y\\
  \enddata
  \tablenotetext{a}{$\Delta\phi_{2}$ and $\Delta\phi_{4}$ are the phase offsets of the $k=2$ and $k=4$ waves relative to the $k=1$ wave.}
\end{deluxetable*}

\subsection{A Comment on the Impact of Tidal Deformation}

It is predicted that tidal forces stretch both binary components into ellipsoids with major axes along the radial direction and introduce ellipsoidal oscillation --- brightness variation due to changes in the projected area \citep[e.g.,][]{Morris1985,Pfahl2008}. The projected area reaches its minimum at the $\phi=0,\,\pi$ phases and maximum at the $\phi=\pi/2,\,3/2\pi$ phases. The brightness variation amplitude will be proportional to the brown dwarf radius change due to tidal deformation \citep{Pfahl2008}:
\begin{equation}
  A\propto f_{\mathrm{BD}}\frac{\Delta R}{R} \sin^{2}i,
  \label{eqn:tidal}
\end{equation}
in which $i$ is the inclination and $f_{\mathrm{BD}}$ is the brown dwarf's flux relative to the white dwarf. In the phase curve, the ellipsoidal oscillation manifests as a $k=2$ wave that is in the opposite phase to those of our observed $k=2$ Fourier components. In a few transiting exoplanet and brown dwarf systems, the tidal deformation of the host stars have been identified in the phase curves \citep[e.g.,][]{Beatty2019,Wong2020}. In the same studies, the planetary deformation signals are diluted by the star/planet flux ratio and fall below the detection limits. In our cases, the  host white dwarfs are too compact to exhibit noticeable deformation. The brown dwarf companions, although are more resistant to deformation than hot Jupiters due to their higher surface gravities, suffer less from the flux diluting effect because of the lower white dwarfs' flux intensities relative to typical planets' host stars, and hence may present detectable ellipsoidal deformation signals.

Using the ROCHE code, which is a part of the LCURVE software \citep{Copperwheat2010}, we find the amounts of deformation of the WD~0137B and EPIC~2122B to be $\Delta R/R=6.6\%$ and $4.5\%$, respectively. $\Delta R/R$ is defined by the standard deviation of radii across all directions divided by the average radius. Inserting these numbers, as well as the inclinations and flux ratios ($i=35^{\circ},\, f_{\mathrm{p}}=7.02\%$ for WD~0137B; $i=56^{\circ},\,f_{\mathrm{p}}=4.5\%$ for EPIC~2122B) into Equation~\ref{eqn:tidal}, we derive the ellipsoidal oscillation amplitudes to be 0.15\% and 0.68\% for WD~0137B and EPIC~2122B, respectively. Compared to the Fourier series fitting results, these amplitudes are 33\% and 18\% of the $k=2$ wave amplitudes.

Two caveats in this estimate are worth noting. First, the observed $k=2$ signals and the ellipsoidal oscillations are in the opposite phases. Therefore, the tidal effect does not explain the high-order Fourier signals, but may slightly reduce the strength of the signal introduced by other effects (discussed later in \S\ref{sec:lc-interpretation}). Second, using Equation~\ref{eqn:tidal} to estimate the oscillation amplitude assumes that the ellipsoid has a homogeneous brightness distribution, which is certainly not the case for the highly-irradiated brown dwarfs. A complete model requires integrating across the heterogeneous ellipsoid, which requires precise knowledge of the surface brightness distribution and is beyond the scope of this paper. Because the exact ellipsoidal oscillation signals are likely insignificant, we will neglect this effect in our analysis.

\subsection{Spectroscopically-resolved Light Curves}

\begin{figure*}[!t]
  \centering
  \includegraphics[height=0.35\textwidth]{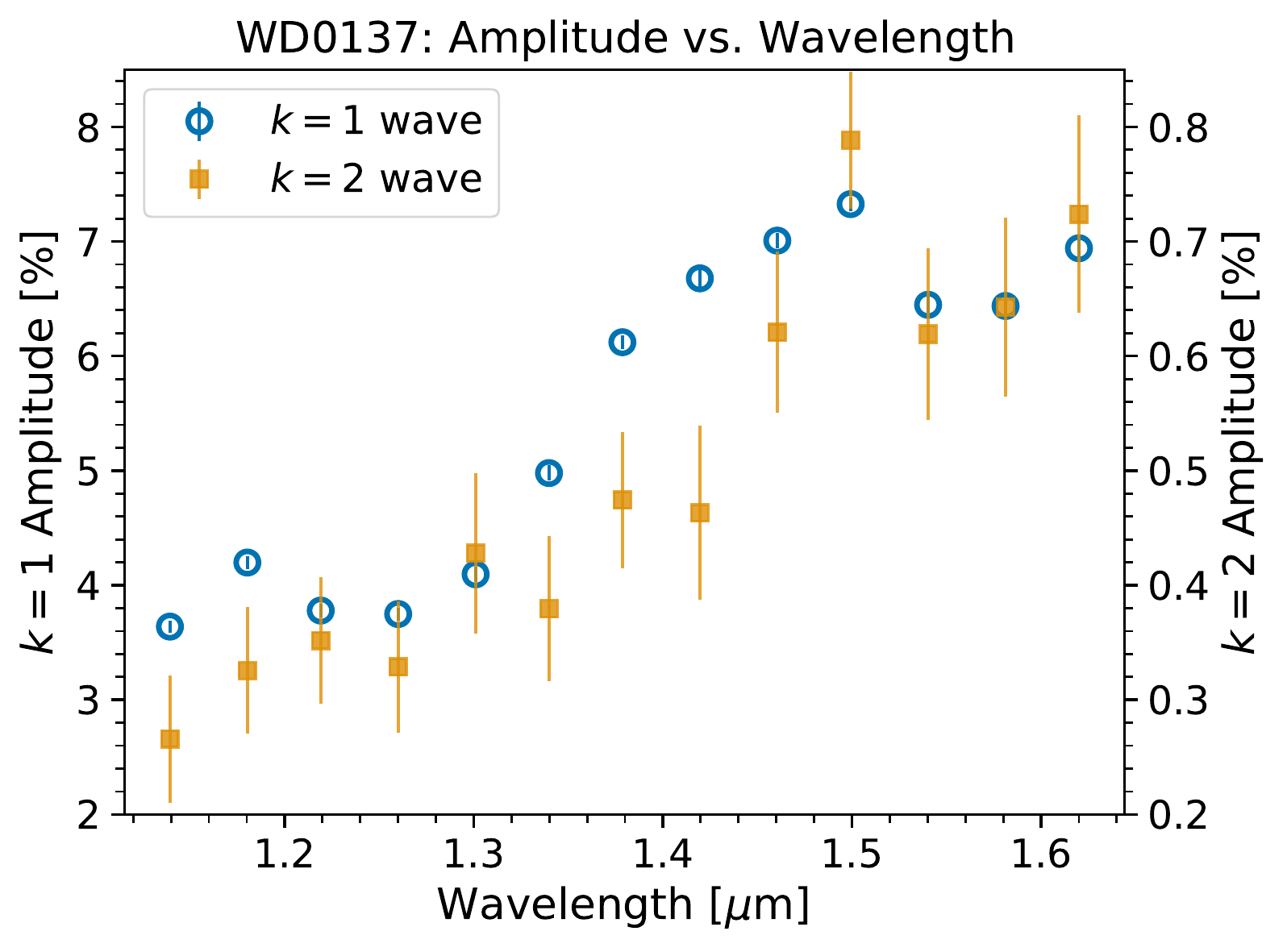}
  \includegraphics[height=0.35\textwidth]{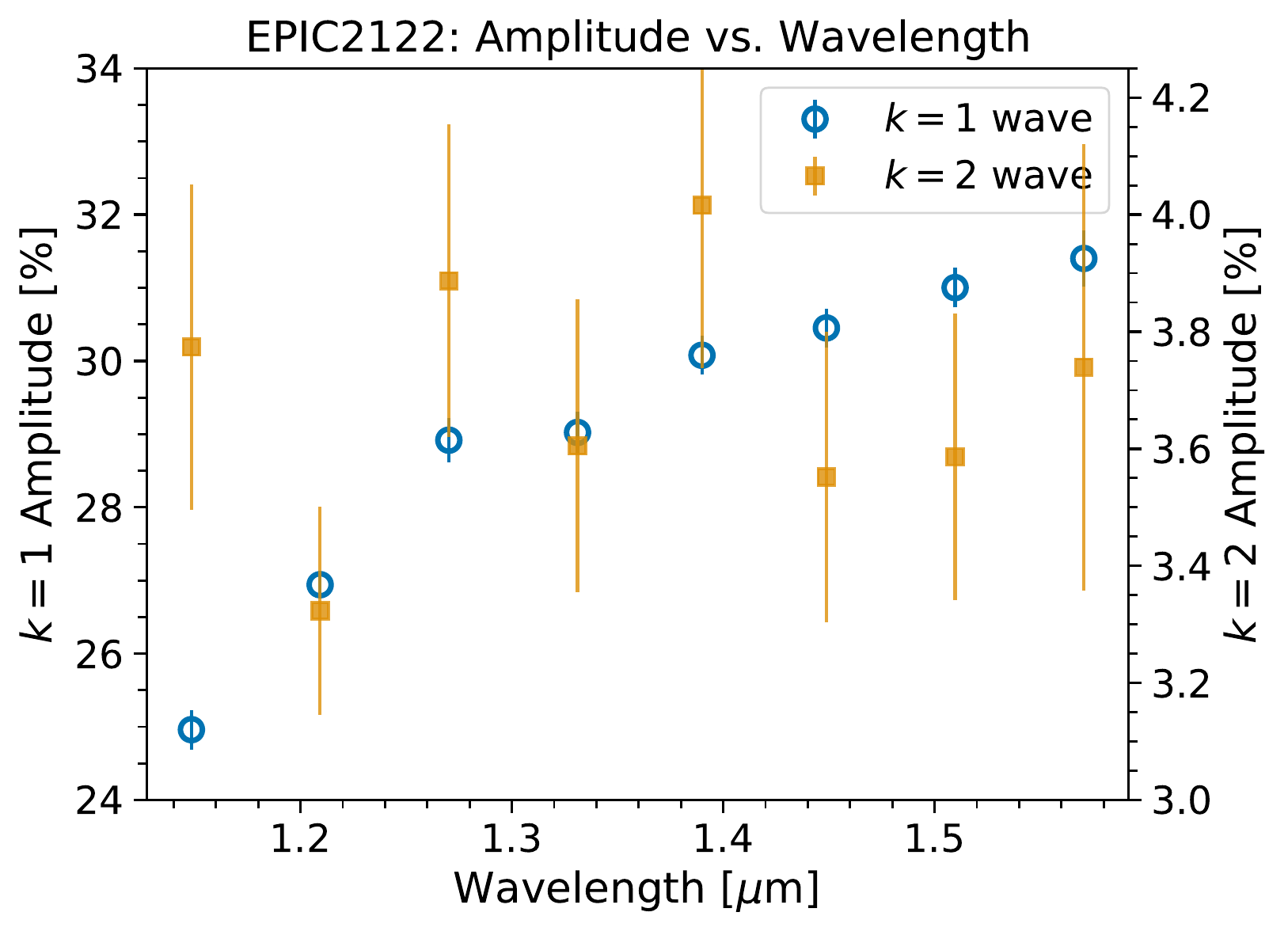}
  \includegraphics[height=0.38\textwidth]{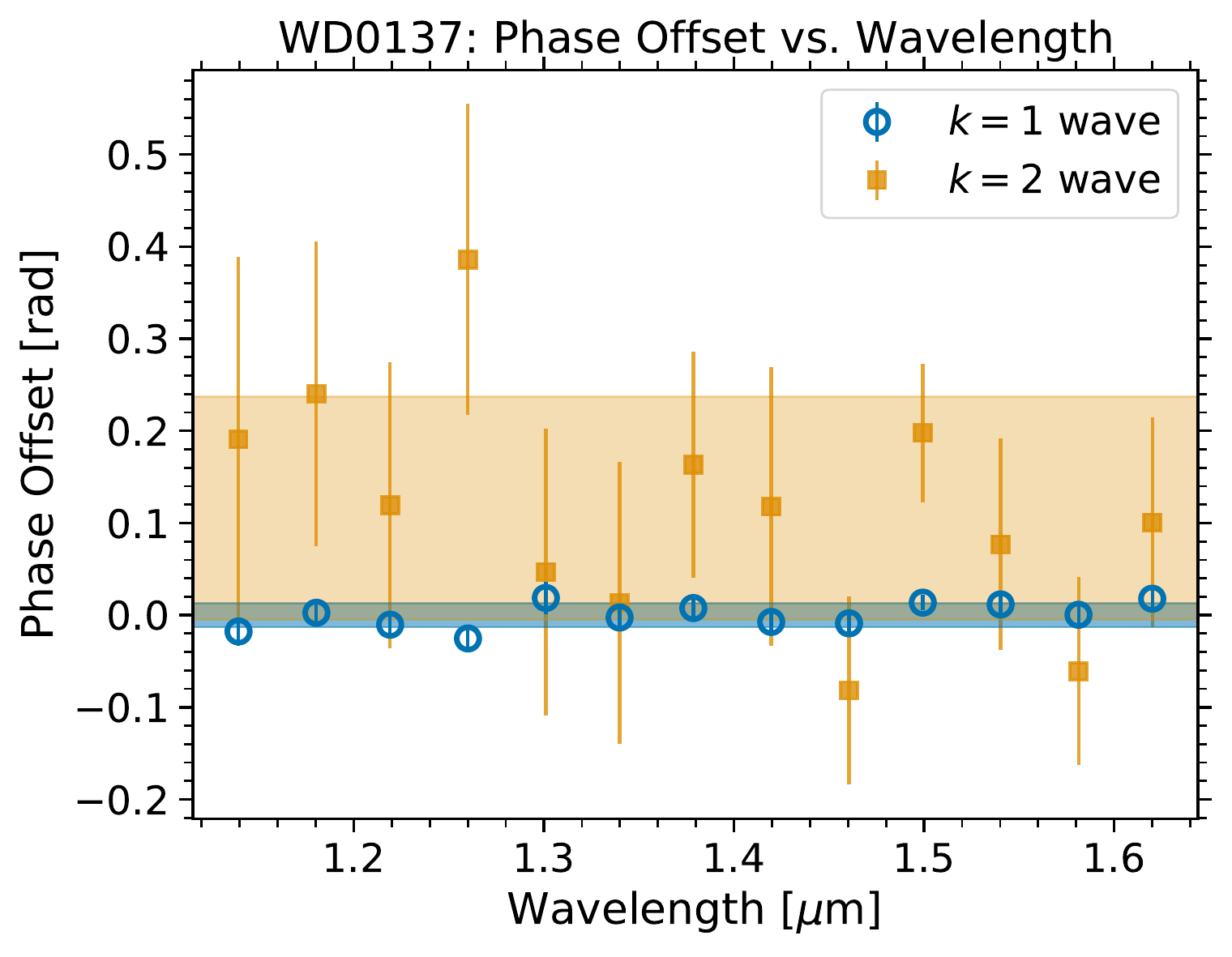}
  \includegraphics[height=0.38\textwidth]{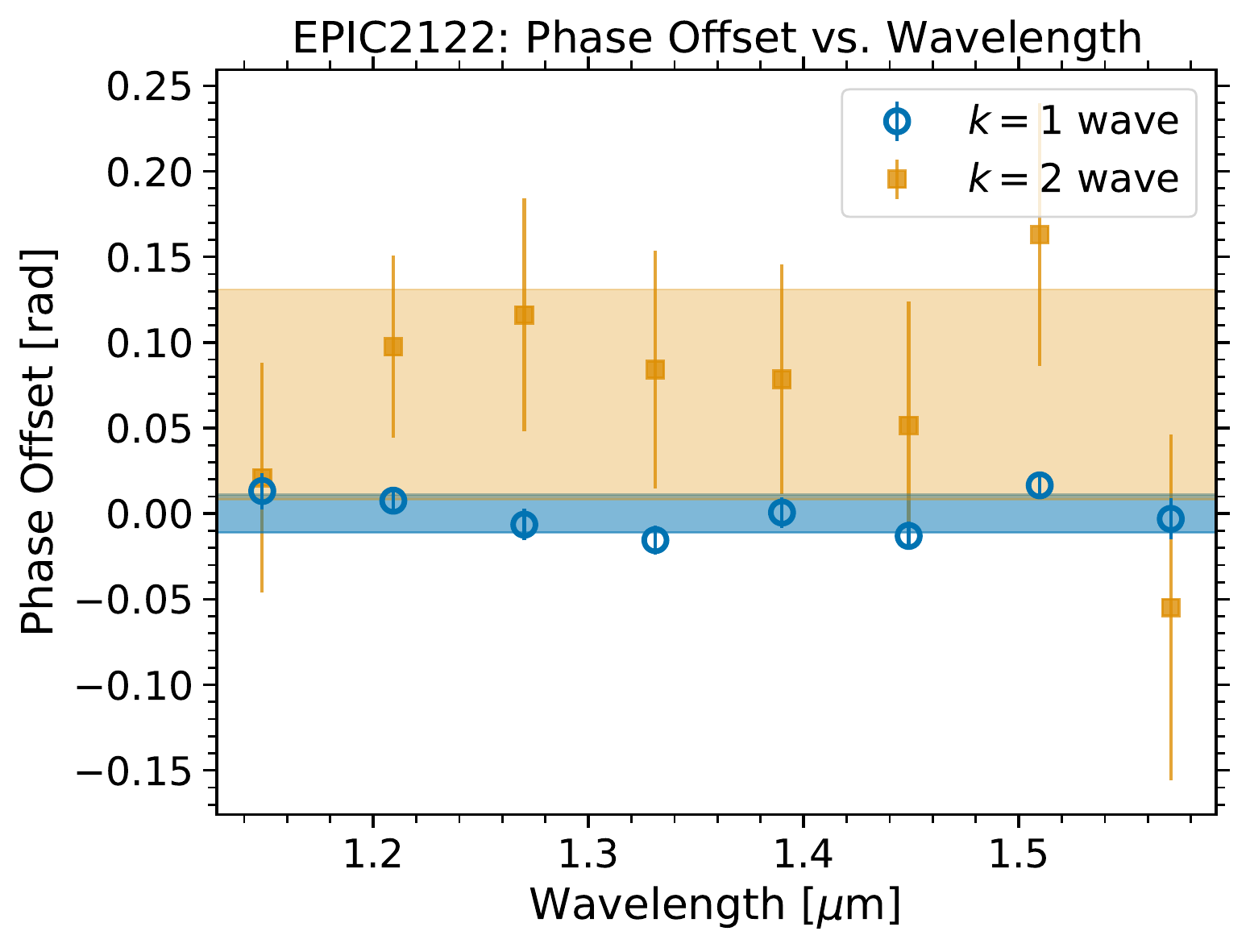}
  \caption{The amplitudes and phase offsets of the $k=1$ (blue circles) and $k=2$ (yellow squares) waves as functions of wavelength for WD~0137 (left) and EPIC~2122 (right). In the amplitude plots, the left and right axes show the scales for the $k=1$ and $k=2$ waves, respectively. For easy visual comparisons, the $k=2$ amplitudes are scaled up by $10\times$ and $8\times$ for WD~0137 and EPIC~2122, respectively. In the phase offset plots, the blue and yellow horizontal shades show the $\pm1\sigma$ ranges of the average phase offsets of the $k=1$ and $k=2$ waves, respectively.}
  \label{fig:spec_lc_fit}
\end{figure*}

To investigate the wavelength-dependence of the modulations, we fit the Fourier series to spectroscopically-resolved light curves. These light curves are integrated in wavelength bins, the sizes of which are determined based on the signal-to-noise ratios (SNR) of the observations. For the higher SNR WD~0137 data, the \SIrange{1.12}{1.65}{\micro\meter} bandpass is split into thirteen \SI{0.04}{\micro\meter} channels, resulting an average SNR of 160 per bin; for the lower SNR EPIC~2122 data, the bin size is \SI{0.06}{\micro\meter} and the number of channels is eight, resulting an average SNR of 120 per bin. The maximum order ($N$) and the period ($P$) of the Fourier series are fixed to the best-fitting values in the broadband light curve analyses ($N=2, P=\SI{1.917}{\hour}$ for WD~0137; $N=4, P=\SI{1.1368}{\hour}$ for EPIC~2122). Independent least-square fits are performed to each light curve to find their best-fitting semi-amplitudes and phase offsets.

Figure~\ref{fig:spec_lc_fit} shows the best-fitting semi-amplitudes and phases of the $k=1$ and $k=2$ waves as functions of wavelengths. In the WD~0137 case , the amplitudes of both $k=1$ and $k=2$ waves increase towards longer wavelengths. From \SIrange{1.12}{1.65}{\micro\meter},  $\mathrm{amp}_{1}$ and $\mathrm{amp}_{2}$  rise from \SI{3.6}{\percent}  to \SI{6.9}{\percent} and from \SI{0.26}{\percent} to \SI{0.72}{\percent}, respectively. These trends are primarily due to the fact that the brown dwarfs cause the modulations and they contribute more to the total emission flux at longer wavelengths. Remarkably, in the wavelength intervals from \SIrange{1.15}{1.20}{\micro\meter} and from \SIrange{1.35}{1.50}{\micro\meter}, which coincide with two water absorption bands, $\mathrm{amp}_{1}$ is significantly higher than what a linear trend would predict. In contrast, $\mathrm{amp}_{2}$ agrees better with a linear trend except in the \SI{1.50}{\micro\meter} channel where the amplitude is higher than the linear trend by $2\sigma$. The phase offset of neither Fourier component shows significant wavelength dependence. The $k=2$ wave is slightly ahead of the $k=1$ wave by $<0.2$~\si{rad} (${\sim}1\sigma$) in most channels, similar to what has been observed in the broadband light curve.

The amplitudes and phase offsets of EPIC~2122s spectroscopic modulations have a somewhat simpler wavelength dependence than those of WD~0137. The $k=1$ wave amplitude increases with wavelength nearly linearly, from \SI{24.9}{\percent} at \SI{1.14}{\micro\meter} to \SI{31.4}{\percent} at \SI{1.62}{\micro\meter}. $\mathrm{amp}_{2}$ stays almost constant at around \SI{3.6}{\percent}. The $k=4$ wave signals are too weak to be detected in the spectroscopically resolved phase curves and are hence left out in Figure~\ref{fig:spec_lc_fit}. Unlike WD~0137, neither Fourier component's amplitude varies in the water  absorption bands. For phase-offset --- similar to WD~0137 and to the broadband observations --- no significant spectroscopic variations are detected.

\subsection{Interpreting the Phase Curves: Mapping WD~0137B and EPIC~2122B}
\label{sec:lc-interpretation}

\begin{figure*}[t]
  \centering
  \includegraphics[height=0.38\textwidth]{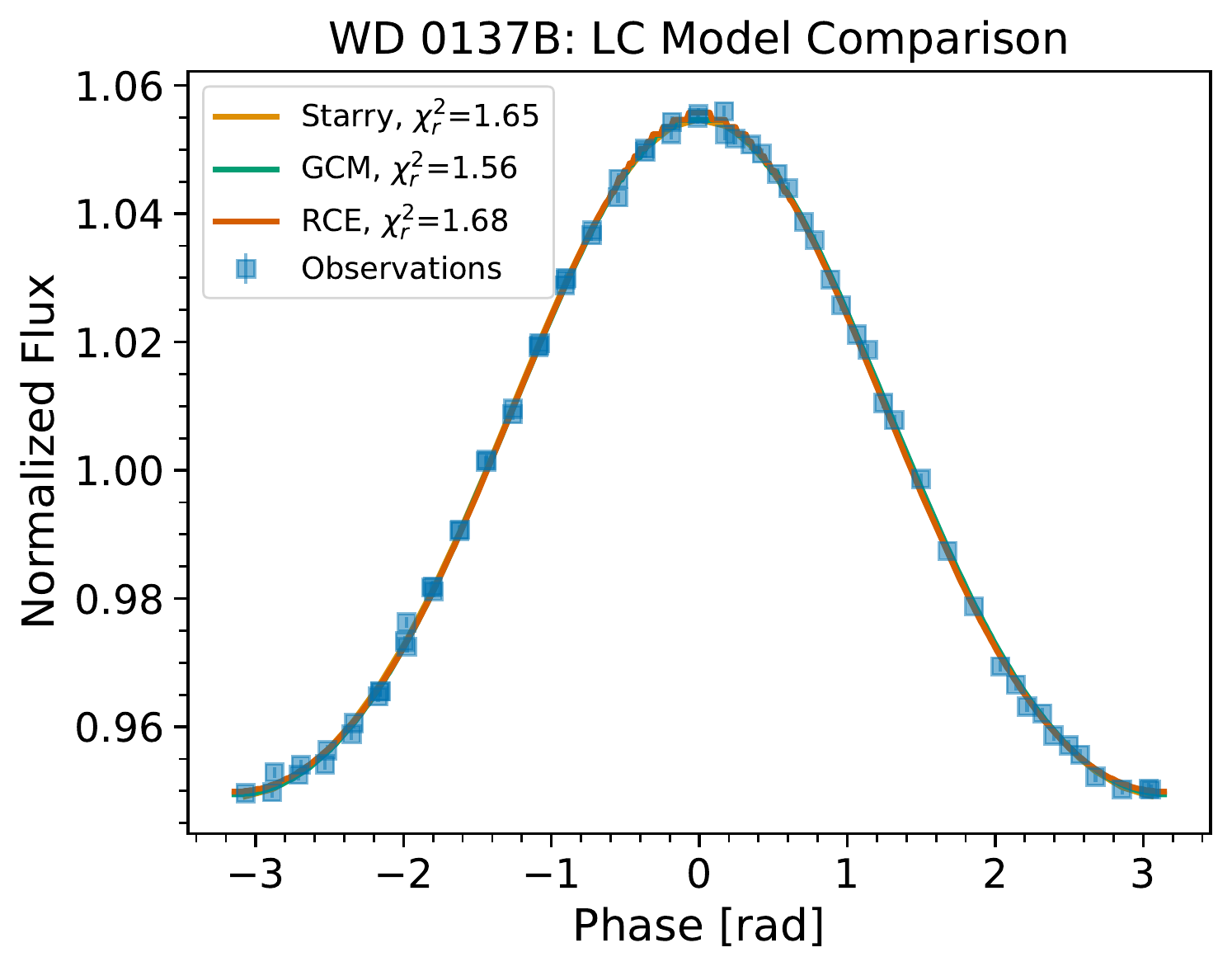}
  \includegraphics[height=0.38\textwidth]{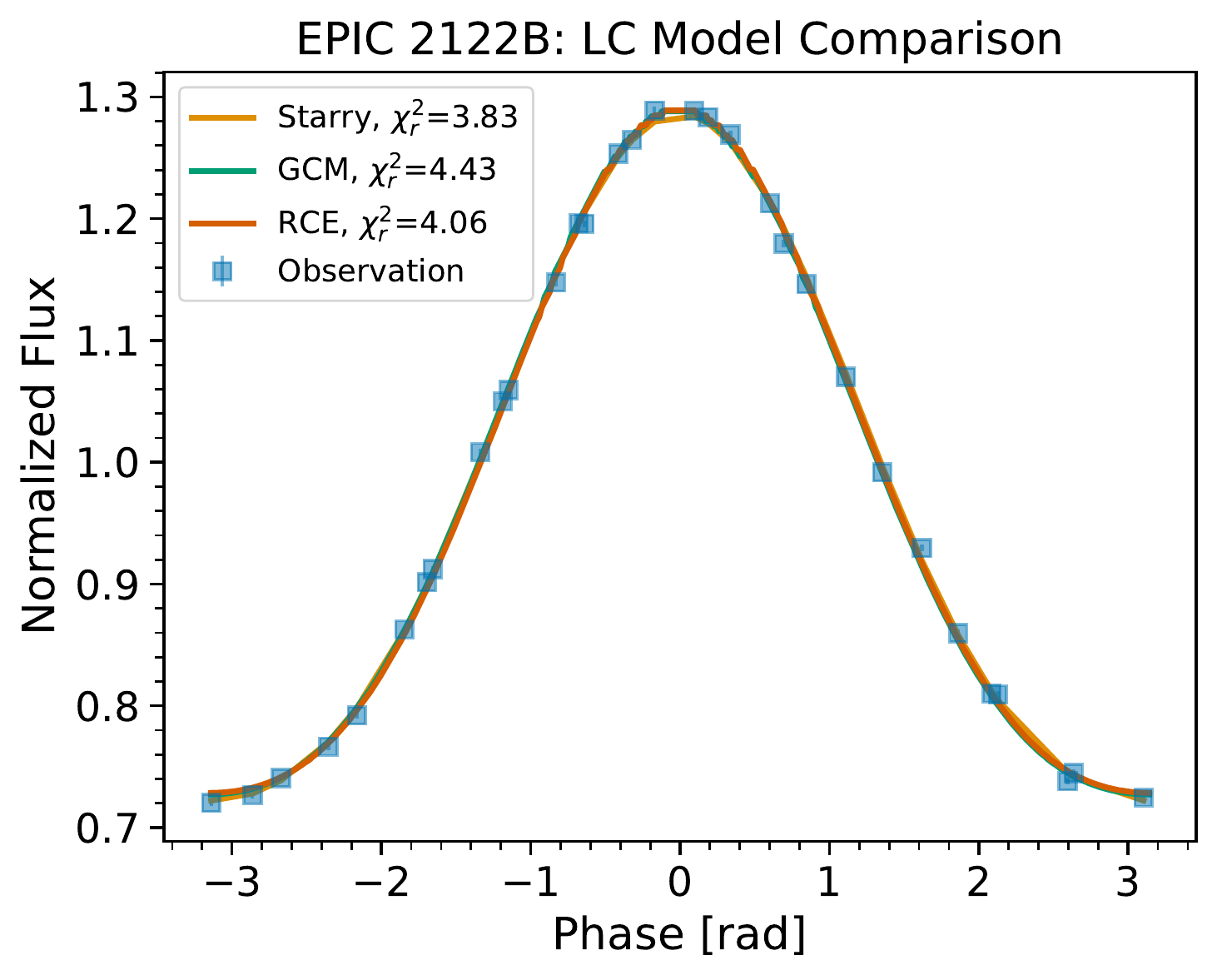}
  \includegraphics[width=0.48\textwidth]{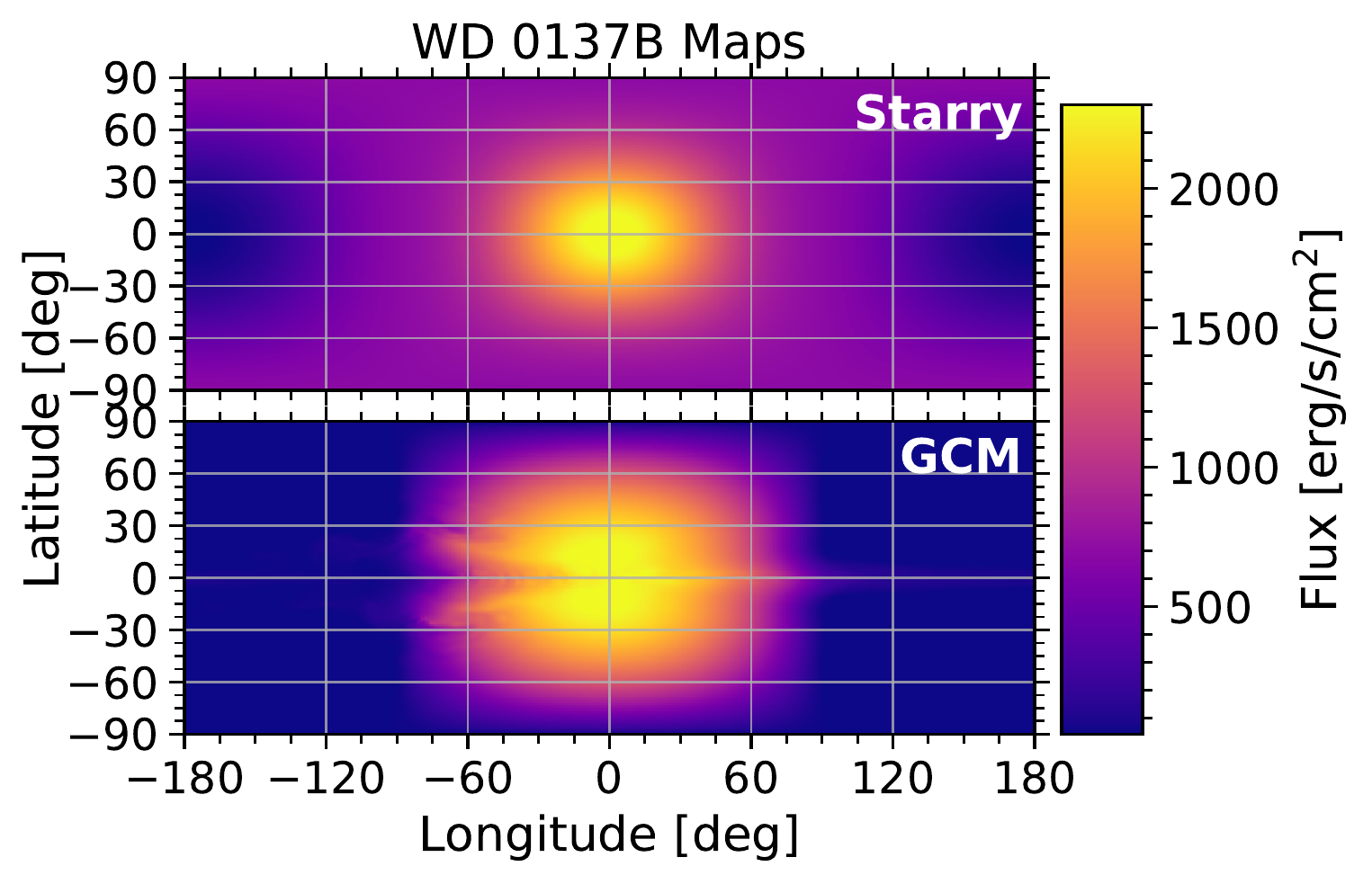}
  \includegraphics[width=0.48\textwidth]{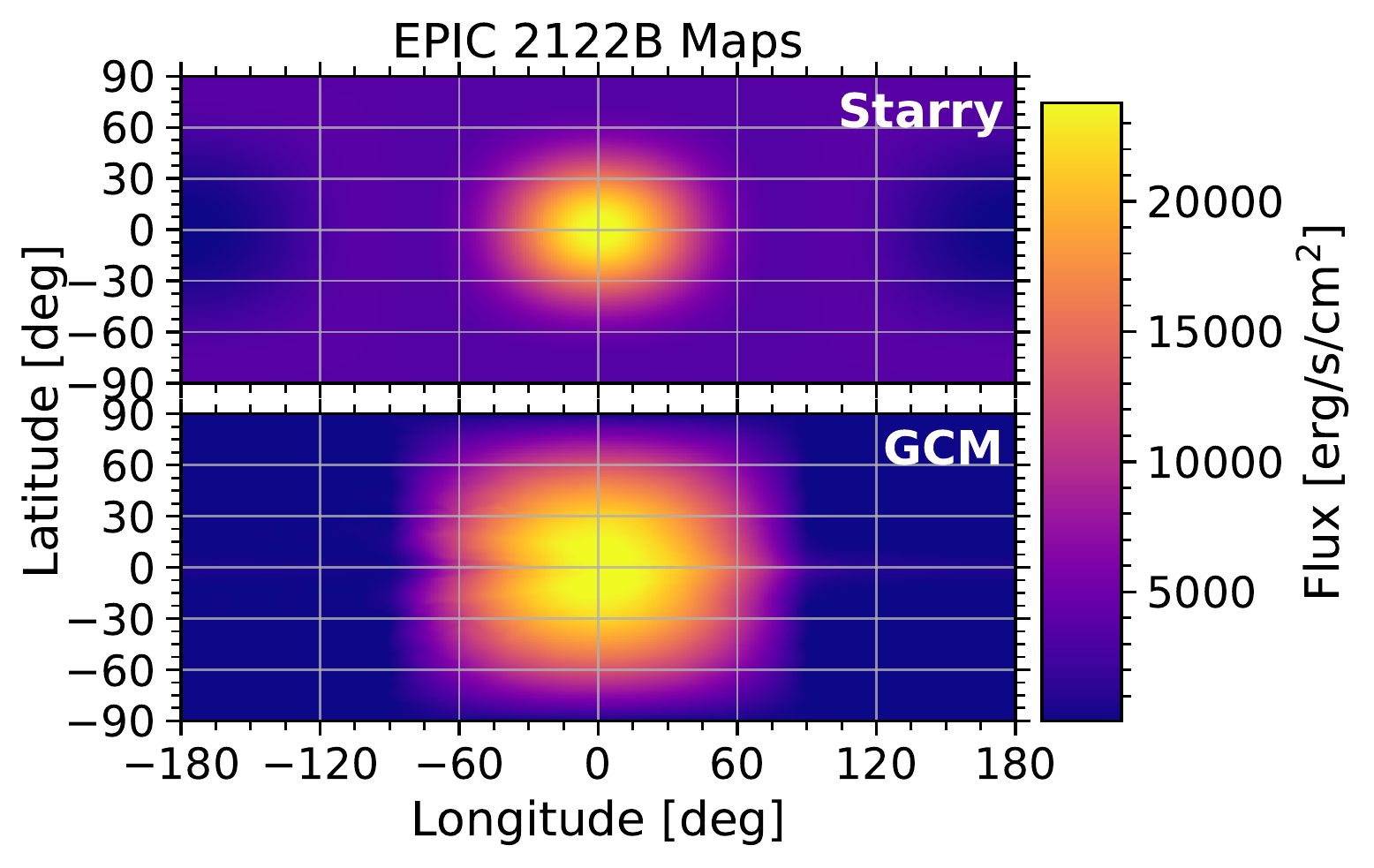}
  \caption{Retrieving top-of-the-atmospheric flux maps from the phase curves of WD~0137B and EPIC~2122B. Upper panels: comparisons of the 1.12--1.65 \micron{}  phase curves to three models: the best-fit \texttt{starry} phase curve (yellow), the scaled GCM phase curve (green), and the phase curve derived based on radiative-convective equilibrium (orange). All three models fit well to the observations and are almost indistinguishable from each other. In the figure legends, the model names are followed by the reduced-$\chi^{2}$ values. Lower panel: the retrieved \texttt{starry} and GCM top-of-the-atmospheric flux maps. While different in details, both maps share the same dominant features: a confined hot spot centered on the brown dwarfs' sub-stellar points.}
  \label{fig:fluxmap}
\end{figure*}  

  \begin{figure}[h]
    \centering
    \includegraphics[width=\columnwidth]{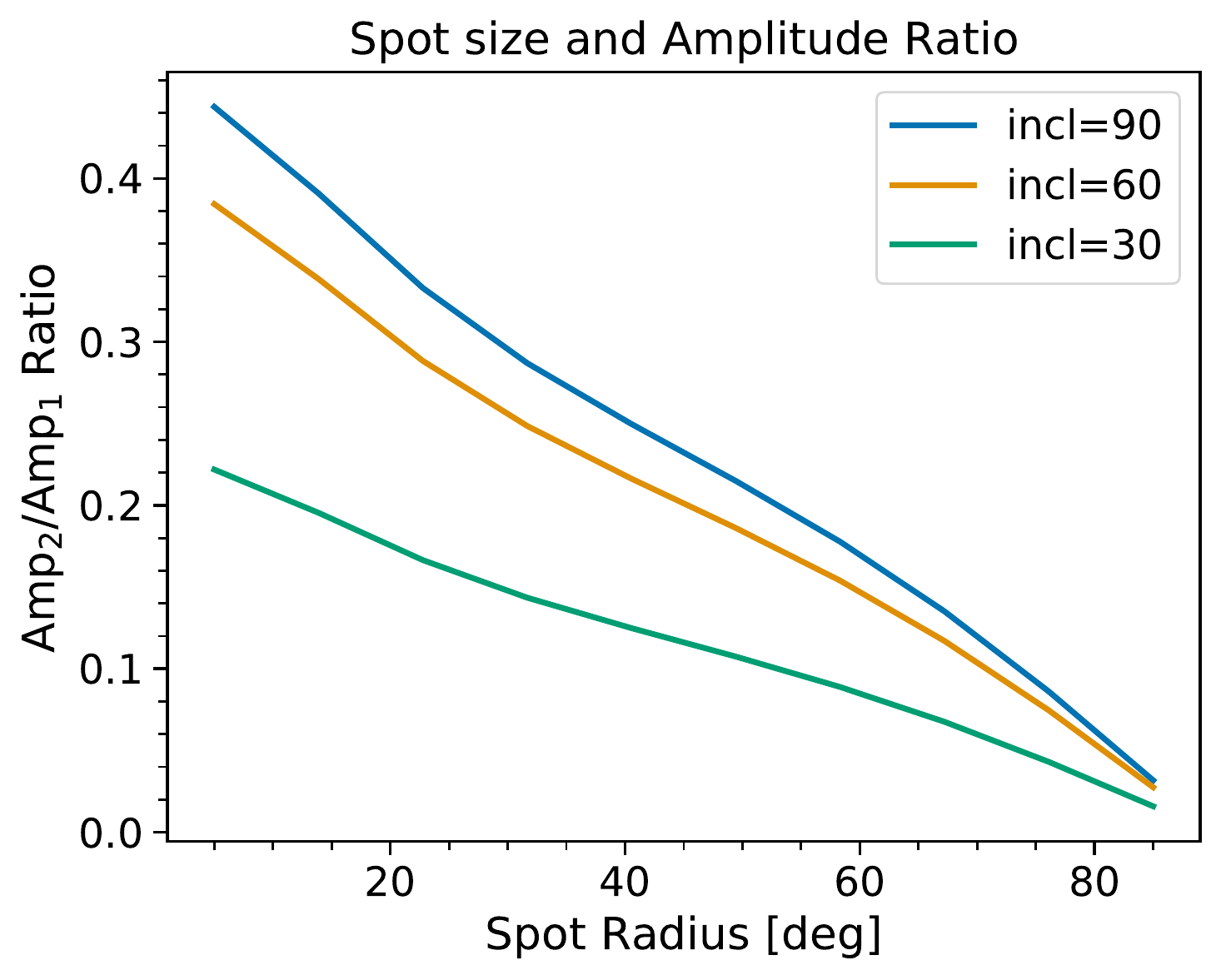}
    \caption{The amplitude ratios between the $k=1$ and $k=2$ waves in the phase curve of a hot spot map. Regardless of viewing geometry ($i=35^{\circ}$ and $56^{{\circ}}$ for WD 0137B and EPIC 2122B, respectively), the smaller the spot, the more significant the $k=2$ wave is. Therefore, higher-order Fourier components constrain the hot spot size.}
    \label{fig:spotsize}
  \end{figure}
  
  To investigate the circulation patterns of the two irradiated brown dwarfs, we first retrieve top-of-the-atmospheric flux maps from the observed phase curves using the \texttt{starry} code \citep{Luger2019}, which  constructs surface maps using spherical harmonic function as bases. Converting light curves to maps is highly degenerate. Therefore, rather than relying on spherical harmonic functions entirely, we adopt physically motivated prior setups to reduce the degrees of freedom of the model and regularize the fits. Based on the two properties of the observed light curves: 1) they peak at $\phi=0$; 2) phase shifts between different wave modes are insignificant (Figure~\ref{fig:spec_lc_fit}), we restrict our maps to only include the $l=1, m=0$ spherical harmonics basis ($Y_{1,0}=\frac{1}{2}\sqrt{\frac{3}{\pi}}\cos \theta$, $\theta$ is longitude) and a flat-top hot spot centered at the substellar point (lat=0, lon=0). The hot spot injection is implemented using \texttt{starry}'s built-in \texttt{spot} function. In this approach, there are four free parameters in map retrieval: 1) the baseline flux; 2) the scaling factor for $Y_{1,0}$; 3) the brightness contrast of the spot; and 4) the angular radius of the hot spot. The inclinations of the brown dwarfs' spin axes are fixed based on literature values: $35^{\circ}$ for WD~0137B \citep{Maxted2006} and $56^{\circ}$ for EPIC~2122B \citep{Casewell2018}. We use MCMC  to find the best-fitting values.

  The \texttt{starry} mapping results are presented in Figure~\ref{fig:fluxmap}. For both WD~0137B and EPIC~2122B, the $Y_{1,0}$+spot model  reproduces the observations very well. Because the integrated phase curve of $Y_{1,0}$ is always a $k=1$ sine wave, the higher-order Fourier terms in the phase curves  must come from the hot spot. As a result, the spot size can be constrained from the amplitude of the higher-order terms (Figure~\ref{fig:spotsize}). For both WD~0137B and EPIC~2122B, the best-fitting spot angular radii are $10^{\circ}{-}15^{\circ}$.

  \subsection{Interpreting the Phase Curves: Comparing Observations with GCMs}

  \begin{figure}[h]
  \centering
  \includegraphics[width=\columnwidth]{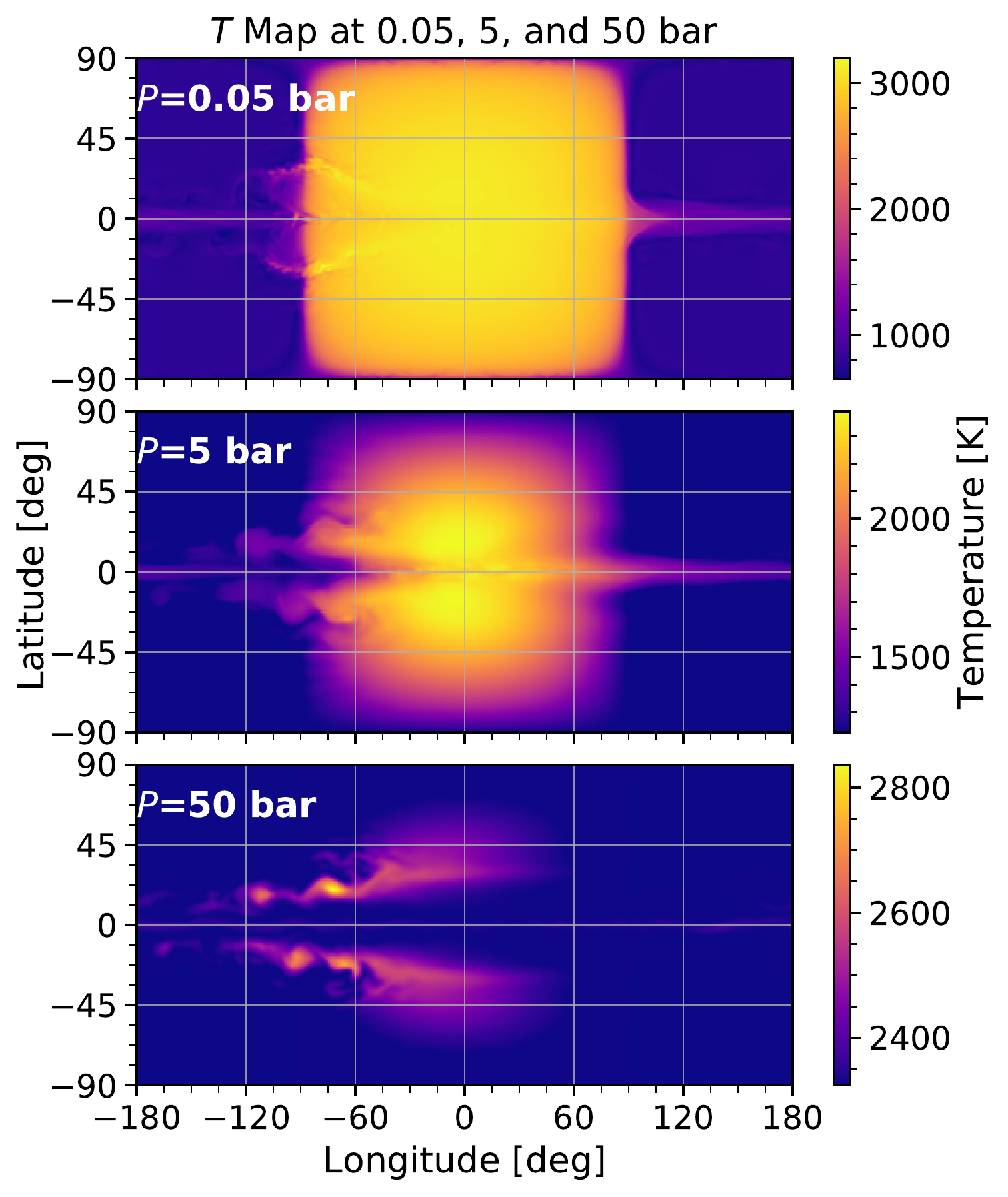}
  \caption{Temperature distributions at 0.05, 5, and 50 bar in the GCM simulation of WD~0137B. Deep into the atmosphere, the temperature is approximately uniformly distributed. Near the top of the atmosphere, the temperature bifurcates into the two extrema of the day- and night-sides. At an intermediate pressure level, the hot spot is the most pronounced.}
  \label{fig:Tmap}
\end{figure}

We then perform general circulation model (GCM) simulations for both brown dwarfs to examine whether the top-of-atmospheric flux maps retrieved from the observations agree with the theoretical predictions. Our models are slightly modified from those used in \citep{Tan2019}. \revise{In this model, the stellar irradiation is partitioned into two channels. This is to approximate the strong effect of the UV irradiation on the day-side thermal structures. Under these setups, the GCMs are controlled by four free parameters: the brown dwarf Bond albedo, the partition fraction, and the two opacities of the two channels.} We describe the details of our GCM simulations in the Appendix \ref{sec:GCM}.

  As shown in Figure~\ref{fig:fluxmap}, for both WD~0137B and EPIC~2122B, the GCM maps qualitatively match the \texttt{starry} retrieval results. In particular, the GCM maps also contain hot spots that are similar in size and position, compared to the ones in the \texttt{starry} maps. The close match between the GCMs and the observations is more clearly presented in the phase curve comparisons. \revise{We note that in our GCMs, the phase curve amplitude can be tuned by adjusting the free parameters, and hence we are interested in comparing the phase curve \emph{shapes} rather than the absolute amplitudes. Therefore, we linearly scale the model phase curves to match the mean and amplitude of the observed ones. } As shown in Figure~\ref{fig:fluxmap}, the scaled models fit to the observed phase curves of both WD~0137B and EPIC~2122B remarkably well.

  Therefore, our observations confirm the circulation properties of tidally-locked, fast rotating, and highly irradiated brown dwarfs as predicted by GCMs simulations \citep{Lee2020,Tan2020b}. Compared to typical hot Jupiters \citep[e.g.,][]{Stevenson2014} and ultra-hot Jupiters \citep[e.g.,][]{Arcangeli2019}, WD~0137B and EPIC~2122B have more than one order of magnitude shorter rotation periods. Such fast rotation rates increase the meridional gradient of the Coriolis parameter, reduce the width of the equatorial jet, and ultimately suppress the day-to-night heat transfer efficiency \citep{Tan2020b}. With a low horizontal energy transfer rate, the circulation is nearly in a radiative-convective equilibrium (RCE). \revise{This is further supported by the excellent phase curve fits of the scaled  models for which RCE is assumed (Figure~\ref{fig:fluxmap}).} In this regime,  a hot spot is naturally confined to the substellar point.

  Based on the temperature maps in the GCM simulations (Figure~\ref{fig:Tmap}), at a high altitude/low atmospheric pressure level ($P<0.1$~bar), the temperature is strongly affected by the incoming irradiation and bifurcates into two extremes. At this level, the hot spot almost occupies the entire dayside. Deep into the atmosphere ($P>10$~bar), the temperature is mostly determined by the internal heat of the brown dwarf, and has nearly a globally homogeneous distribution. At the intermediate pressure level ($P\approx1$~bar), the hot spot covers a portion of the day-side hemisphere. As shown in Figure~\ref{fig:spotsize}, the smaller the spot, the more significant the high-order Fourier terms are presented in the phase curves, and hence these components are more easily detected in light curves that probe the intermediate pressure levels. These are exactly the pressure levels that the continuum of the WFC3/G141 spectrum is sensitive to. On the contrary, the water absorption bands and the wavelengths of the Spitzer/IRAC channels probe the low pressure level (high altitude) \citep[see Figure 3 of ][]{Lothringer2020}. As a result, the $k=2$ wave is less significant in the \SI{1.4}{\micro\meter} phase curve of WD~0137B, and a single-component sine wave fits the Spitzer phase curves well \citep{Casewell2015}. This result motivates a broadband spectroscopic time-resolved observations of these brown dwarfs that can continuously probe 0.1 to 1 bar pressure levels, which will map how temperature gradients evolve vertically and reveal the three-dimensional atmospheric structures.

  \section{Phase-Resolved Spectra of the Brown Dwarfs}
  \label{sec:spec}
\newcommand{\teffwd}{\ensuremath{T_{\mathrm{eff, WD}}}\xspace}
\newcommand{\teffbd}{\ensuremath{T_{\mathrm{eff, BD}}}\xspace}
\newcommand{\teffbdhot}{\ensuremath{T_{\mathrm{eff, BD, hot}}}\xspace}
\newcommand{\teffbdcool}{\ensuremath{T_{\mathrm{eff, BD, cool}}}\xspace}
\newcommand{\loggwd}{\ensuremath{\log g_{\mathrm{WD}}}\xspace}
\newcommand{\loggbd}{\ensuremath{\log g_{\mathrm{BD}}}\xspace}

\subsection{Modeling and Subtracting the White Dwarf Spectra}

\begin{figure*}[t]
  \centering
  \includegraphics[width=0.48\textwidth]{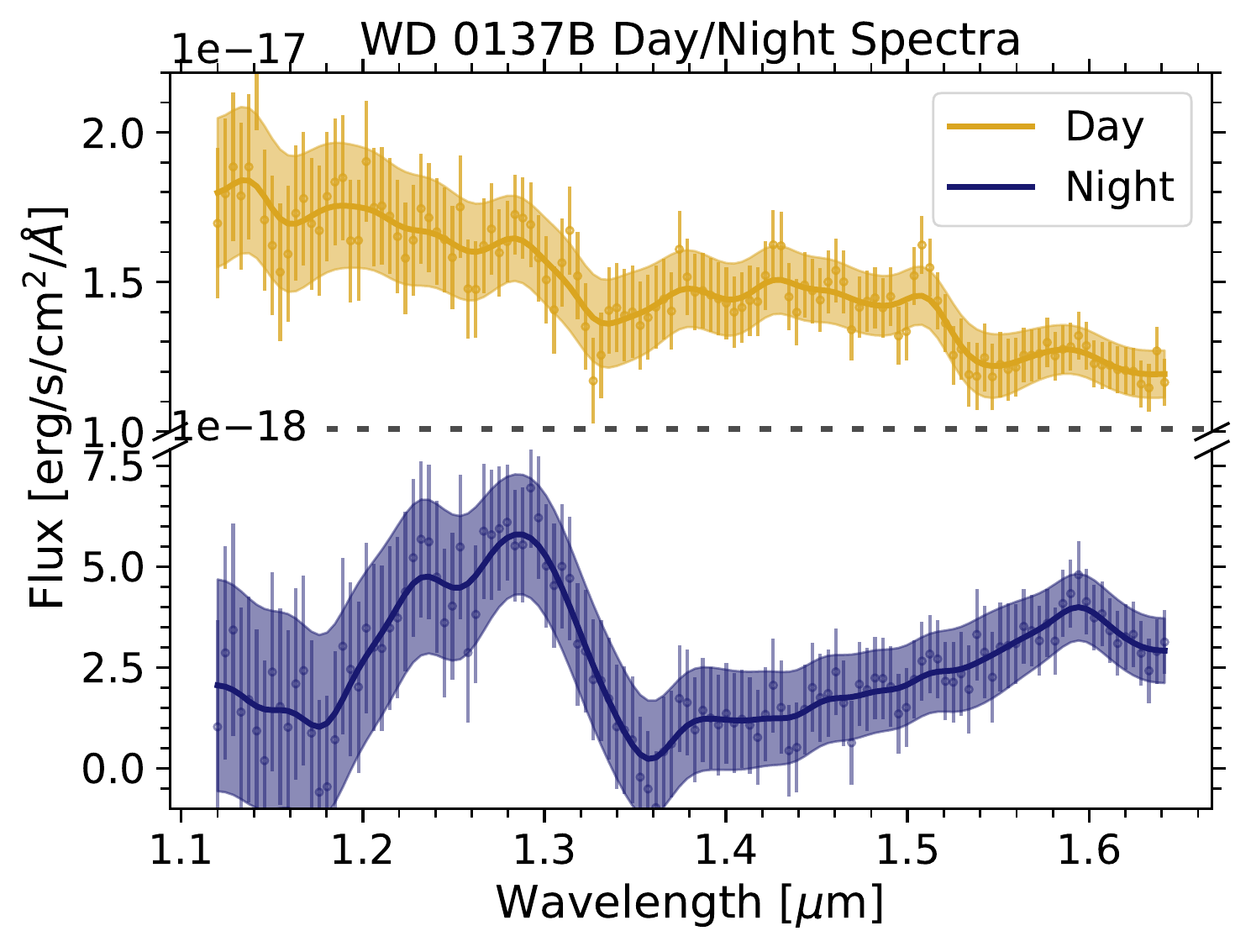}
  \includegraphics[width=0.48\textwidth]{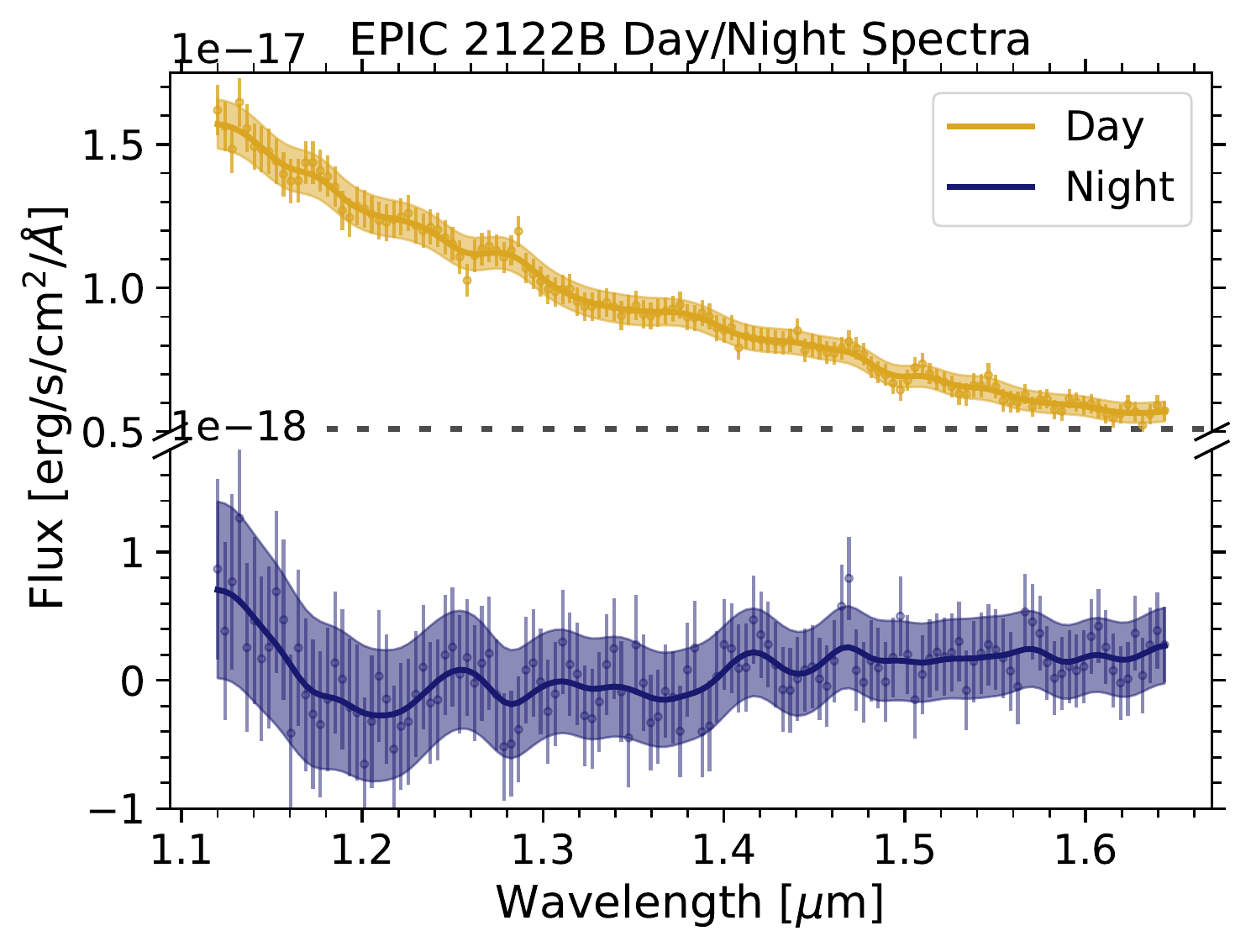}
  \caption{The Day/Night Spectra of WD 0137B (left) and EPIC 2122B (right). The yellow and purple dots represent the observed day- and night-sides spectra, respectively; the error bars show their $1\sigma$ uncertainties. The solid lines and swashes are observations and $1\sigma$ uncertainties convolved with a Gaussian kernel with a width equal to the spectral resolution of the G141 grism ($R\sim130$ at \SI{1.4}{\micro\meter}), illustrating the actual resolving power of our observations.  The $y$-axes are discontinuous to properly visualize the flux density and demonstrate their wavelength-dependence.}
  \label{fig:daynightspec}
\end{figure*}

\begin{figure*}[t]
  \centering
  \includegraphics[height=0.6\textwidth]{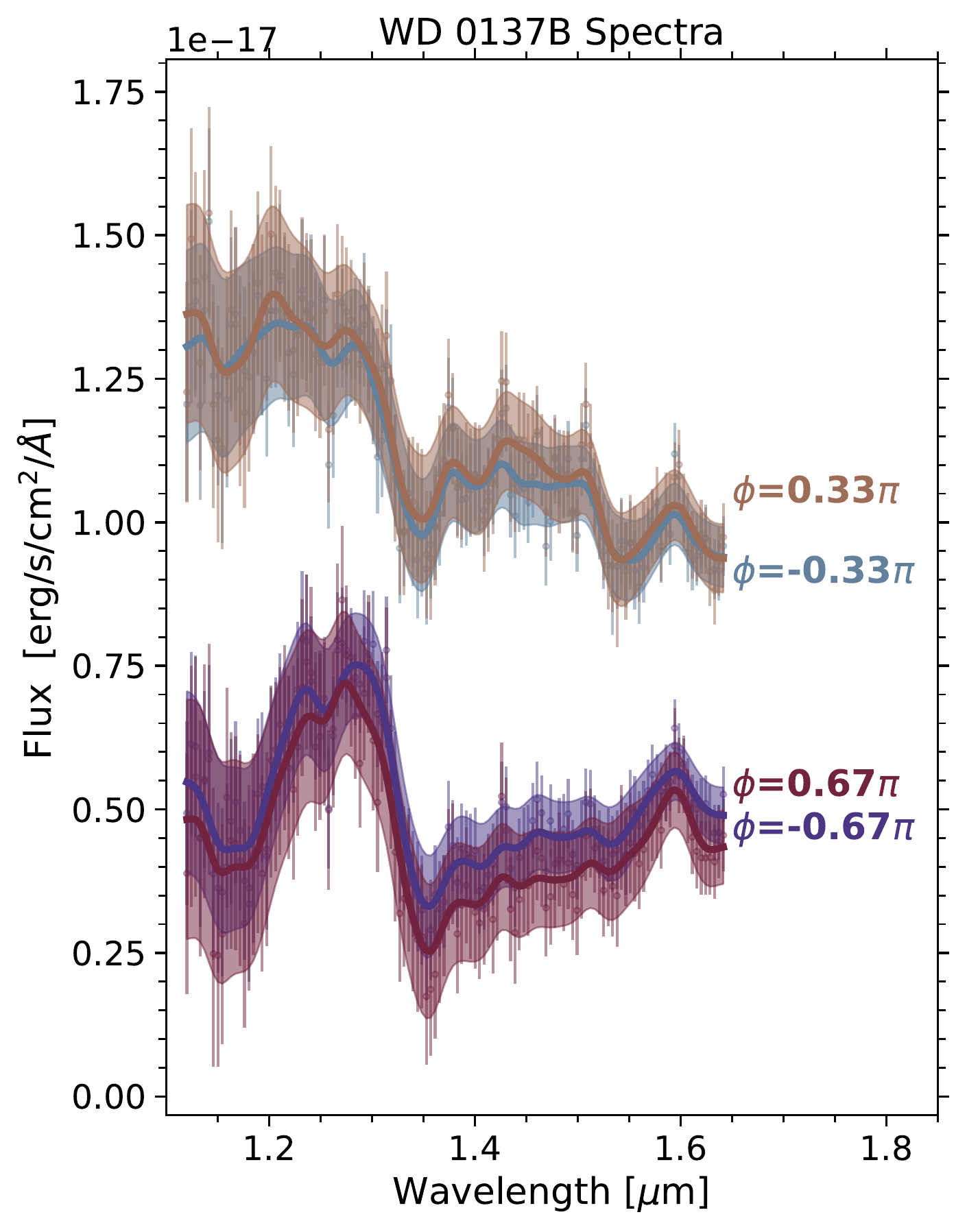}  
  \includegraphics[height=0.6\textwidth]{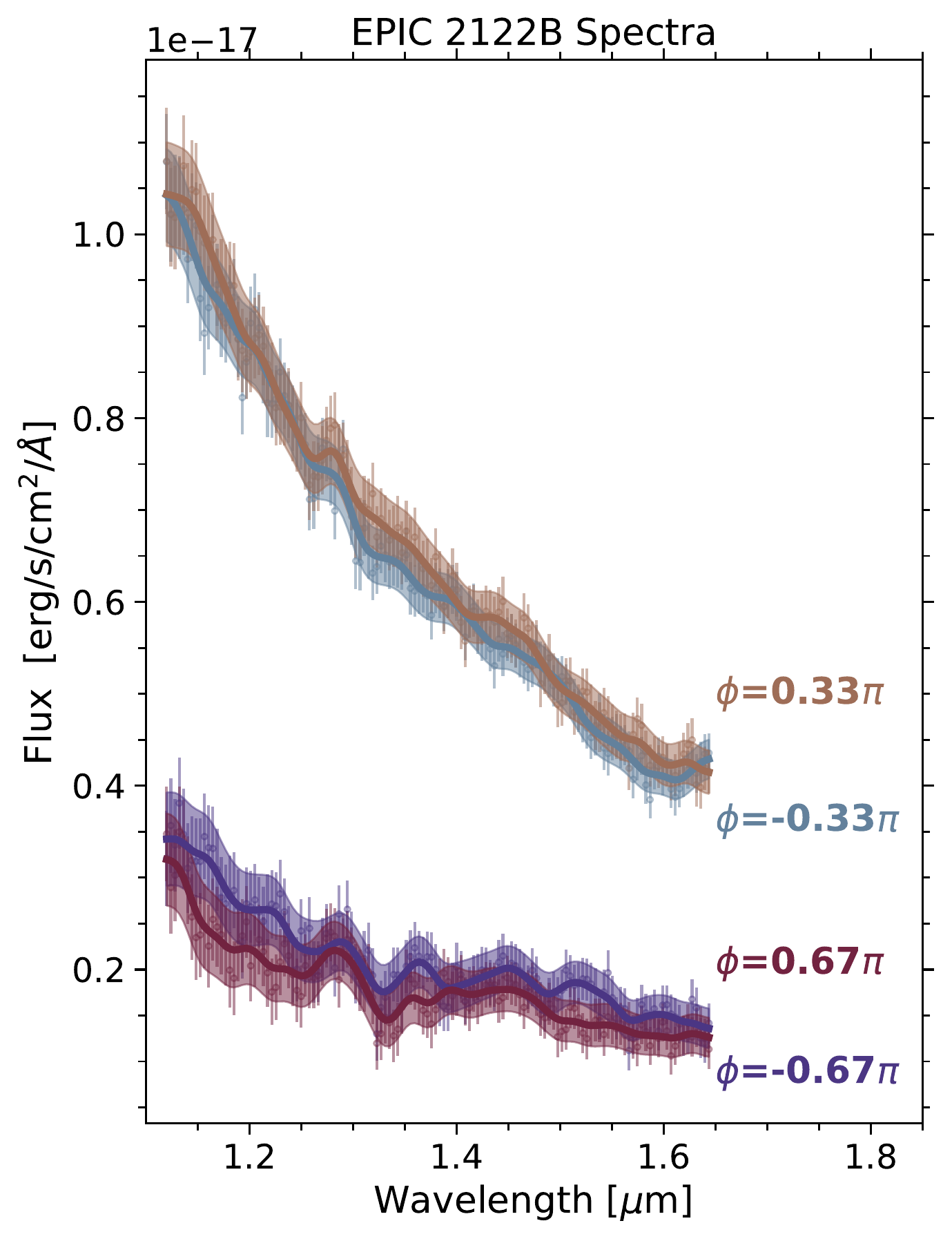}
  \caption{Spectra of the brown dwarfs in WD~0137 (left) and EPIC~2122 (right) in four representative rotational phases between the day- and night-sides ($\phi=\pm\pi/3$ and $\pm2\pi/3$). The spectra are color-coded by their phases with light colors indicating the day-side and dark colors indicating the night-side. Same as Figure~\ref{fig:daynightspec}, dots and error bars show the immediate observational results; lines and swashes show spectra smoothed with the instrument's spectral resolution unit. All spectra are plotted in absolute flux densities without normalization or vertical offsets applied. The differences between spectra are real flux differences.}
  \label{fig:bdspec}
\end{figure*}

The observed WD+BD spectra (upper subplots in Figure~\ref{fig:spectralsequence}) consist of two components: a constant white dwarf spectrum ($S_{\mathrm{WD}}$) and a variable brown dwarf spectrum ($S_{\mathrm{BD}}(\phi) $)
\begin{equation}
  \label{eq:4}
  S_{\mathrm{WD+BD}}(\phi) = S_{\mathrm{WD}} + S_{\mathrm{BD}}(\phi).
\end{equation}
To obtain phase-resolved brown dwarf spectra for substellar  atmospheric investigations, we subtract white dwarf model templates from the observations.
Both WD~0137A and EPIC~2122A have pure hydrogen atmospheres, resulting in nearly featureless spectra that have been accurately modeled. In the \citet[e.g.,][]{Koester2010} grid that we adopted,  two free parameters --- \teffwd and \loggwd --- define the spectral template. We also include a linear scaling factor ($a_{\mathrm{WD}}$)  as the third free parameter. \teffwd and \loggwd of WD~0137A and EPIC~2122A have been precisely determined in previous studies using high-resolution spectra \citep[][]{Maxted2006, Casewell2015, Longstaff2017, Casewell2018}, and the scaling factor can be constrained using archival photometry. \revise{We combine these prior information by fitting synthetic photometry integrated from the model spectra to archival photometry with  Markov Chain Monte Carlo (MCMC, implemented by \texttt{emcee} \citealt{Foreman-Mackey2012}). By conducting this fit in a Bayesian manner, we incorporate the tight constraints of \teffwd and \loggwd and properly propagate the variant and covariant uncertainties to the template subtraction results.}

 In this fit, the best-fitting values of $a_{\mathrm{WD}}$, \teffwd{}, and \loggwd{}  maximize the posterior probability (expressed in the $\log$-scale), which is defined by
\begin{equation}
  \begin{split}
  \label{eq:6}
  \log \mathrm{P}_{\mathrm{WD}}(\theta) = \log \mathrm{Pr}_{\teffwd} + \log \mathrm{Pr}_{\loggwd} + \log \mathcal{L}(\theta)
  \end{split}
\end{equation}
in which the log-likelihood function $\log \mathcal{L}(\theta)$ is 
\begin{equation}
  \label{eq:2}
  \log \mathcal{L}(\theta) \propto -\frac{\chi^{2}}{2},
\end{equation}
and $\theta$ represents a set of $a_{\mathrm{WD}}$, \teffwd{}, and \loggwd{}.
$\mathrm{Pr}_{\teffwd}$ and $\mathrm{Pr}_{\loggwd}$ are the prior probability distributions of the effective temperature and surface gravity of the white dwarf. We assume them to be normally distributed (WD~0137A: $\teff{\sim}\mathcal{N}(16500,\,500)$~K, $\logg{\sim}\mathcal{N}(7.50,\,0.02)$, \citet{Maxted2006}; EPIC~2122A $\teff{\sim}\mathcal{N}(24450,\,150)$~K, $\logg{\sim}\mathcal{N}(7.63,\,0.02)$, \citet{Casewell2018}). The $\chi^{2}$ is the chi-squares difference between model synthetic and observed photometry, for which we include $V$, $\mathrm{VST}/g$, and $G$ bands for WD~0137A and $B$, $\mathrm{VST}/g$, and $G$ bands for EPIC~2122A. We exclude the $U$-band observations due to the large extinction uncertainty and the infrared data due to high brown dwarf flux contamination. Both binaries show modulations in these bands but the phases at which the photometric data were taken are unknown. Therefore, we consider the modulations as part of the photometric uncertainty: the adopted errors are the relative semi-amplitudes and the reported uncertainties combined in quadrature. Each MCMC fit contains 64 walkers and 600 iterations, with the first 300 discarded as part of the ``burn-in'' processes.

After the MCMC runs, we proceed by subtracting the best-fit WD models to isolate the brown dwarf components. The uncertainties in the brown dwarf spectra are a combination of the variance intrinsic to the observations (i.e., photon noise, readout noise, and dark current) and those propagated from the uncertain white dwarf properties. On average, variances in the white dwarf flux contribute to 49\% and 6.7\% of the total uncertainties in the spectra of WD~0137B and EPIC~2122B, respectively. In Figure \ref{fig:daynightspec}, the day- and night-side spectra of WD~0137B and EPIC~2122B are presented. The day- and night-side spectra are the averages of the observations of phases of $\phi\in[-0.3, 0.3]$~rad and  $\phi\in[\pi-0.3, \pi+0.3]$~rad, respectively. These ranges ensure that at least three observations with minimal spectral variations are included in each of the combinations and secure the precisions of the derived day/night spectra. The day-side spectra of both brown dwarfs are featureless slopes. As for the night-side,  a strong \SI{1.4}{\micro\meter} water absorption feature is present in WD~0137B, but no significant features are detected in EPIC~2122B due to poor SNRs. Between the two extrema, both brown dwarfs have nearly an order-of-magnitude flux differences, and EPIC~2122B has a stronger overall flux variation than WD~0137B. In Figure \ref{fig:bdspec}, brown dwarf spectra at four representative phases between the day- and night-sides ($\phi=\pm\pi/3$ and $\phi=\pm2\pi/3$) are also provided. In both brown dwarfs, the overall flux and spectral shapes evolve dramatically from day- to night-sides. In addition, a strict east-west symmetry is presented in both cases: spectra of the $+1/3\pi$ and $+2/3\pi$ agree with those of the $-1/3\pi$ and $+2/3\pi$ within $1\sigma$.

\subsection{Semi-Empirically Fitting the Phase-Resolved Spectra}
To investigate the  spectroscopic phase variations in the brown dwarf spectra, we conduct a joint analysis to fit the white dwarf and brown dwarf components together. We adopt a semi-empirical model to fit the combined white dwarf and brown dwarf spectra. This approach is motivated by the two-dimensional retrieval approach developed for hot Jupiter phase curves \citep{Feng2020}, in which the heterogeneous hot Jupiter atmospheres are approximated as linear combinations of two patches. Similarly, our model is expressed as a linear combination of  the white dwarf ($S_{\mathrm{WD}}$) and two brown dwarf spectral bases: the day-side $S_{\mathrm{dayside}}$ and the night-side $S_{\mathrm{nightside}}$. In the previous subsection, the white dwarf spectrum and its scaling factor have been tightly constrained, and the brown dwarf bases have also been empirically determined.  We define the day-side scaling coefficient as $f(\phi)$ and thus express the night-side scaling coefficient as $1 - f(\phi)$. So the semi-empirical model can be expressed as
  \begin{equation}
    \label{eq:7}
    \begin{split}
      &S_{\mathrm{WDBD}}(\phi) = a_{\mathrm{WD}}S_{\mathrm{WD}}(\teffwd, \loggwd)\\
      &+ f(\phi) S_{\mathrm{day-side}} + (1-f(\phi)) S_{\mathrm{night-side}},
      \end{split}
  \end{equation}

  We fit each WFC3 spectra to this model to find the best-fitting $a_{\mathrm{WD}}, \teffwd, \loggwd,$ and $f(\phi)$.  Same as the white dwarf fits, the fittings are conducted by \texttt{emcee} to maximize the posterior probability
  \begin{equation}
    \log \mathrm{P}_{\mathrm{WDBD}}(\theta) = \log \mathrm{Pr}(a_{\mathrm{WD}},\,\teffwd,\,\loggwd)  + \log \mathcal{L}(\theta).
  \end{equation}
  In this equation, $\mathrm{Pr}(a_{\mathrm{WD}},\,\teffwd,\,\loggwd)$ is the joint posterior distribution derived in the previous subsection. $\theta$ represents a set of $a_{\mathrm{WD}}$, \teffwd, \loggwd, and $f(\phi)$.  $\log \mathcal{L}(\theta)$ is proportional to the $\chi^{2}$ difference between the observations and the WD+BD model.

  \begin{figure*}[!t]
    \centering
    \includegraphics[height=0.38\textwidth]{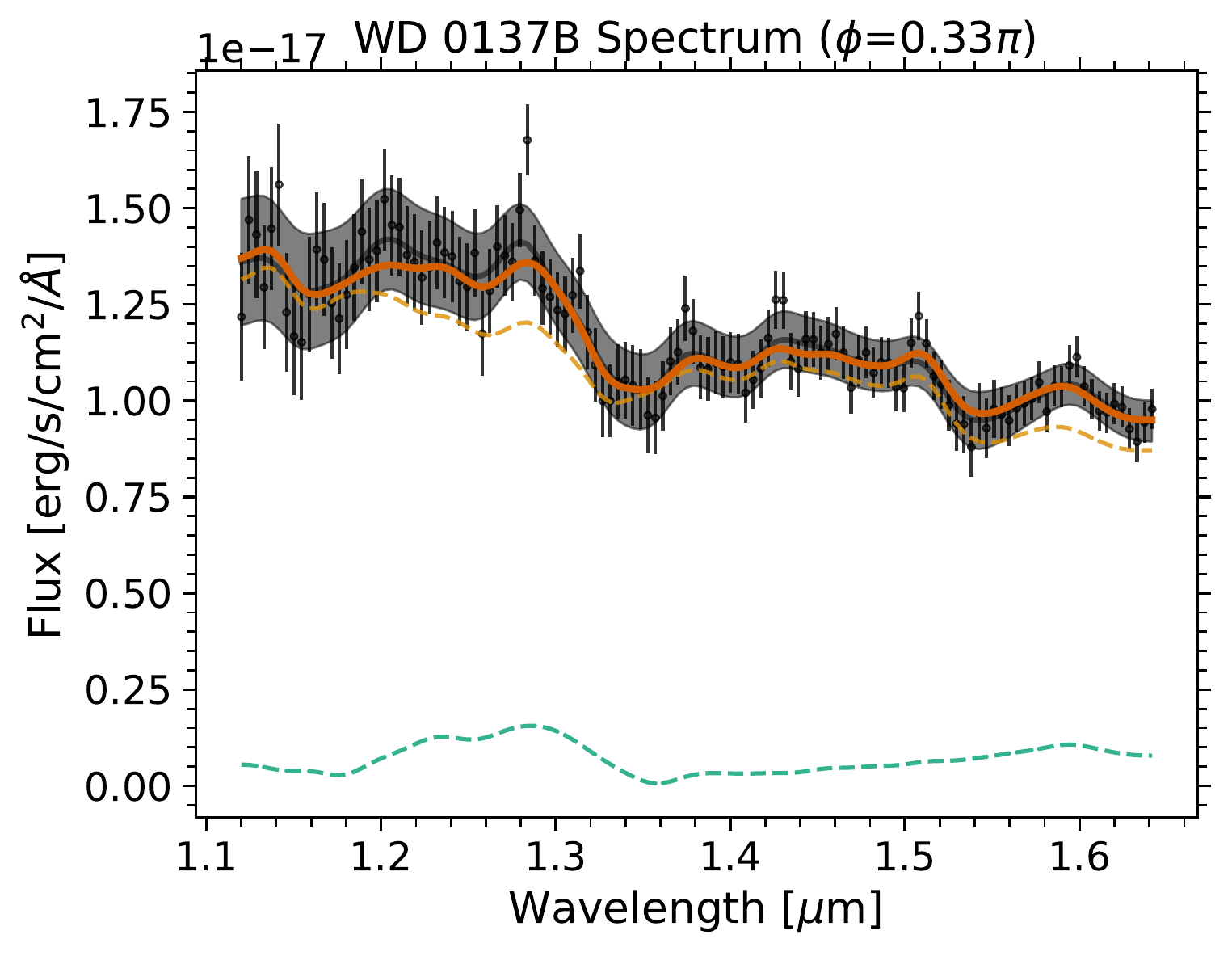}
    \includegraphics[height=0.38\textwidth]{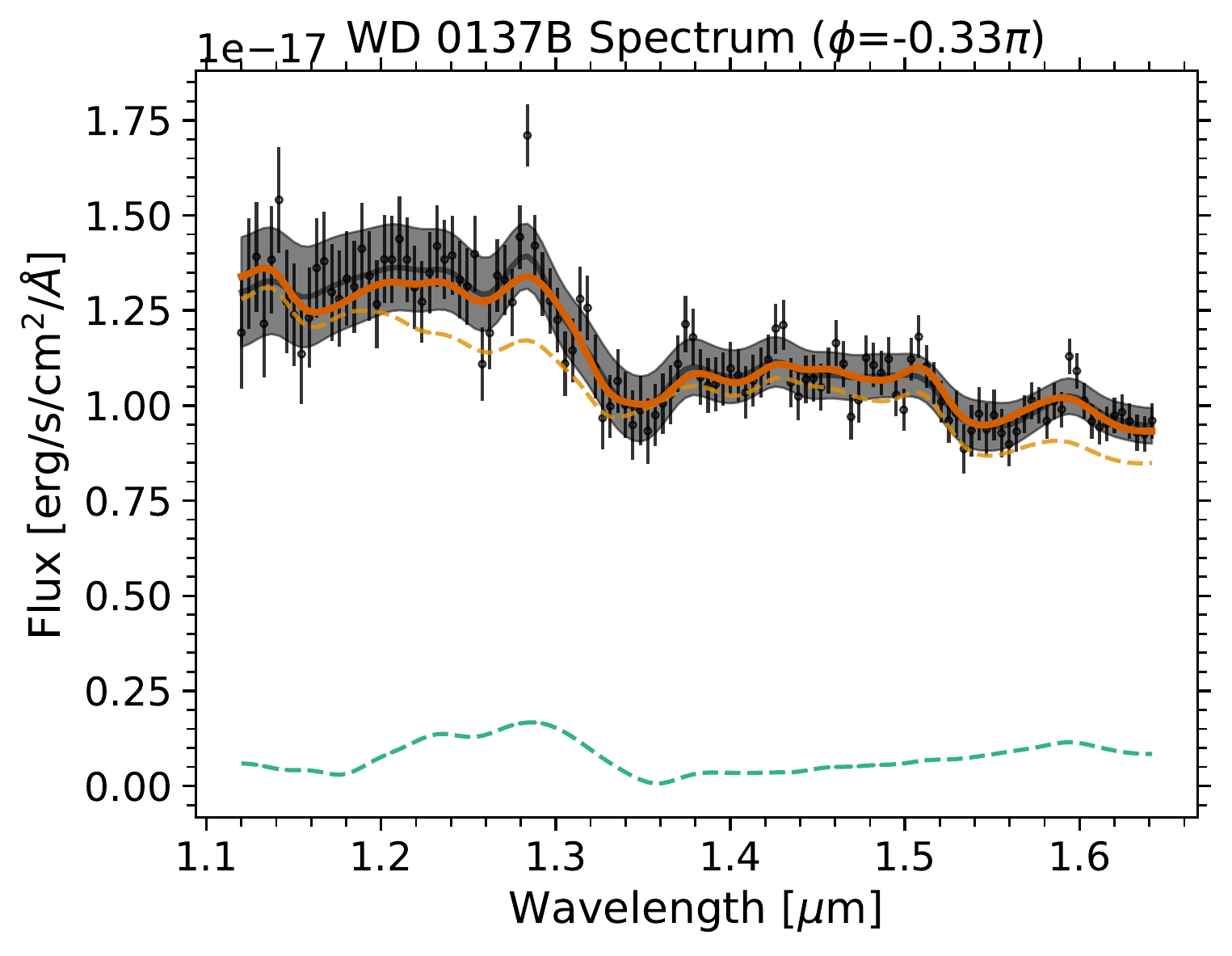}
    \includegraphics[height=0.38\textwidth]{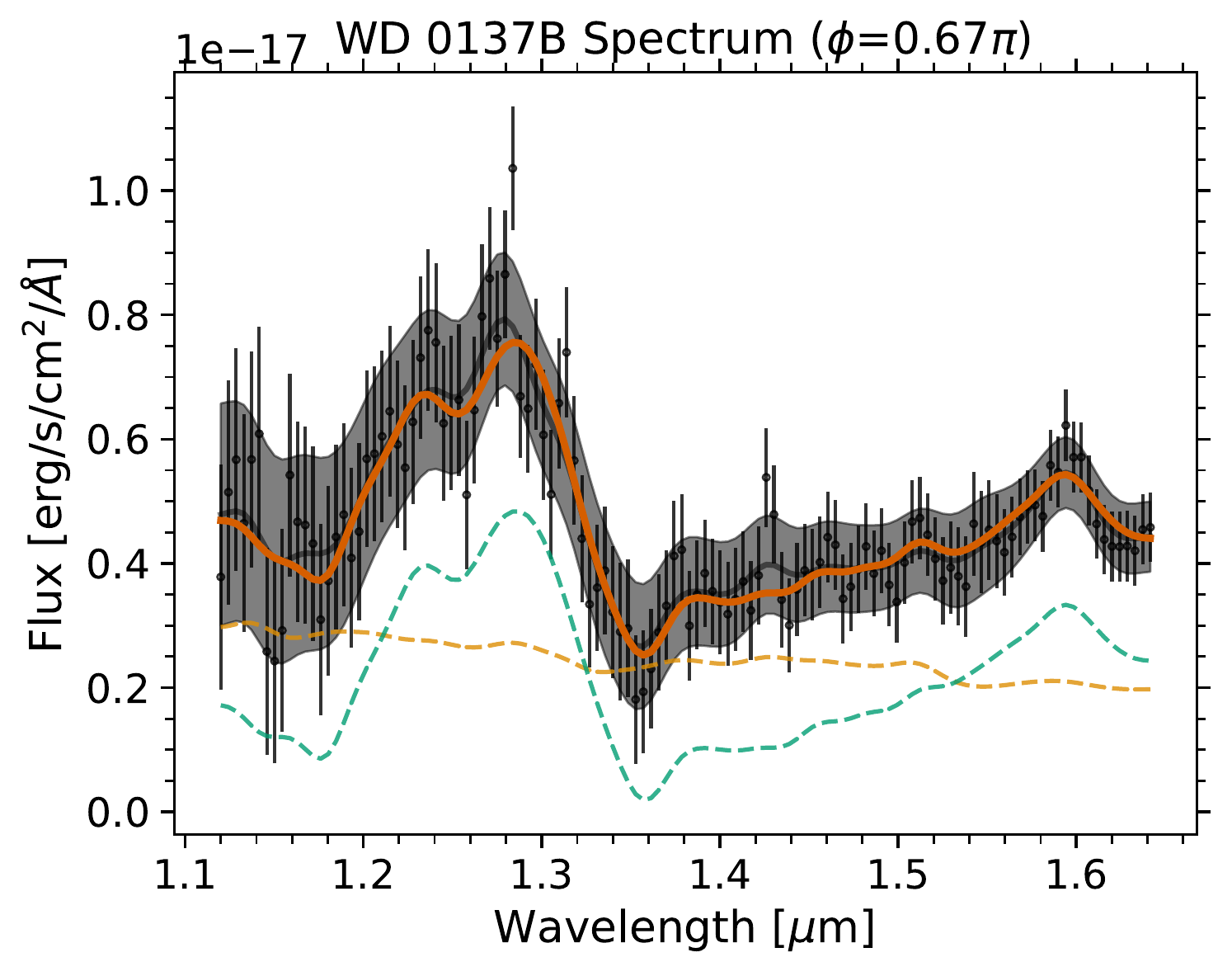}
    \includegraphics[height=0.38\textwidth]{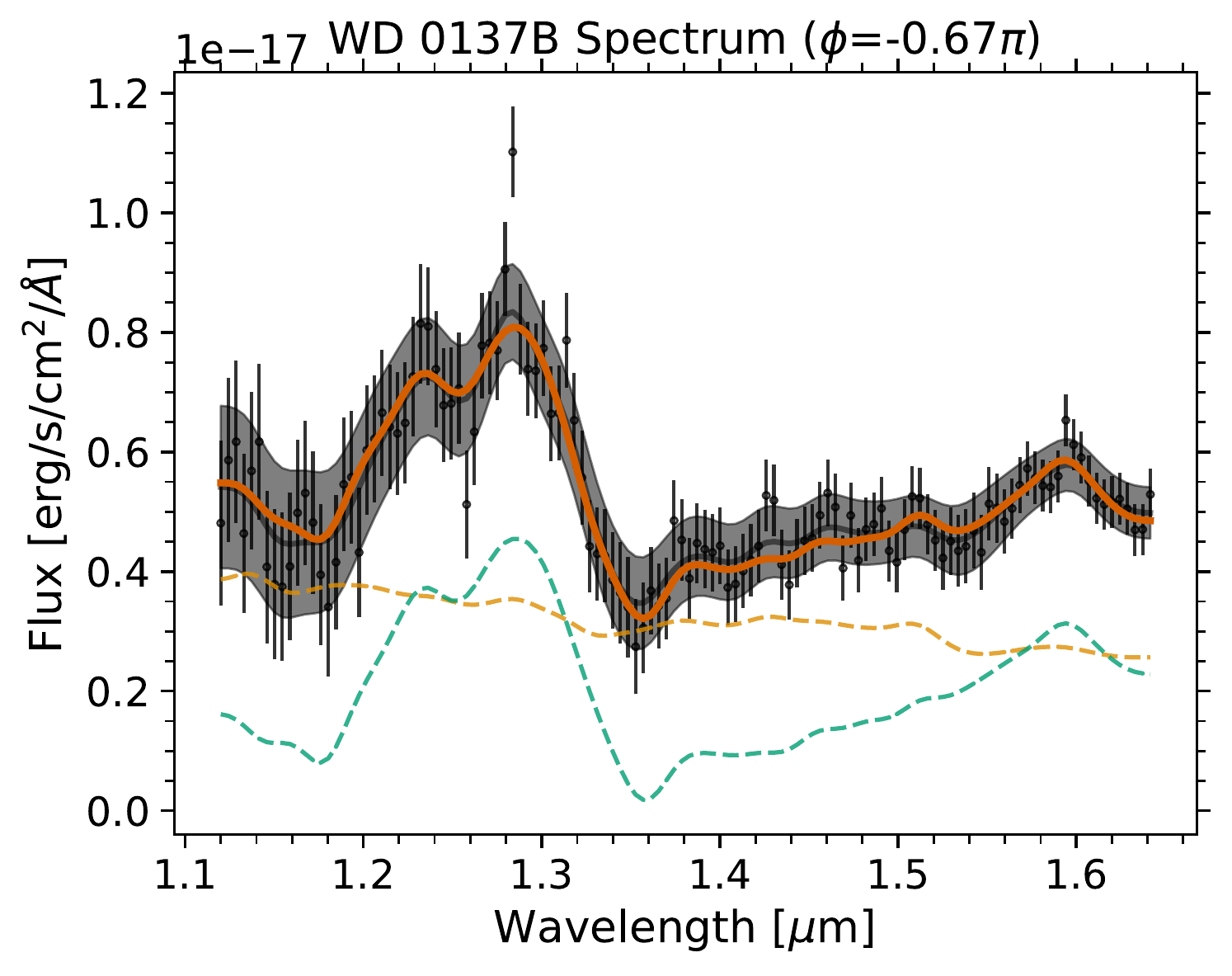}
   \includegraphics[width=0.4\textwidth]{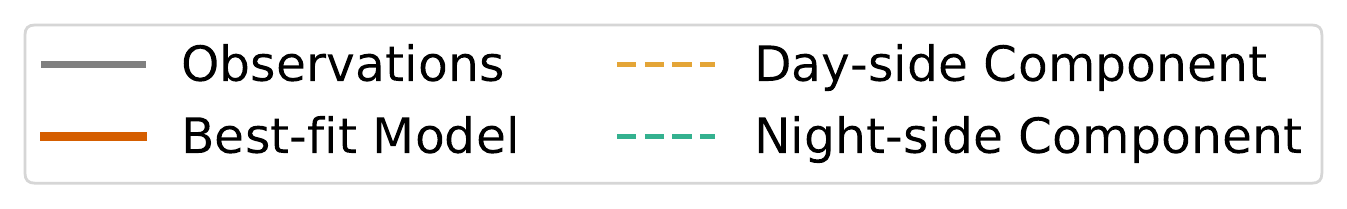}
    \caption{WD+BD spectral model fitting results for WD~0137 at four representative phases between the day- and night-sides at $\phi=\pm1/3\pi$ and $\phi=\pm2/3\pi$. The best-fitting white dwarf components are subtracted from all spectra. In plotting the observational results, the conventions are the same as those in Figure~\ref{fig:daynightspec}: black dots and errorbars are immediate results; lines and swashes are spectra and uncertainties convolved with a Gaussian kernel to reflect the actual spectral resolving power. The orange lines are the best-fitting brown dwarf models, which are the sum of the scaled day-side component (yellow dashed lines) and the scaled night-side component (green dashed lines). In every case, this semi-empirical model fits the observations well.}
    \label{fig:WD0137specfit}
  \end{figure*}

  \begin{figure*}[!t]
    \centering
    \includegraphics[height=0.38\textwidth]{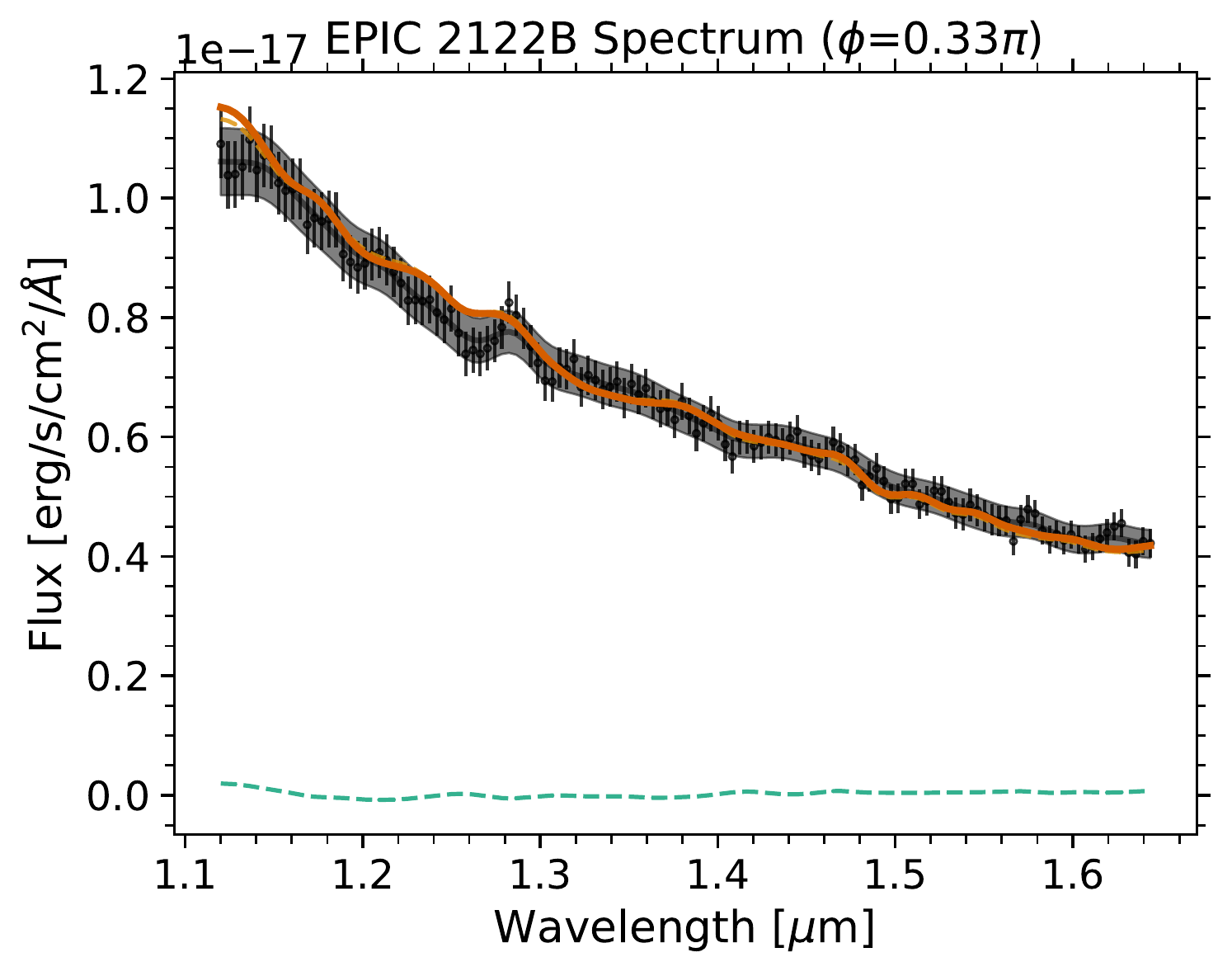}
    \includegraphics[height=0.38\textwidth]{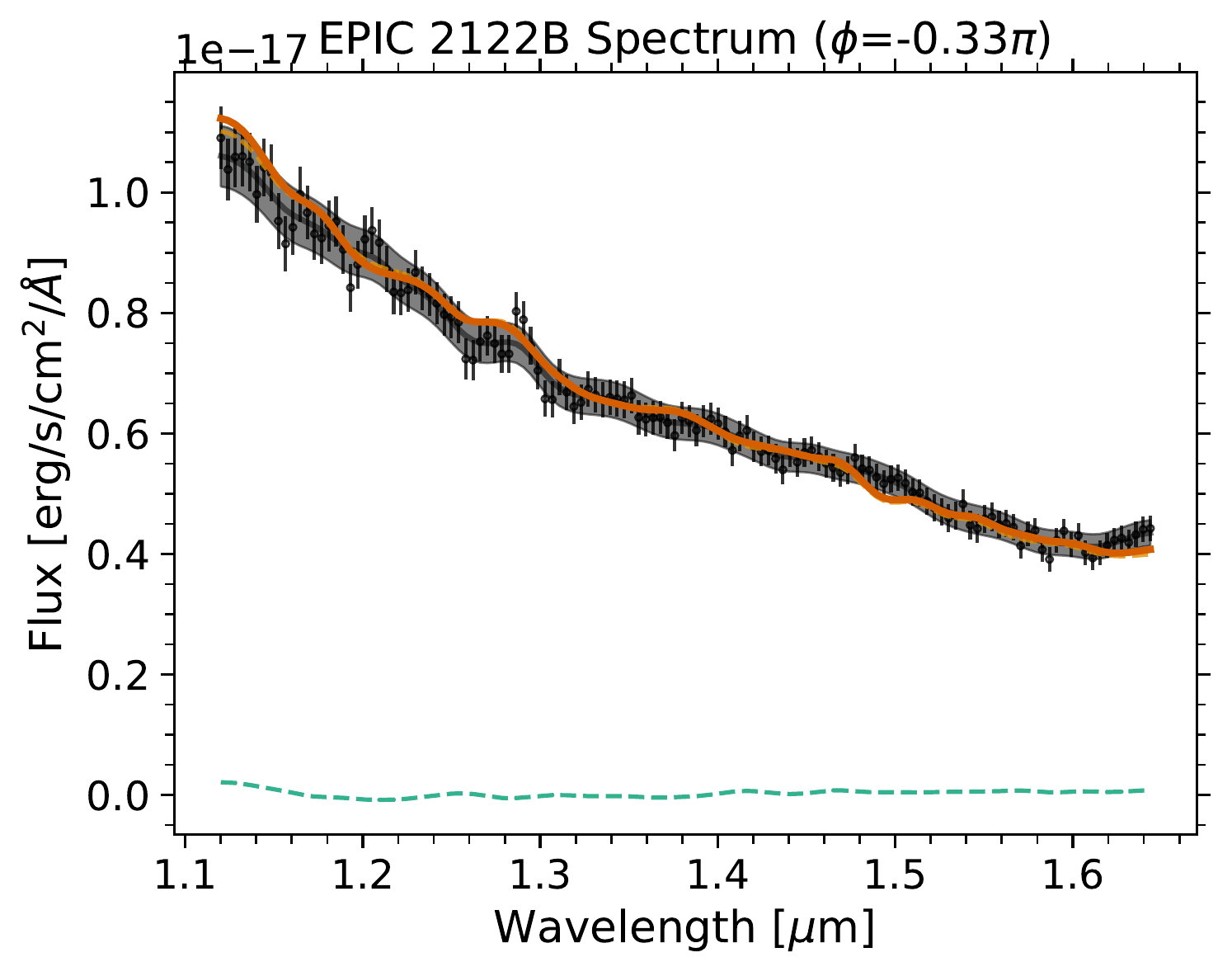} 
    \includegraphics[height=0.38\textwidth]{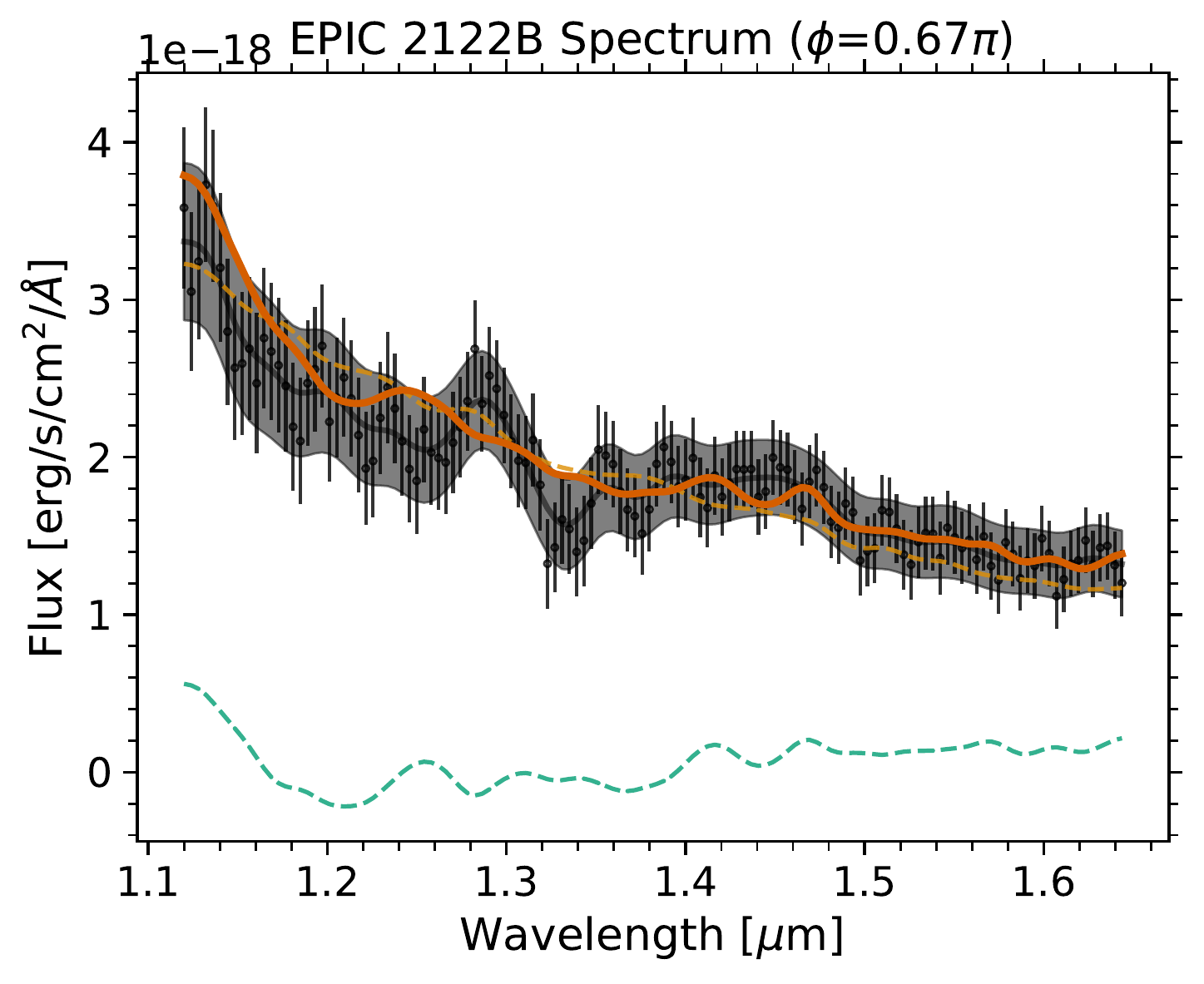}
    \includegraphics[height=0.38\textwidth]{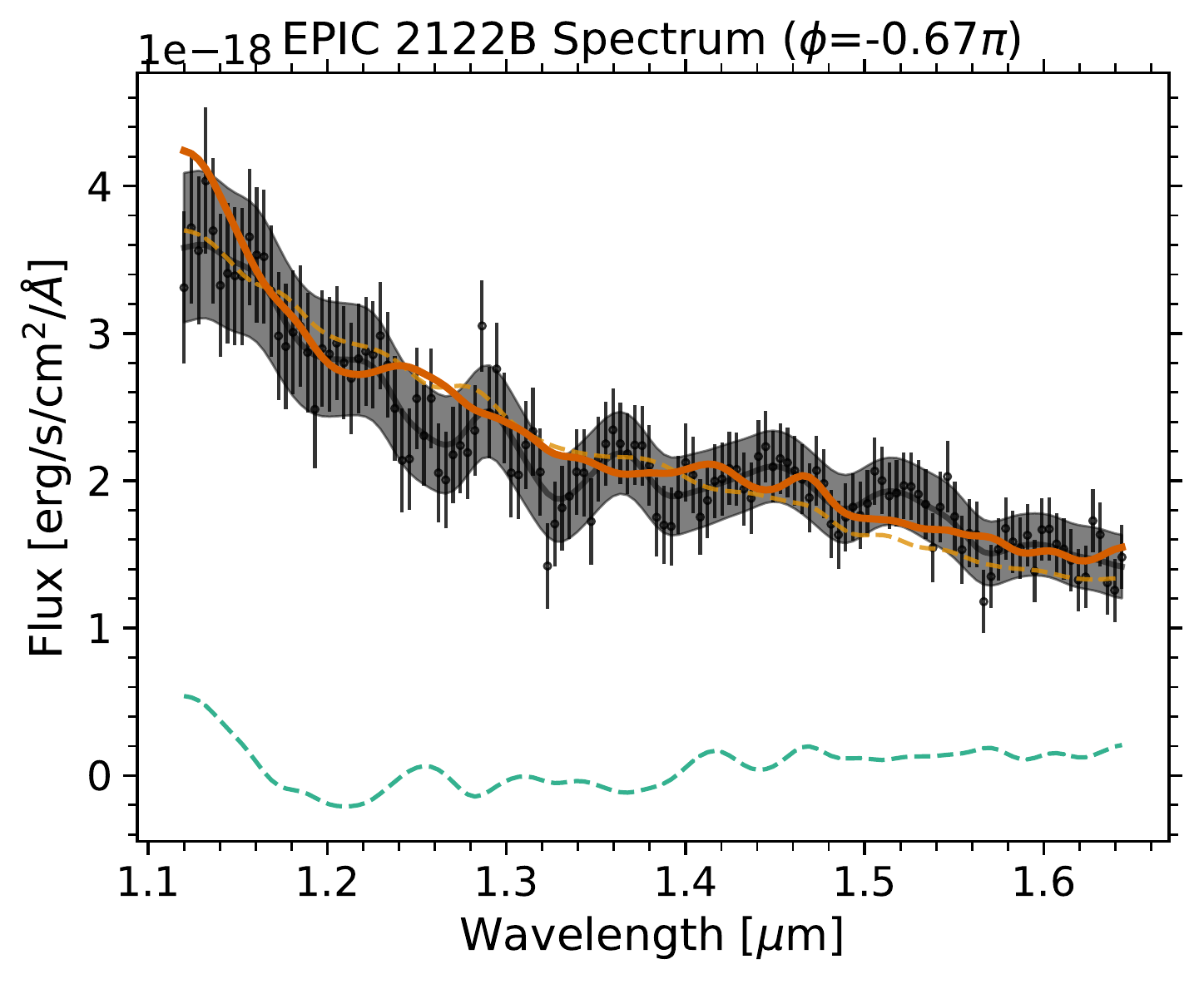}
    \includegraphics[width=0.4\textwidth]{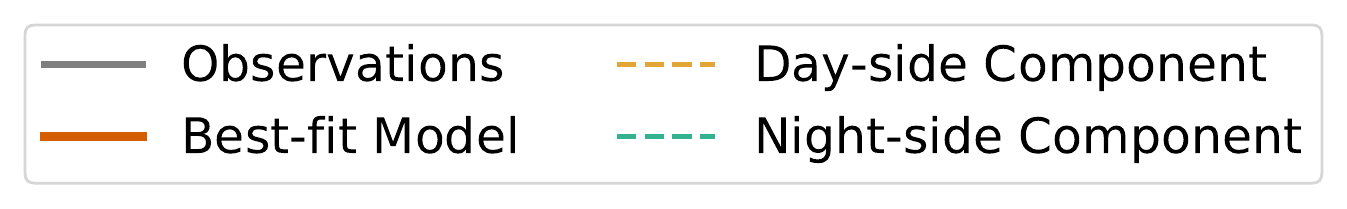}
    \caption{Similar to Figure~\ref{fig:WD0137specfit} but for EPIC~2122.}
    \label{fig:EPIC2122specfit}
  \end{figure*}
  
  In both cases, this semi-empirical model fits the observations  very well at every orbital/rotational phase. Figures \ref{fig:WD0137specfit} and \ref{fig:EPIC2122specfit} show the fitting results of spectral at four representative phases for WD~0137 and EPIC~2122, respectively. In every case, the model agrees with the observations within $1\sigma$. The excellent fit confirms the premise that the phase-resolved spectra of the irradiated brown dwarfs can be accurately reduced into a two-dimensional linear space spanned by the day- and night-side spectra. This remarkable result significantly simplifies the analysis of the spectral time series. We no longer need to examine every spectrum but rather conduct detailed investigations of the day- and night-side spectra and the contribution factors.

  \begin{figure}[!h]
    \centering
    \includegraphics[width=\columnwidth]{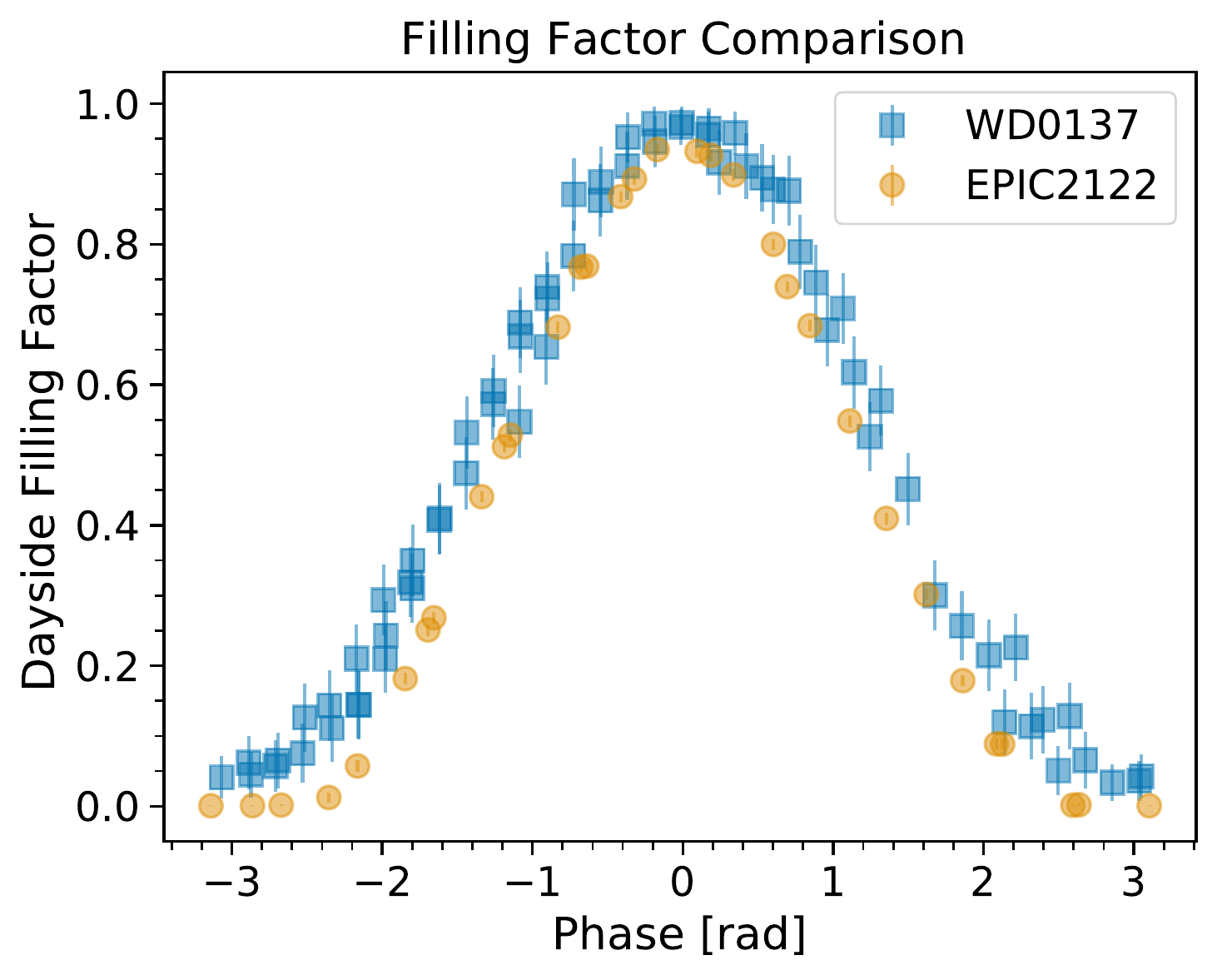}
    \caption{Comparing the day-side spectral filling factors as functions of rotational phase between WD~0137B (blue squares) and EPIC~2122B (yellow circles). Both have nearly sinusoidal shapes, but the EPIC~2122B curve has a flatter valley and a more narrow peak.}
    \label{fig:fillingfactor}
  \end{figure}

  In Figure \ref{fig:fillingfactor},  phase-resolved day-side contribution $f(\phi)$ that best fit to WD~0137B and EPIC~2122B are compared. They both have nearly sinusoidal shapes. However, the EPIC~2122B curve has a flatter night-side valley and a more narrow peak. This difference may suggest that the temperature longitudinal gradient is steeper in the day-side of EPIC~2122B than WD~0137B.

  \subsection{Day- and Night-sides Brightness Temperatures of the two Brown Dwarfs}
\newcommand{\TB}{\ensuremath{T_{\mathrm{B}}}\xspace}
  \begin{figure*}[t]
    \centering
    \includegraphics[height=0.38\textwidth]{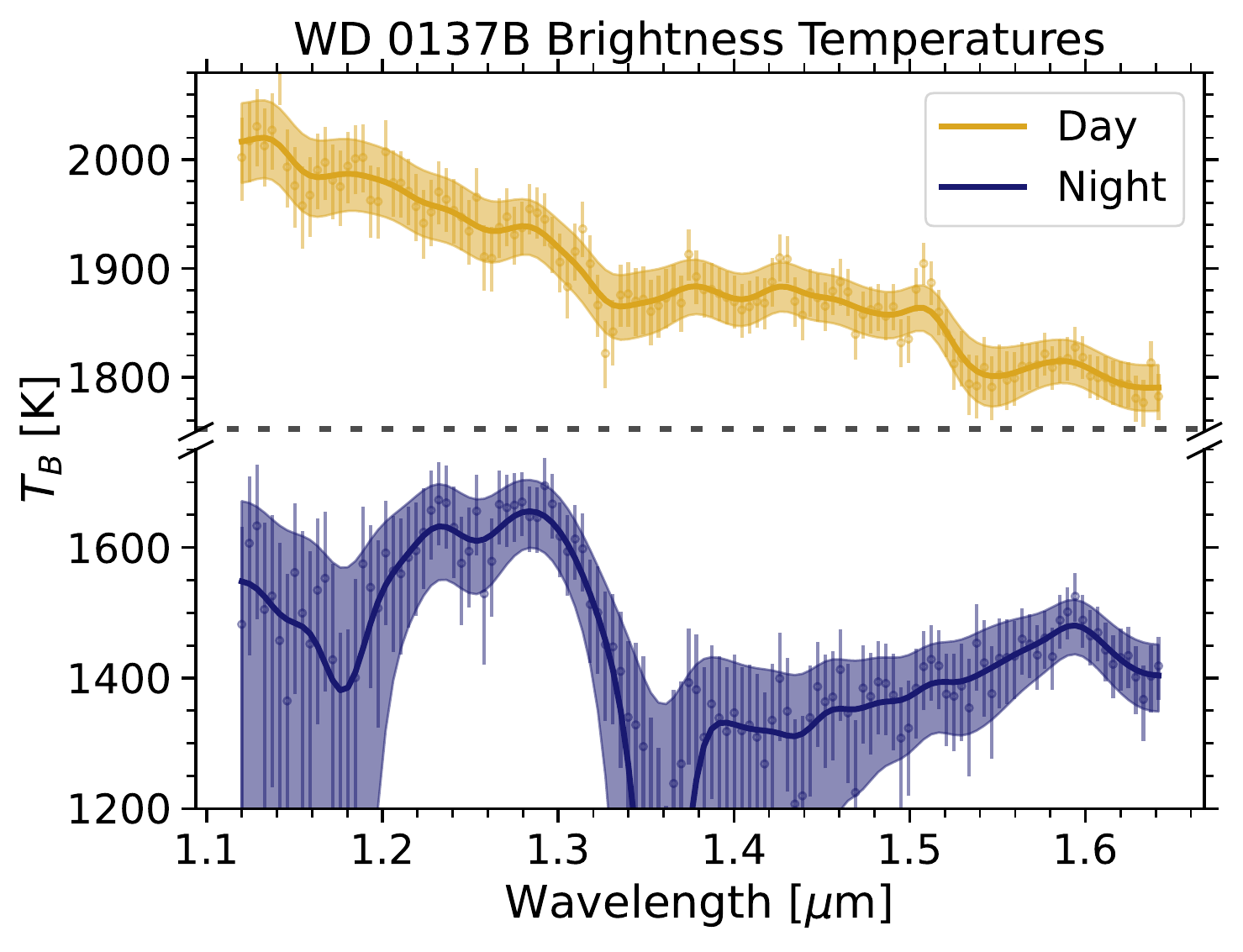}
    \includegraphics[height=0.38\textwidth]{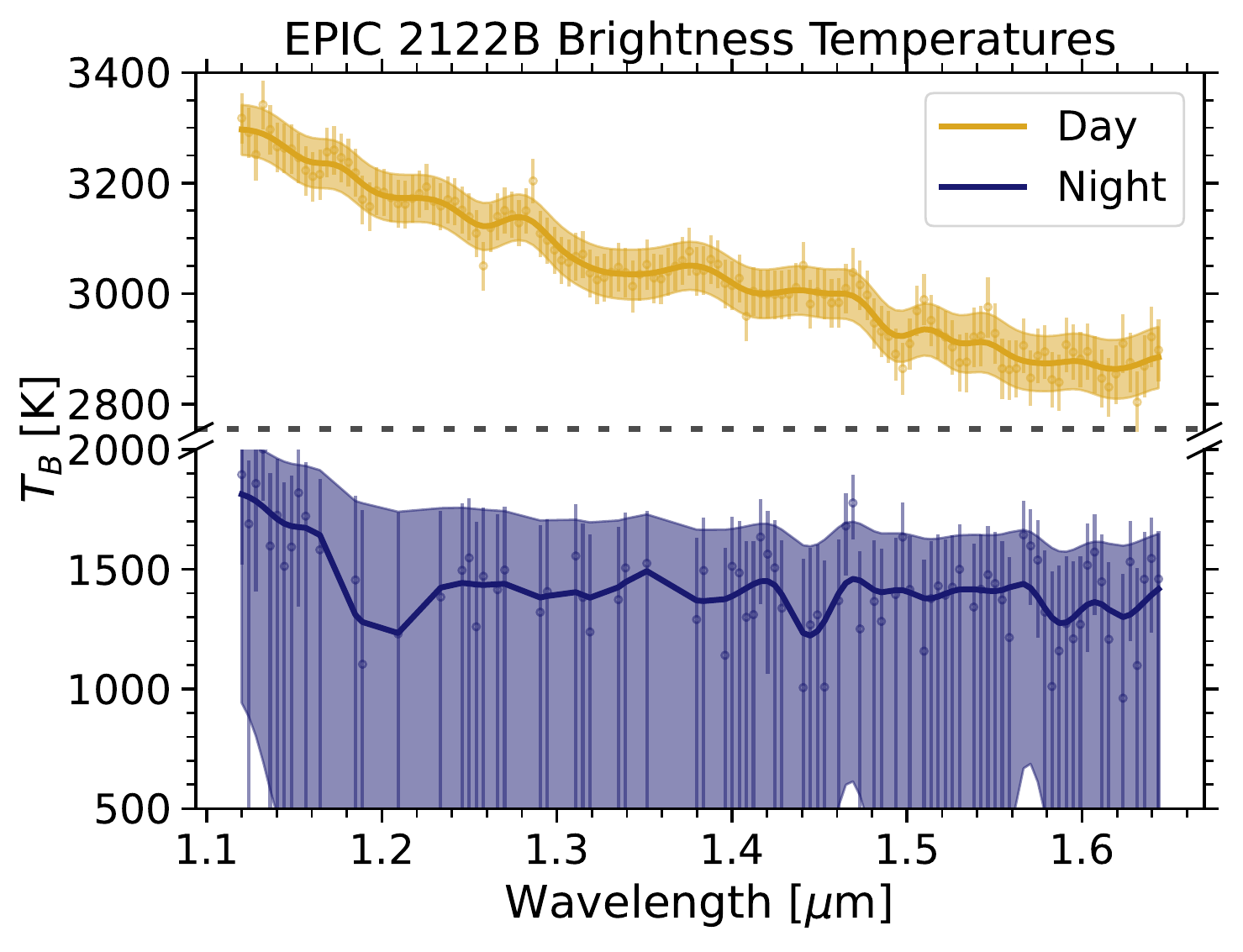}
    \caption{The day- and night-sides brightness temperatures of WD 0137B (left) and EPIC 2122B (right). The plotting styles and conventions follow Figure~\ref{fig:daynightspec}: the yellow and purple dots are \TB of day- and night-sides, respectively; error bars are $1\sigma$ uncertainties; lines and swashes are observations convolved with a Gaussian kernel to demonstrate the instrument's spectral resolving power. The $y$-axes are discontinuous to properly zoom into the temperature ranges and demonstrate their wavelength-dependence.}
    \label{fig:brightnessT}
  \end{figure*}

  Because the day- and night-sides can serve as the linear bases that represent the entire phase curves sufficiently and accurately to the limits of the current data,  we focus the analysis on these two spectra. First, we characterize the spectra by deriving the wavelength specific brightness temperatures (\TB); in the next subsection we compare the spectra to substellar models.

  Expressing emission spectra as \TB offers a straightforward view of the atmospheric vertical structure and is widely used brown dwarf and exoplanet studies \citep[e.g.,][]{Ackerman2001, Morley2014}, as well as previous works on our targets \citep{Casewell2015, Casewell2018}. Combined with a thermal ($T\mbox{-}P$) profile, \TB can be mapped to a pressure level where the emission flux emerges. This mapping enables a dissect of the vertical structures of the substellar atmospheres in which the pressure level of the photosphere varies by more than an order of magnitude as a function of wavelength due to strong molecular absorption bands. To facilitate comparisons with other irradiated atmospheres (see \S\ref{sec:day-night-contrast}), we present the day- and night-sides spectra of WD 0137B and EPIC 2122B in brightness temperatures. It is worth noting that our wavelength range is not in the Rayleigh-Jeans tail, so the Planck function, rather than the Rayleigh-Jeans law, should be used in deriving \TB.

We define \TB as a hemispherically-averaged measure. So its value satisfies the equation
\begin{equation}
  \label{eq:5}
  f(\lambda) = \frac{\pi\,R_{\mathrm{BD}}^{2}}{d^{2}}\,B(\TB),
\end{equation}
in which $f(\lambda)$ is the observed brown dwarf flux density at wavelength of $\lambda$, $d$ is the distance of the system, $R_{\mathrm{BD}}$ is fixed to $1\,R_{\mathrm{Jup}}$ and $B(\TB)$ is the Planck function.  At wavelength $\lambda$, we invert Equation~\ref{eq:5} to find $T_{\mathrm{B},\lambda}$. The results are shown in Figure~\ref{fig:brightnessT}. In Table~\ref{tab:temperatures}, we list the \TB of the two brown dwarfs in several representative bands (G141 broadband, F127M, F139M, and F153M). In all four cases (day/night of WD~0137B and EPIC~2122B), \TB demonstrates strong wavelength dependence. In the day-sides of both brown dwarfs, $\TB$ decreases monotonically towards longer wavelengths, from \SI{2010}{\kelvin} at \SI{1.12}{\micro\meter} to \SI{1780}{\kelvin} at \SI{1.65}{\micro\meter} in WD~0137B and from \SI{3290}{\kelvin} to \SI{2870}{\kelvin} in EPIC~2122B. In the night-sides, WD~0137B's $\TB$ has non-monotonic variations, which corroborates the water absorption in its night-side spectrum. It peaks at the center of $J$-band (\SI{1.3}{\micro\meter}) with a value of \SI{1650}{\kelvin}, drops below \SI{1400}{\kelvin} at \SI{1.4}{\micro\meter}, and then rises above \SI{1450}{\kelvin} at \SI{1.6}{\micro\meter}. As for the night-side of EPIC~2122B, its more noisy spectra lead to a wider $\TB$ range: from \SIrange{500}{1500}{\kelvin}. These variations of $\TB$ demonstrate that the G141 spectra probe a wide range of the brown dwarfs' atmospheric pressure levels where temperatures can vary by a few hundred \si{\kelvin}.

\citet{Casewell2015} derived the day- and night-side $\TB$ of WD~0137B in multiple bands, including a few that have overlapping wavelength ranges with our observations: $T_{\mathrm{B,\,day}}=2418_{-329}^{+201}$~\si{\kelvin} in $J$, $T_{\mathrm{B,\,day}}=1585\pm329$\,\si{\kelvin} in $H$, and $T_{\mathrm{B,\,night}}=2085_{-769}^{+287}$\,\si{\kelvin} in $J$. In comparison, our day-side measurements are lower in the $J$-band ($T_{\mathrm{B,\,day}}=1935\pm28$\,\si{\kelvin}, using the F127M band result as an approximation) but higher in the $H$-band ($T_{\mathrm{B,\,day}}=1833\pm23$~\si{\kelvin}, using the F153M band result as an approximation). We find a somewhat less wavelength-dependent $\TB$ than \citet{Casewell2015}. Because the white dwarf templates are almost identical, the inconsistent brightness temperatures are likely caused by different photometric calibrations. However, it is worth noting that the day/night brightness temperature contrasts in the $J$-band are consistent between the two observations.  For the rest of the paper, we will focus on discussing the WFC3-based measurements. It is also worth noting that the night-side $\TB$ of WD~0137B's in Spitzer \SI{3.6}{\micro\meter} and \SI{4.5}{\micro\meter} bands listed in \citet{Casewell2015} are in agreement with the recently identified uniform night-side temperatures of hot Jupiters \citep{Keating2019,Beatty2019}.

  \subsection{Day-side Spectral Model Comparisons}

  \begin{figure*}[t]
    \centering
    \includegraphics[width=0.48\textwidth]{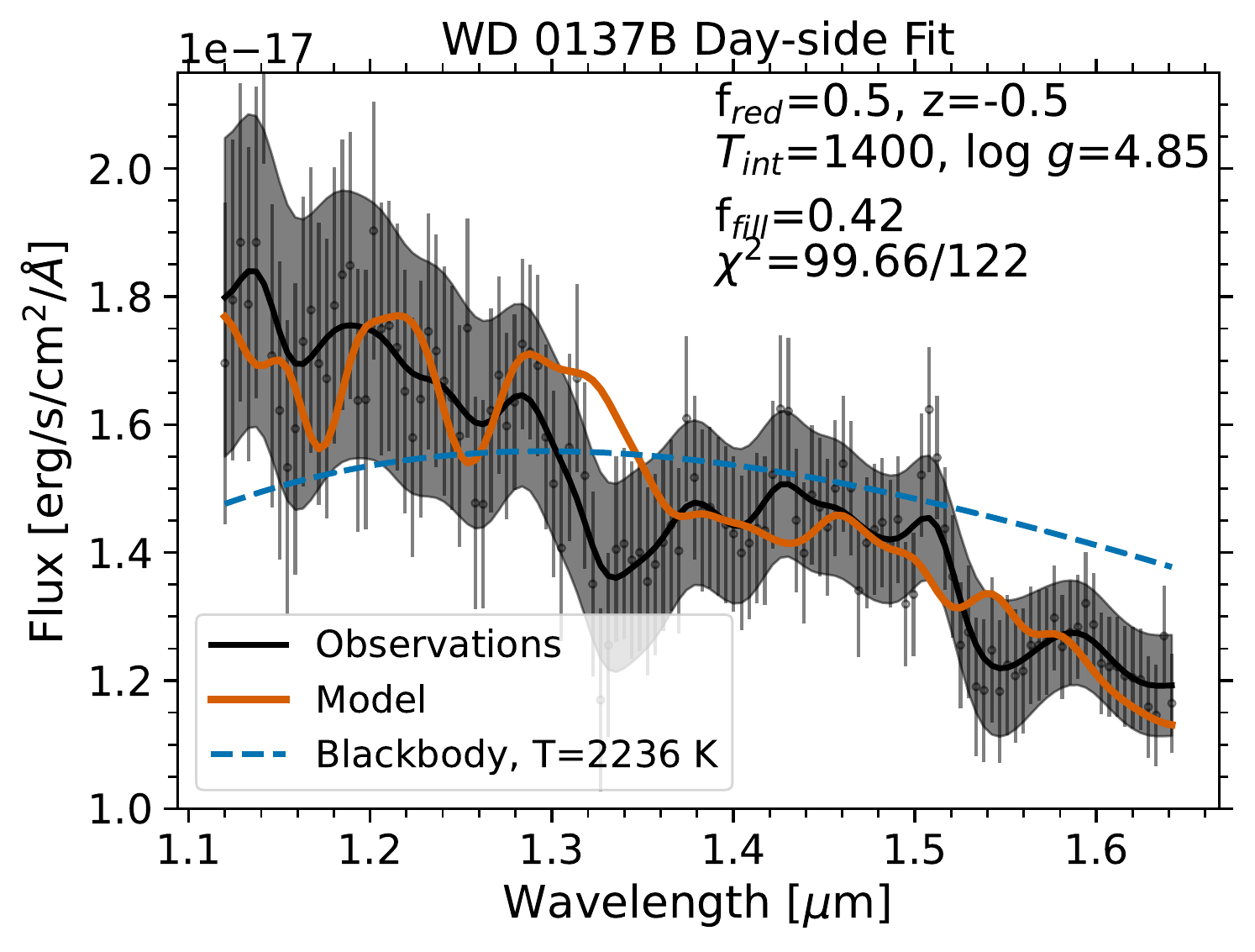}
    \includegraphics[width=0.48\textwidth]{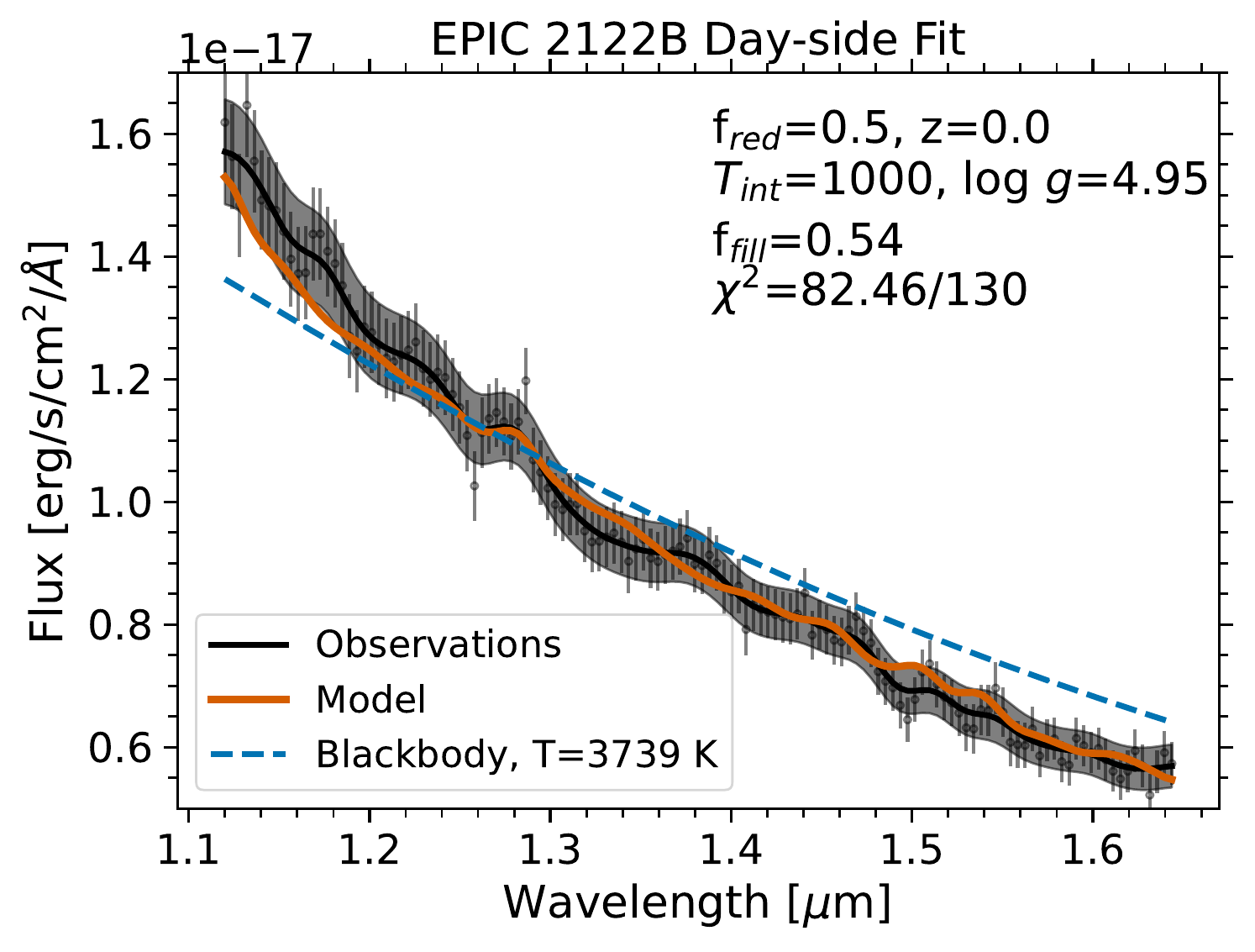}
    \caption{Fitting the irradiated atmospheric models to the day-side spectra of WD~0137B (left) and EPIC~2122B (right). In both panels, observational data are plotted in the same conventions as in Figure~\ref{fig:daynightspec}. The best-fitting models are represented by orange solid lines. In both cases, the irradiated atmospheric models fit the observations well, resulting in reduced-$\chi^{2}$ values close to unity. Irradiation makes the model spectra steeper than the blackbody (blue dashed line) and thus fit the observations better.} 
    \label{fig:daysideFit}
  \end{figure*}

  \begin{figure*}[t]
    \centering
    \includegraphics[height=0.38\textwidth]{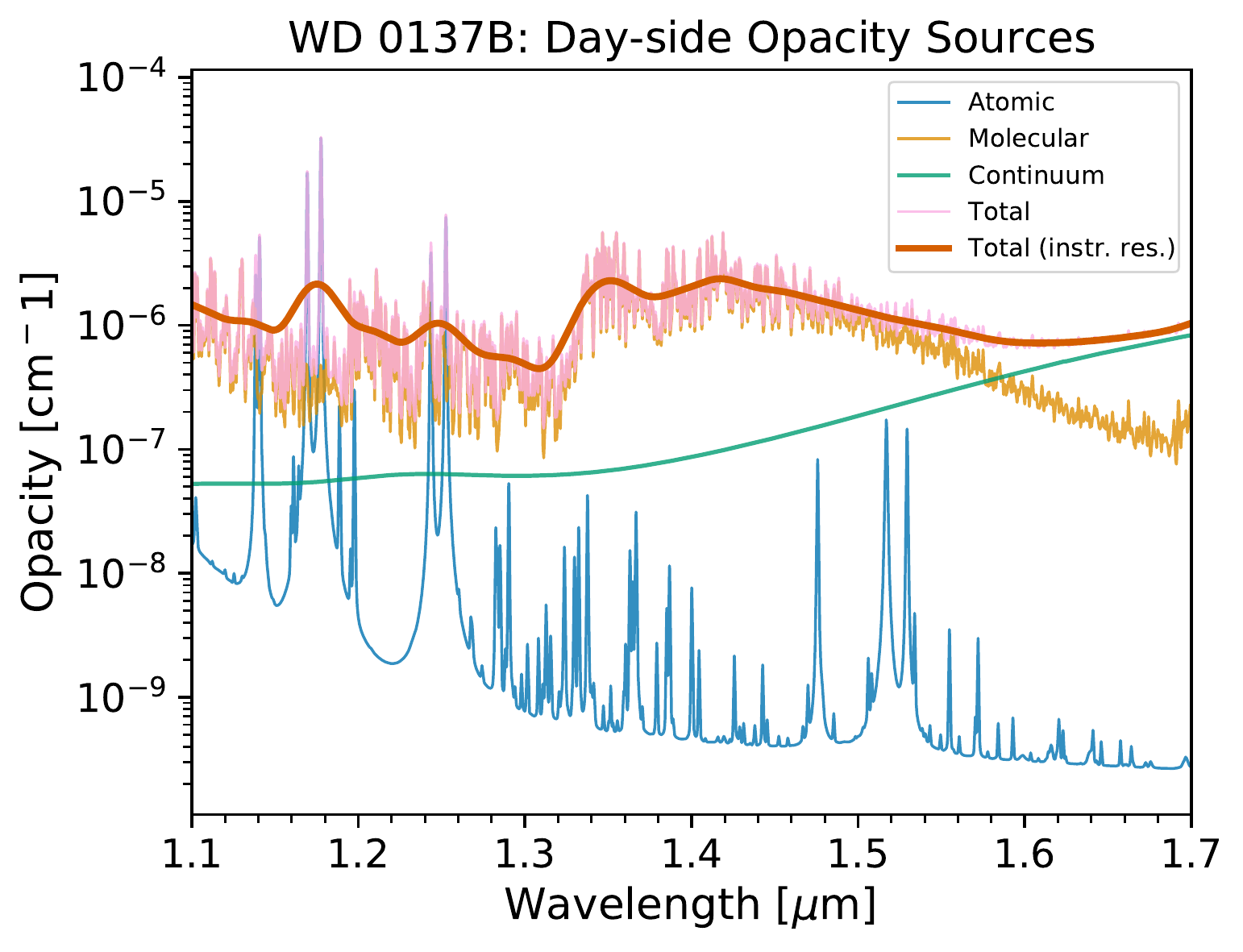}
    \includegraphics[height=0.38\textwidth]{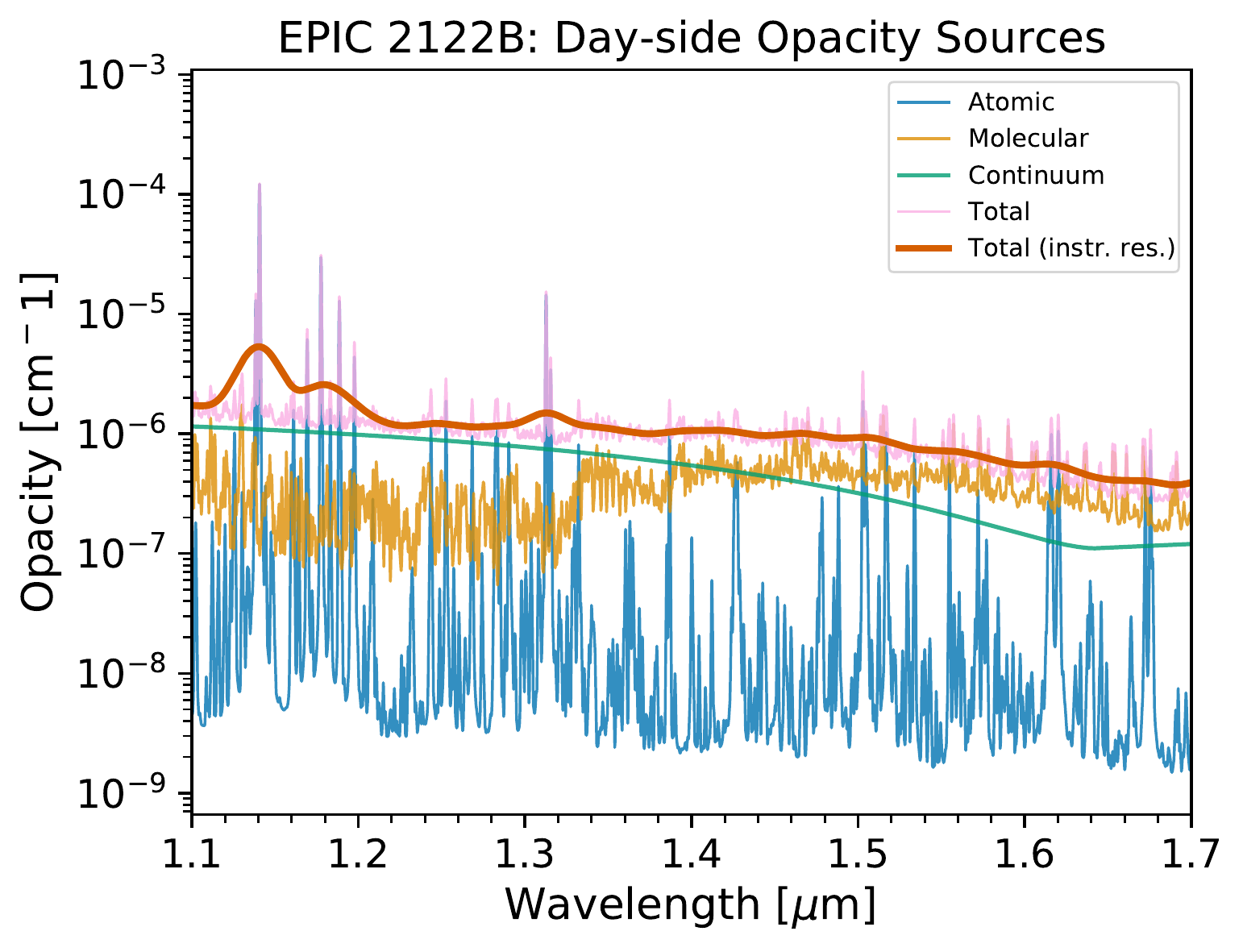}
    \includegraphics[height=0.42\textwidth]{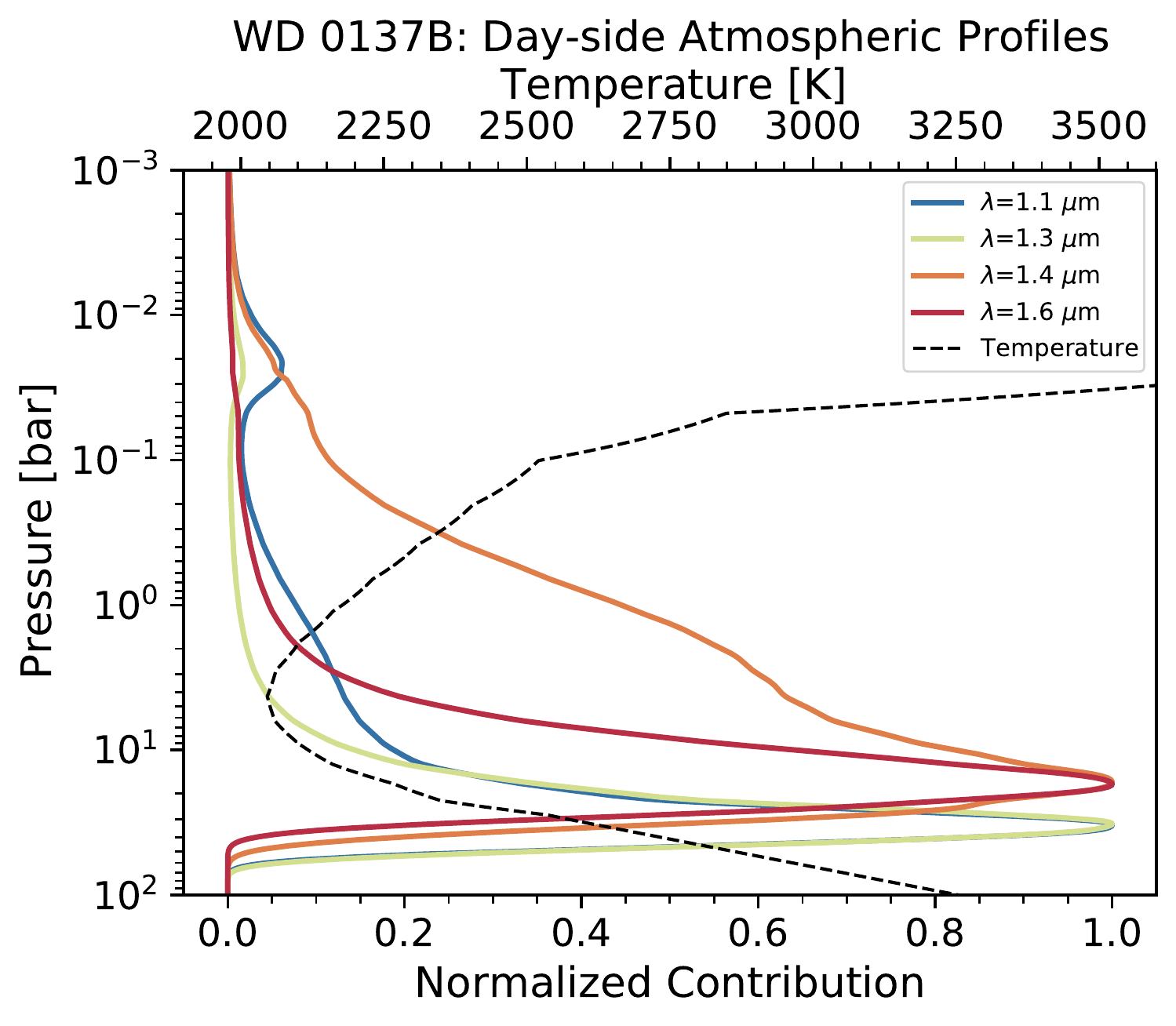}
    \includegraphics[height=0.42\textwidth]{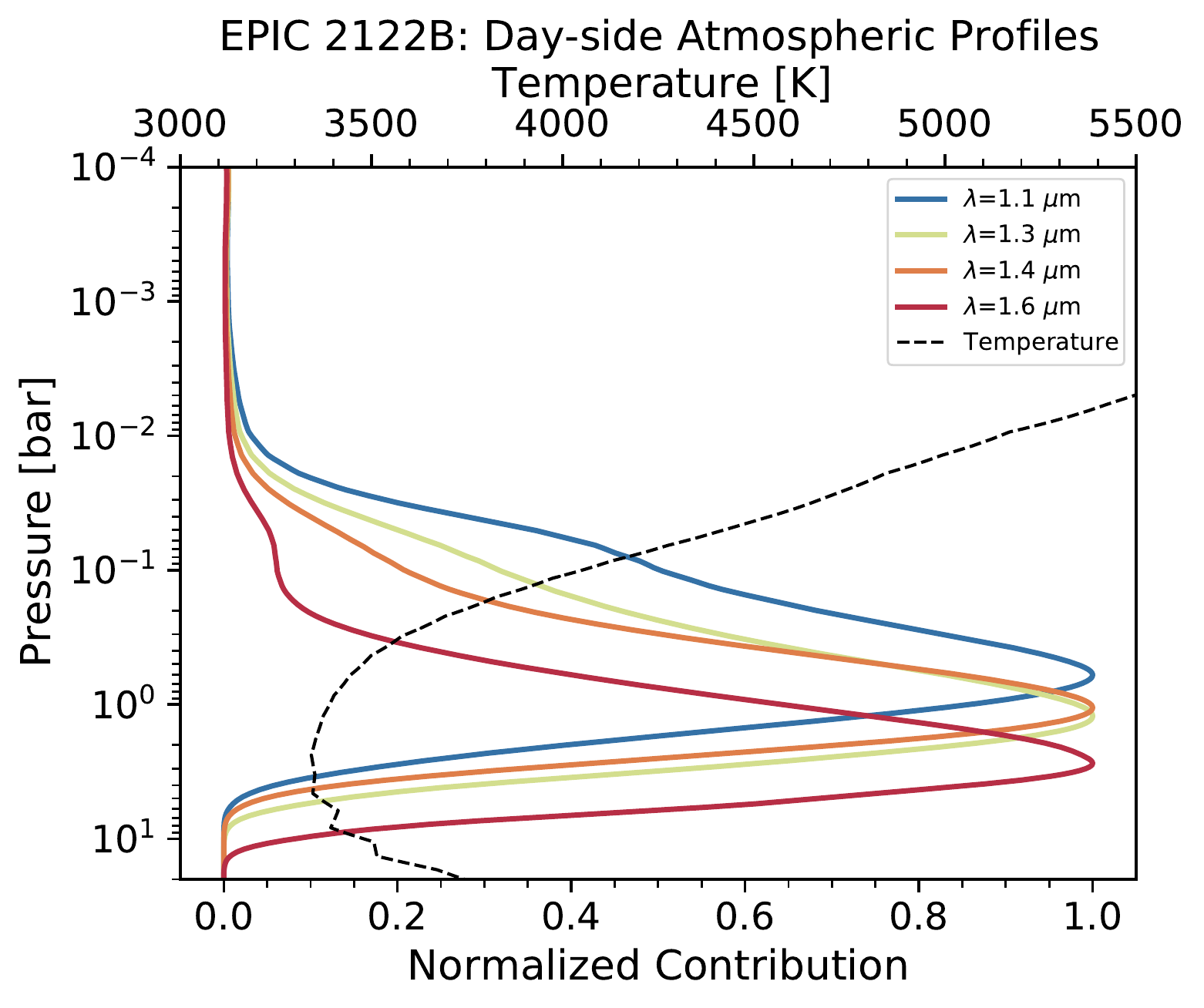}
    \caption{The opacity sources and atmospheric profiles of the best-fitting day-side models of WD~0137B (left) and EPIC~2122B (right). The upper panels demonstrate the opacity contributions from various sources. Atomic, molecular, continuum (mostly H$_{2}$-H$_{2}$/H$_{2}$-He CIA for WD~0137B and H$^{-}$ for EPIC~2122B), and total opacities are shown in blue, yellow, green, and pink lines. The thick red lines show the total opacities down-sampled to G141's spectral resolution ($R\sim130$ at \SI{1.4}{\micro\meter}).  WD~0137B's opacities are measured at $P=\SI{6}{\bar}$ and EPIC~2122B's opacities are measured at $P=\SI{1}{\bar}$. These pressure levels correspond to the approximate positions where the contribution functions at the \SI{1.4}{\micro\meter} H$_{2}$O band peak. The flux contribution functions and thermal profiles are presented in the lower panels. The blue, green, orange, and red solid lines show the normalized flux contribution functions at wavelengths of 1.1, 1.3, 1.5, and 1.6~\si{\micro\meter}. The black-dashed lines are the thermal profiles. In the day-sides of both objects, temperature inversions occur in the pressure range of \SIrange{1}{10}{bar}.}
    \label{fig:atmo_profiles}
  \end{figure*}

\newcommand{\fred}{\ensuremath{f_{\mathrm{red}}}\xspace}
\newcommand{\Tint}{\ensuremath{T_{\mathrm{int}}}\xspace}

  We seek models that best fit to the observed spectra. For the day-side spectra, we adopted the irradiated atmospheric spectral grids, which are based on the model presented in \citet{Lothringer2020}. The grids for WD~0137B and EPIC~2122B are derived separately to accommodate for the different strengths and spectral energy distributions of the white dwarf's irradiation, and the irradiation's subsequent effect on the thermal profiles and photochemistry in the brown dwarfs' atmospheres. Each grid is defined by four parameters: metallicity ($Z$), surface gravity (\loggbd{}), interior temperature (\Tint), and the heat redistribution fraction (\fred) that represents the redistribution of the heat from day-side to night-side (see Equation (1) in \citealt{Lothringer2020}\footnote{This equation defines the quantity $T_{\mathrm{irr}}$, which is the effective temperature of the brown dwarf if there is no internal heat, by combining a heat redistribution factor $f$, the brown dwarf Bond albedo $A$, the orbital distance $a$, the white dwarf radius $R_{\mathrm{WD}}$, and the white dwarf effective temperature $T_{\mathrm{eff,\,WD}}$. Those authors adopted $A=0$, $a=0.65R_{\odot}$, $R_{\mathrm{WD}}=0.186R_{\odot}$, and $T_{\mathrm{eff,\,WD}}=\,$\SI{16500}{\kelvin} for WD~0137; $A=0$, $a=0.44R_{\odot}$, $R_{\mathrm{WD}}=0.017R_{\odot}$, and $T_{\mathrm{eff,\,WD}}=$\,\SI{24900}{\kelvin} for EPIC~2122.}). The metallicity is sampled at three values $Z=0,\,\pm0.5$. The surface gravity grid sample points differ slightly between WD~0137B and EPIC~2122B models to account for their different masses: $\loggbd=4.85,\, 5.10,\,\mbox{and }5.35$ for WD~0137B and $\loggbd=4.95,\, 5.20,\,\mbox{and }5.45$ for EPIC~2122B, corresponding to $M_{\mathrm{BD}}=28.6,\,50.8,\,\mbox{and }\,90.3\,\mjup$ and $M_{\mathrm{BD}}=36.0,\,63.9,\,\mbox{and }\,113.7\,\mjup$, respectively (assuming $R=1\,R_{\mathrm{Jup}}$ for both brown dwarfs). The interior temperature is defined as the effective temperature of a non-irradiated brown dwarf and sampled at $\Tint=1400\,\mbox{and }\,1800$~\si{\kelvin} for WD~0137B and $\Tint=1000,\,1500,\,\mbox{and }\,2000$~\si{\kelvin} for EPIC~2122B. The heat redistribution fraction grids include three values: $\fred=0.125, 0.25, \mbox{ and } 0.5$. In the zero albedo scenario, $\fred=0.25$ corresponds to absorbed heat evenly redistributed across the sphere and $\fred=0.5$ corresponds to absorbed heat evenly redistributed within the day-side hemisphere. The $\fred=0.125$ case is an \emph{ad hoc} representation of a high albedo day-side.

  In fitting the grid models, we include a filling factor ($f_{\mathrm{fill}}$) as a free parameter. It represents the proportion of the projected area where the day-side emission comes from, and to account for the possibility that the observed day-side emission is dominated by a hot spot that only occupies part of the day-side hemisphere. The inclusion of the filling factor is further supported by the study of \citet{Taylor2020} who found that a single thermal profile with a filling factor performs as well as the more complex multi-profile approach in the retrieval of a heterogeneous atmosphere composed of high-contrast hot and cool components. For each grid point, we optimize $f_{\mathrm{fill}}$ such that it minimizes the residuals and record the $\chi^{2}$ values. Then, $\chi^{2}$ values at all grid points are compared and  the model that has the overall least $\chi^{2}$ is the best-fitting one. The best-fitting model for WD~0137B's day-side spectrum has $Z=-0.5$, $\logg=4.85$, $\Tint=\SI{1400}{\kelvin}$, $\fred=0.5$, and $f_{\mathrm{fill}}=0.43$, although increasing $Z$ to 0 or $\logg$ to 5.10 also results in reasonably good fits. All models with $\fred=0.5$ and $Z\ge0$ fit EPIC~2122B's day-side  well, and the best one has $Z=0$, $\logg=4.95$, $\Tint=\SI{1000}{\kelvin}$, $\fred=0.5$, and $f_{\mathrm{fill}}=0.54$. For references, we also fit Planck functions with $f_{\mathrm{fill}}$ fixed to the best-fitting values in the irradiated model fits and show the results in Figure~\ref{fig:daysideFit}. The best-fitting blackbody temperatures are \SI{2236}{\kelvin} and \SI{3739}{\kelvin} for WD~0137B and EPIC~2122B, respectively. Non-irradiated spectral grids (Sonora cloudless grid \citealt{Marley2018} and BT Settl \citealt{Allard2012}) are experimented with, but all model spectra result in poor fits.

  Despite the fact that these two brown dwarfs receive different irradiation, the day-side spectral fitting results of WD~0137B and EPIC2122~B share multiple commonalities. In both cases, \fred{} is the most critical parameter determining the fit qualities, and $\fred=0.5$, the upper bound in the grid, is required to reproduce the steep slopes in the observations. This value corresponds to the day-side only redistribution scenario. This result is in agreement with what we find from the band-integrated phase curves: circulation is inefficient in transporting heat from day- to night-side in these fast rotating objects. For both brown dwarfs, the best-fitting $f_{\mathrm{fill}}$ values are around 50\%. This result, again, is in agreement with what the phase curves suggest: there is a hot spot that partially covers the day-side hemisphere. Additionally, because the distance between the binary components is on the same order as the brown dwarf radius, the incident angle at the limb regions of the brown dwarf is greater than those in the parallel beam incidence scenario. This leads to an even smaller amount of energy received by the limb, further reducing its contribution to the day-side spectra. Based on the best-fitting results, the day-side effective temperatures ($T_{\mathrm{eff,\,day}}=(T_{\mathrm{irr}}^{4} + T_{\mathrm{int}}^{4})^{1/4}$, $T_{\mathrm{irr}}$ is the irradiation temperature defined in \citealt{Lothringer2020}) are \SI{2430}{\kelvin} and \SI{4050}{\kelvin} for WD~0137B and EPIC2122B, respectively. 

  The reason that the irradiated atmospheric models perform better than non-irradiated models or blackbodies can be well illustrated by Figure~\ref{fig:atmo_profiles}, in which we break down the opacity contributions from atomic, molecular, and continuum sources, as well as present the flux contribution functions and thermal profiles of the best-fitting models. Due to their difference in irradiation and temperatures, the opacities sources differ significantly between the two brown dwarfs. In the cooler WD~0137B, molecular opacities dominate most of the \SIrange{1.1}{1.7}{\micro\meter} wavelength range, except at a few alkaline lines where atomic opacities peak and the $\lambda>\SI{1.6}{\micro\meter}$ region where the continuum opacities (from H$_{2}$-H$_{2}$/H$_{2}$-He collision-induced absorption, CIA) become significant. In the hotter EPIC~2122B, the continuum opacities contribute the most at shorter wavelengths ($\lambda<\SI{1.4}{\micro\meter}$) but their importance drops below molecular opacities at longer wavelengths ($\lambda>\SI{1.4}{\micro\meter}$). These differences result in a divergence in their flux contribution functions: in WD~0137B, longer wavelengths probe lower pressure levels (higher altitude); in EPIC~2122B, longer wavelengths probe higher pressure levels (lower altitude). There is another crucial difference between the day-sides of the two brown dwarfs: while thermal inversions are predicted in both day-sides, WD~0137B's day-side flux in the G141 bandpass is mostly from below the inversion but EPIC~2122B's day-side flux is mostly from above the inversion. Combining the effects of the contribution functions and the thermal profiles, we find that the shorter wavelength emission is from a hotter region (and vice versa) in both brown dwarfs. Therefore, the day-side brightness temperature should decrease with wavelength, and this is exactly what we observed in WD~0137B and EPIC~2122B (Figure~\ref{fig:brightnessT})! However, the blackbody and non-irradiated models predict either a flat \TB trend or a decrease of \TB in molecular bands, contrary to the observed trend. We attribute this considerable disagreement to incorrect opacities and thermal profiles of those models in describing the brown dwarfs' day-sides.

  The reason that the day-side brightness temperatures and effective temperatures differ are twofold. First, our brightness temperatures are hemispherically averaged quantities. However, the effective temperatures, which are derived from the spectral model fitting, characterize the hot spots, which dominate the emission, even though they occupy only fractions of the day-side surface area. Therefore, it is not surprising that the effective temperatures are higher than the band-averaged brightness temperatures, because the former represents only the hotter part of the hemisphere. Similarly, $f_{\mathrm{fill}}$ is also applied to the Planck spectra in Figure~\ref{fig:daysideFit}, hence on the shorter wavelength side ($\lambda < 1.25$~\si{\micro\meter}), the blackbody models have lower flux than the observed spectra despite their higher temperatures. Second, brightness temperatures characterize the pressure levels where the emission at a specific wavelength comes from, while effective temperatures describe the source function. In the irradiated atmospheres that contain significant molecular and continuum absorption sources, a joint use of the two quantities provides a more complete view of the atmospheric thermal structure (Figure~\ref{fig:atmo_profiles}).

  \subsection{Night-side Spectral Model Comparisons}

  \begin{figure*}[t]
    \centering
    \includegraphics[width=0.48\textwidth]{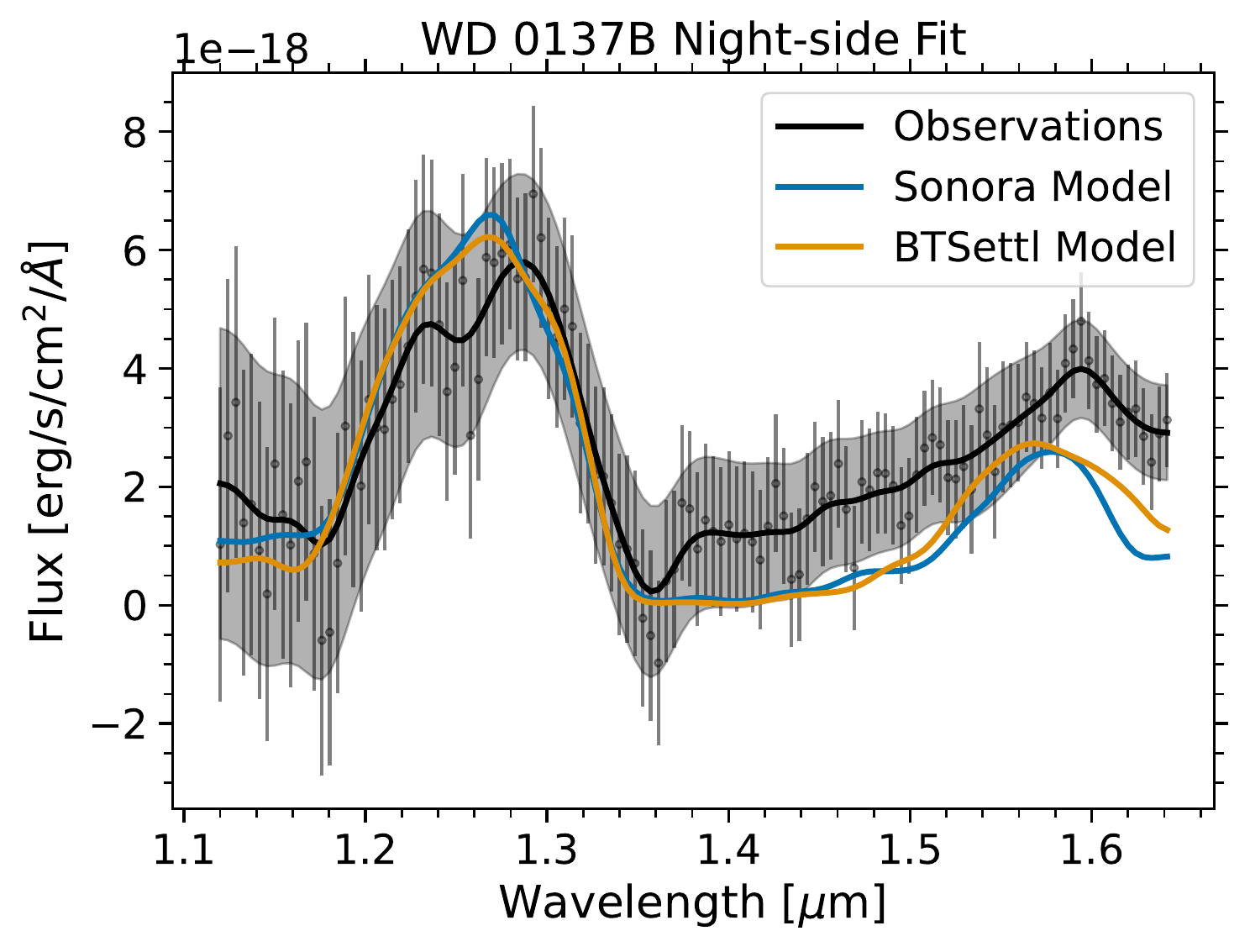}
    \includegraphics[width=0.48\textwidth]{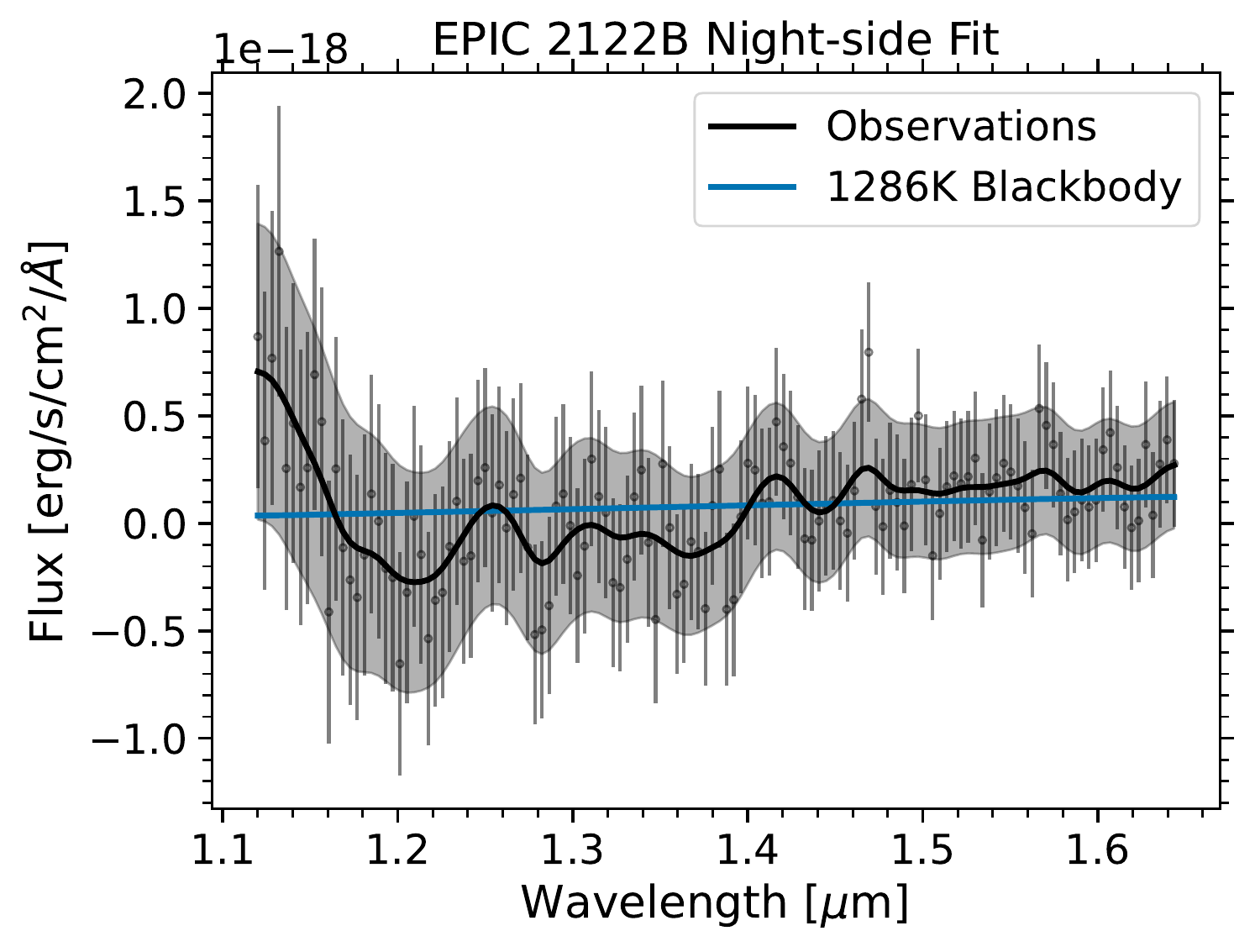}
    \caption{\revise{The night-side spectra of WD~0137B (left) and EPIC~2122B (right) and their best-fitting models. Same as in Figure~\ref{fig:daysideFit}, the observations are shown in black solid lines,  the $1\sigma$ uncertainties are in black shades, and the models are in colored lines. The left panel compares the BT Settl and Sonora models with the nigh-side spectrum of WD~0137B. In the right panel, a ${\sim}1300$~\si{\kelvin} blackbody fits reasonably well to the low SNR night-side spectrum of EPIC~2122B.}}
    \label{fig:nightsideFit}
  \end{figure*}

  For the night-side spectrum of WD~0137B, we experimented with two non-irradiated and clear atmosphere grids: the BTSettl \citep{Allard2012} and Sonora \citep{Marley2018} models. In this fit, we fix the scaling of the model to $R_{\mathrm{BD}}^{2}/d^{2}$, in which $R_{\mathrm{BD}}=1\,R_{\mathrm{Jup}}$ and $d$ is derived from the Gaia DR2 parallaxes \citep{Gaia2018}, leaving the effective temperature ($\teff$) and surface gravity ($\logg$) as the only free parameters. We optimize these parameters using a least-$\chi^{2}$ criterion, and find that the observed spectra favor a \SIrange{1000}{1100}{\kelvin} \teff{} and a high \logg{} of 5.5 for both grids.  In the left panel of Figure~\ref{fig:nightsideFit}, the best-fitting BTSettl and Sonora models are compared with the WD~0137B's night-side spectrum. The models fit well in the $J$-band but overestimate the water absorption depth at \SI{1.4}{\micro\meter}. This mismatch suggests that the temperatures at high altitude are too low in the models compared to the brown dwarf, resulting in deeper water bands than reality.

  Due to the low SNR of EPIC~2122B's night-side spectrum, we only fit a blackbody model and find a best-fitting \teff{} of ${\sim}1300$~\si{\kelvin}. This result is within the range of its night-side brightness temperatures. 

    \renewcommand{\arraystretch}{1.2}
  \begin{deluxetable*}{lccccc}

\tablecolumns{6}
\tablewidth{0pt}
\tablecaption{Day- and Night-sides Temperatures of WD 0137B and EPIC 2122B \label{tab:temperatures}}
\tablehead{
  \colhead{Object} & \multicolumn{4}{c}{$T_{B}$ [K]} & \colhead{\teff~[K]} \\
\colhead{} & \colhead{1.12 to 1.65~\si{\micro\meter}} & \colhead{1.23 to 1.32~\si{\micro\meter}}  &\colhead{1.34 to 1.43 \si{\micro\meter}} & \colhead{1.49 to 1.58 \si{\micro\meter}} &  \colhead{}}
\startdata
WD 0137B day & $1875\pm25$ & $1935\pm28$ & $1878\pm26$ & $1833\pm23$ & 2430\\
WD 0137B night & $1493^{+55}_{-72}$ & $1639^{+53}_{-64}$ & $1344^{+96}_{-161}$ & $1419^{+56}_{-69}$ & $1050\pm100$ \\ \hline
EPIC 2122B day & $3030\pm47$ & $3107\pm43$ & $3013\pm45$ & $2908\pm51$ & $4050$\\
EPIC 2122B night & $1600^{+200}_{-500}$ & $1400^{+300}_{-1400}$ & $1500^{+200}_{-700}$ & $1600^{+200}_{-500}$ & $1300^{+300}_{-500}$ \\ 
\enddata
\end{deluxetable*}

\section{Discussion}
\label{sec:discussion}

\subsection{Comparing Day/Night Contrasts of Irradiated Substellar Objects}
\label{sec:day-night-contrast}

  \begin{figure*}[!t]
    \centering
    \includegraphics[height=0.38\textwidth]{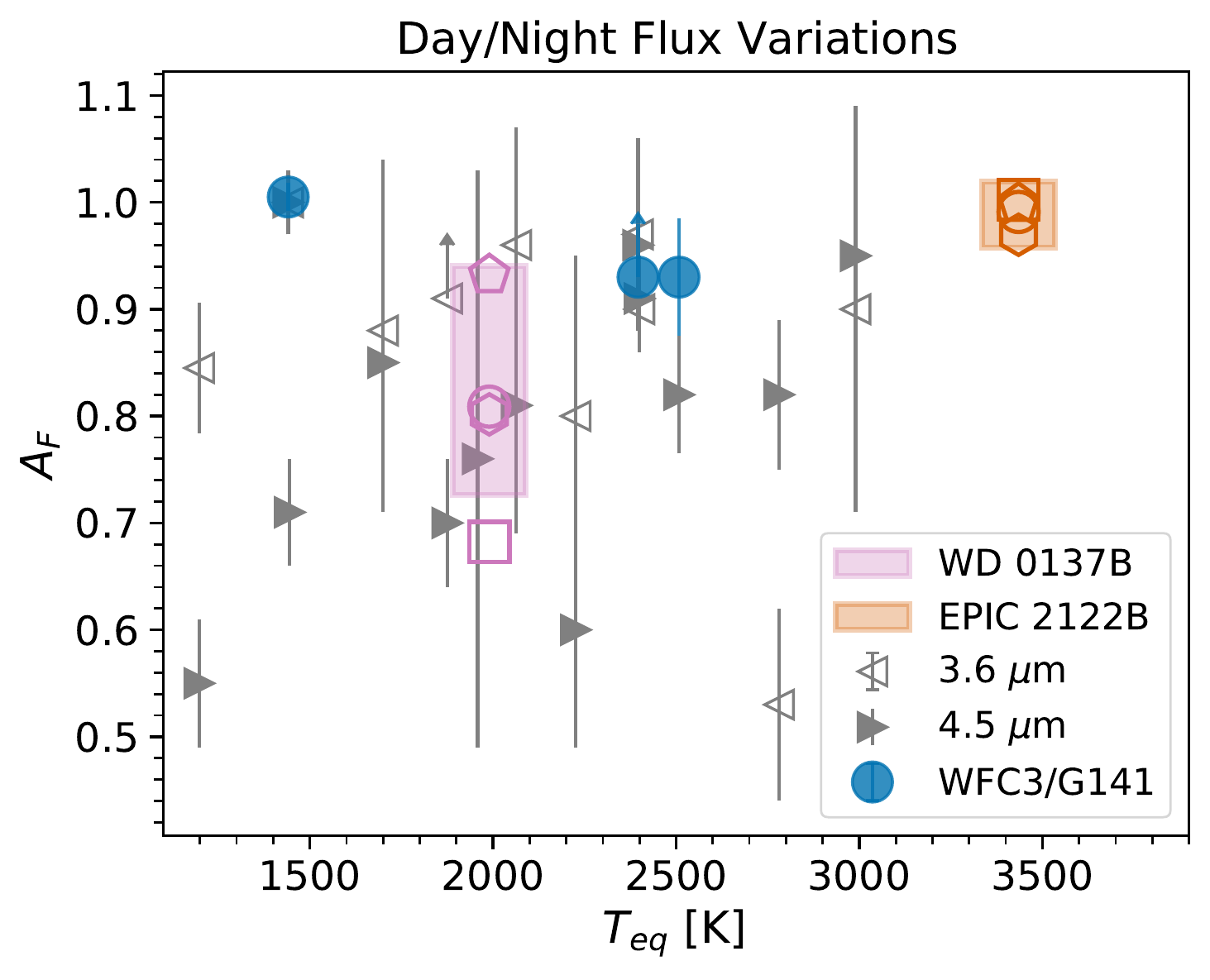}
    \includegraphics[height=0.38\textwidth]{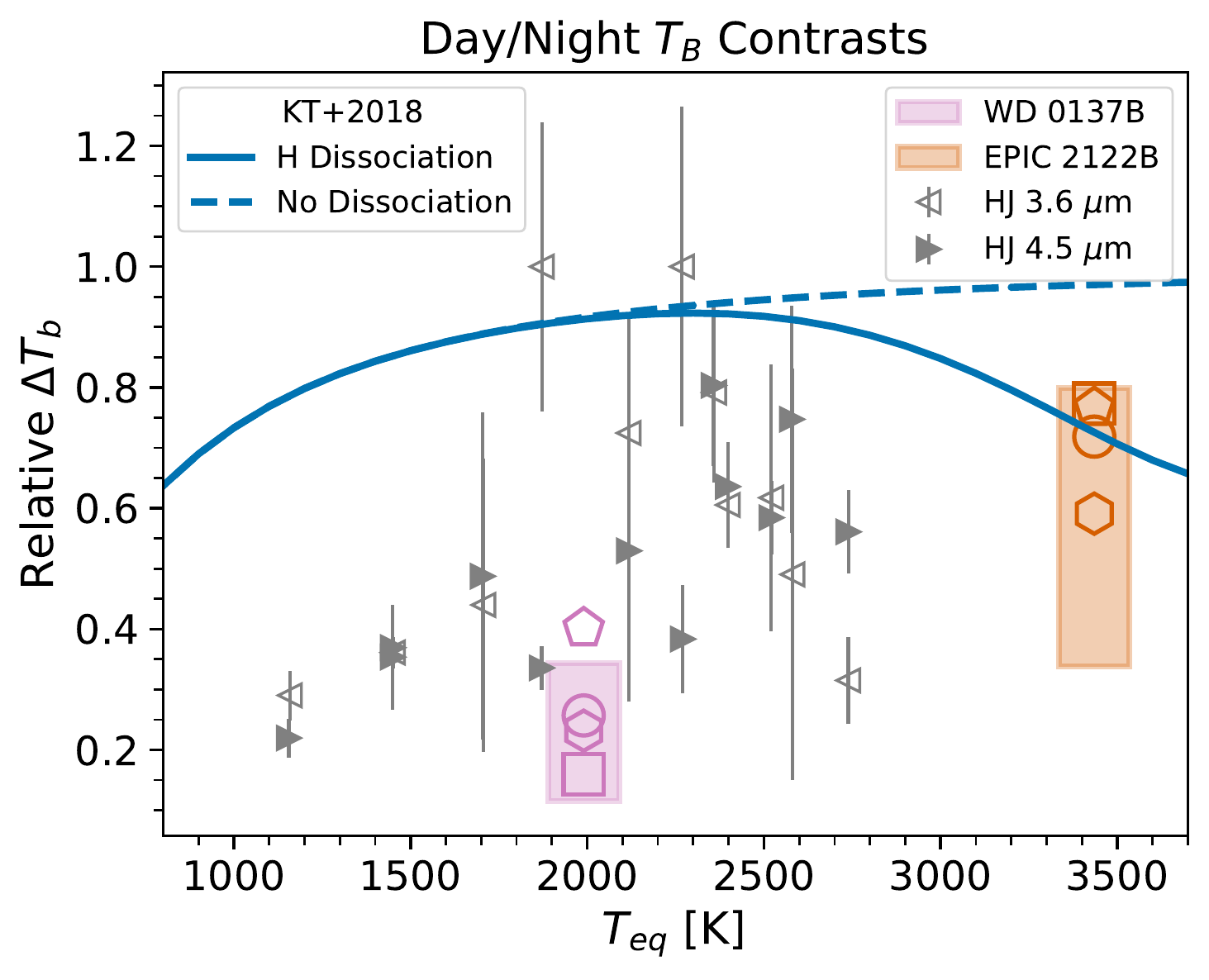}
    \includegraphics[width=0.8\textwidth]{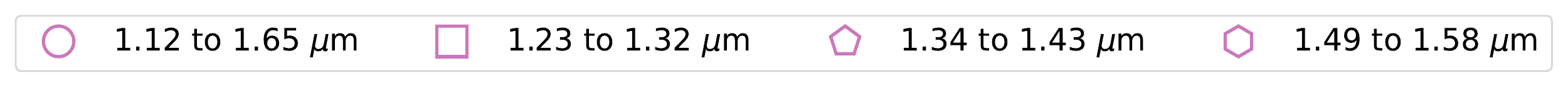}
    \caption{Comparing the day/night flux (left) and brightness temperature contrasts (right) of  WD~0137B and EPIC~2122B with hot Jupiters. The left panel includes relative flux contrasts of Hot Jupiters measured in Spitzer/IRAC \SI{3.6}{\micro\meter} (open triangles) and \SI{4.5}{\micro\meter} (filled triangles) and HST/WFC3/G141 (blue circles). The flux contrasts of WD~0137B and EPIC~2122B are represented as rectangular patches with the vertical lengths representing the standard deviation of $A_{F}$ within the span from \SIrange{1.12}{1.65}{\micro\meter}. Several representative band-averaged flux and brightness contrasts are over-plotted to demonstrate the wavelength-dependent variations. The right panel demonstrate the brightness temperature contrasts using identical markers to those in the left panel. Theoretical $\Delta \TB\mbox{-}T_{\mathrm{eq}}$ relations based on \citet{Komacek2018} are over-plotted in solid and dashed lines, representing whether or not Hydrogen dissociation and recombination are considered in the models.}
    \label{fig:Tdaynight}
  \end{figure*}

  In Figure \ref{fig:Tdaynight}, we compare the day/night contrasts between WD~0137B and EPIC~2122B and hot Jupiters. Two sets of measurements are adopted in this comparison: the relative flux contrast, defined as $A_{F}=(F_{\mathrm{Day}} - F_{\mathrm{Night}}) / (F_{\mathrm{Day}})$, and the relative brightness temperature contrast, defined as $\Delta \TB=(T_{\mathrm{B,\,Day}} - T_{\mathrm{B,\,Night}}) / (T_{\mathrm{B,\,Day}})$. In describing day/night contrasts, $A_{F}$ is directly derived from observations and $\Delta \TB$ is a close approximation (which may require minor correction, see e.g., \citealt{Taylor2020}) of the true temperature contrasts, and hence both variables are popular in exoplanet and brown dwarf literature \citep[e.g.,][]{Perez-Becker2013,Komacek2017,Parmentier2017,Keating2019,Beatty2019,Parmentier2020}.  For hot Jupiters, the $A_{F}$ data are from the collections in \citet{Parmentier2017} with updates from \citet{Parmentier2020}; the $\Delta \TB$ data are from \citet{Beatty2019}. Most of the hot Jupiter measurements are taken with Spitzer/IRAC in its \SI{3.6}{\micro\meter} and \SI{4.5}{\micro\meter} channels with the exception of three WFC3 flux contrast measurements of WASP-43b \citep{Stevenson2014}, WASP-18b \citep{Arcangeli2019}, and WASP-103b \citep{Kreidberg2018}. For WD~0137B and EPIC~2122B, their $A_{F}$ and $\Delta \TB$ are both wavelength-dependent within the G141 bandpass. Therefore, we present their spreads in Figure~\ref{fig:Tdaynight}: the rectangles demonstrate the $\pm1\sigma$ ranges. In addition, a few representative band-averaged values (G141 broadband, HST F127M, F139M, and F153M) are also over-plotted.

  Based on atmospheric circulation models,  the day/night contrast is positively correlated with the effective temperature, because the radiative cooling timescale is the primary factor that determines the day/night temperature difference \citep{Komacek2016}. High equilibrium temperature objects have shorter radiative cooling timescales and therefore have greater day/night temperature contrasts, and vice versa. Nevertheless, additional factors such as frictional drags,  rotation rates \citep{Komacek2016, Komacek2017}, and night-side clouds \citep{Parmentier2020} can also affect the temperature contrast. Trends found in brightness temperatures of hot Jupiters measured in the Spitzer bandpasses support this prediction \citep{Keating2019,Beatty2019}, albeit the large scatters and uncertainties in those measurements (see the gray triangles in Figure \ref{fig:Tdaynight}).

  The $A_{F}$ and $\Delta \TB$ of WD~0137B and EPIC~2122B illustrate a more complicated picture. Rather than support or refute the hot Jupiter trends, the measurements of the irradiated brown dwarfs highlight a different characteristic of day/night contrasts: they are highly wavelength-dependent. Across the G141 bandpass, both $A_{F}$ and $\Delta T_{B}$ have a wide spread and their exact value depends on the specific wavelength. Given the fact that different wavelengths trace different pressure levels of the atmospheres and the day- and night-side thermal profiles vary, this wavelength dependence is not surprising, but a consequence of the three-dimensional nature of the irradiated atmospheres. Spectral variations in $A_{F}$ and $\Delta \TB$ should be common among hot Jupiters and irradiated brown dwarfs (as predicted in \citealt{Parmentier2020}), and the primary reason that they exhibit so strongly in WD~0137B and EPIC~2122B is the high SNR of our observations. The broad ranges of $A_{F}$ and $\Delta \TB$ increase the complexity in comparing day/night contrasts among irradiated objects and testing model predictions against observations. Based on the results of WD~0137B and EPIC~2122B, we strongly advocate for specifying the wavelengths and bandpasses of $A_{F}$ and $\Delta \TB$ when discussing the day/night contrasts of irradiated objects.

In the atmosphere of EPIC~2122B, hydrogen dissociation/recombination can accelerate  day-to-night heat transport and decrease the day/night temperature contrasts \citep{Bell2018,Komacek2018}. In the right panel of Figure~\ref{fig:Tdaynight}, we overlay the theoretical $\Delta \TB\mbox{-}T_{\mathrm{eq}}$ relationships derived using the formalism in \citet{Komacek2018} and assuming the rotation period of EPIC~2122B ($P=\SI{1.14}{\hour}$). Even for such a rapid rotator, the theory still predicts an enhanced day-night heat transport rate when hydrogen dissociation and recombination are considered. This results in a decreasing trend in $\delta \TB-T_{\mathrm{eq}}$ when $T_{\mathrm{eq}}>\SI{2500}{\kelvin}$.  Unfortunately, we are not able to incorporate this effect into our GCMs  because the simulations become numerically unstable when it is included in the extremely fast rotating models. While the lack of a GCM simulation prevents a detailed and spectroscopically resolved comparison between the model and observations, the general agreement of the observed $\Delta \TB$ of EPIC 2122B and the predicted trend (Figure~\ref{fig:Tdaynight}) motivates implementing hydrogen dissociation/recombination as part of GCMs.

  \subsection{Tracking the Irradiated Brown Dwarfs on Substellar Color-Magnitude Diagram}

  \begin{figure*}[t]
    \centering
    \includegraphics[width=0.72\textwidth]{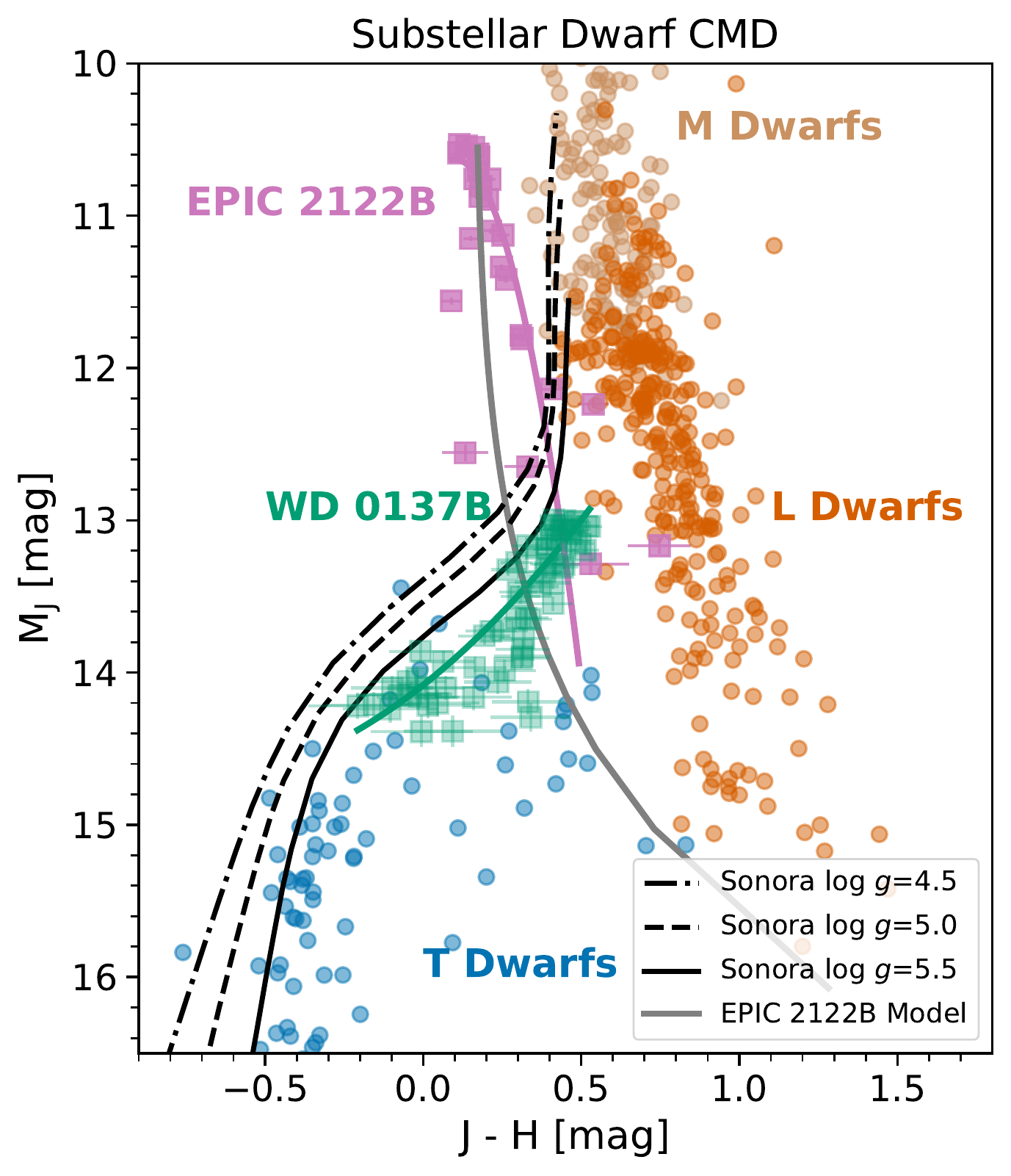}
    \caption{The evolution of WD~0137B and EPIC~2122B on the substellar color-magnitude diagram. We use the synthetic F127M and F153M photometry of WD~0137B and EPIC~2122B to estimate the J and H magnitude and over-plot their tracks (green and magenta squares) onto the substellar CMD established using data from \citet{Best2020}. Several measurements of EPIC~2122B near its night-side are discarded because of low SNRs.  For the WD~0137B and EPIC~2122B tracks, we also overlay the best-fitting second-degree polynomials with solid lines to highlight the overall trends. The substellar CMD is color coded by spectral types: M in brown, L in red, and T in blue. For comparison, we over-plot the Sonora cloudless CMD tracks in three surface gravities (dash-dotted line: $\log g=4.5$; dashed line: $\log g=5.0$; solid line $\log g=5.5$).  The WD~0137B track connects the early-L and mid-T regions in approximately a straight line that is parallel to the Sonora tracks in the same color range. The EPIC~2122B track is parallel to the M to mid-L evolution, but has a bluer color. The gray solid line shows a semi-empirical model for EPIC~2122B. It is constructed by conducting synthetic photometry on a series of spectral mixtures composed of its best-fitting day- and night-side models. }
    \label{fig:CMD}
  \end{figure*}

  We derive the synthetic photometry of WD~0137B and EPIC~2122B from the G141 spectra and compare it with the substellar $M_{H}$-$J-H$ CMD. Because the G141 grism does not cover the entire $H$-band, we use the \textit{HST} F127M and F153M filters instead of the $J$ and $H$ filters. The slight differences caused by different filter throughputs do not affect the subsequent discussion. The absolute photometry is obtained by applying the distance modulus derived from its GAIA DR2 parallax \citep[9.7786 mas for WD 0137B and 2.5851 mas for EPIC 2122B, ][]{Gaia2018}. Figure~\ref{fig:CMD} shows the color-magnitude tracks of WD~0137B and EPIC~2122B overlaid onto the substellar CMD (isolated brown dwarf data are from \citealt{Best2020}).
  
  The effective temperatures of WD~0137B in its day and night-side match those of isolated brown dwarfs of early-L and mid-T types, respectively. The L/T transition \citep{Kirkpatrick2005} occurs between these two spectral types. This  transition is not only a change of course in the color-magnitude diagram (Figure~\ref{fig:CMD}), but also conveys essential information about  substellar atmospheres. The change from the cloudy L-type atmospheres to clear T-type atmospheres due to condensation fronts submerging below the $\tau=1$ layer in cooler effective temperatures or cloud break-up is the most favored explanation for this transition \citep[e.g.,][]{Burgasser2002,Saumon}. Nevertheless,  a change from unstable to stable chemical reactions involving CO and CH$_{4}$ has also been proposed to contribute to this sharp color variation \citep{Tremblin2016}. Between the day and night-sides of WD~0137B, there are regions in its atmosphere where the effective temperatures are favorable for formation and dissipation of clouds \citep[e.g.,][]{Ackerman2001}, as well as disequilibrium chemistry. If these processes also happen in WD~0137B, we expect its day/night color-magnitude evolution to match the CMD of isolated brown dwarfs.

  On the CMD,  WD~0137B's track  starts from the blue edge of the early-L region and directly connects to the mid-T region without showing any  color variations akin to the L/T transition. It deviates from the trend empirically established by isolated brown dwarfs but follows to the theoretical cloudless model tracks \citep{Marley2018} shown as black lines in Figure~\ref{fig:CMD}. This implies that the atmosphere of WD~0137B does not contain longitudinally variable clouds. The cloud opacity and disequilibrium chemistry that operate in isolated brown dwarf atmospheres, if present at all, are not the key drivers of the color changes in WD~0137B. This stark difference between WD~0137B and non-irradiated brown dwarfs in color-magnitude space demonstrates that strong irradiation fundamentally affect clouds, chemistry, and thermal structures in substellar atmospheres, which has also been shown for hot Jupiters \citep[e.g.][]{Fortney2008,Parmentier2016,Beatty2019}.

  As for EPIC~2122B, its CMD track is parallel to the M dwarf to mid-L dwarf evolution but has a bluer NIR color. This color difference is due to the irradiation that causes EPIC~2122B's day-side spectrum to have a steep slope and contain strong flux at short wavelengths (the right panel of Figure~\ref{fig:daysideFit}). As the day-side spectrum's filling factor decrease, EPIC~2122B's CMD track naturally evolves towards the red and faint direction. Based on the spectral model fitting results (\S\ref{sec:spec}), we construct an \emph{ad hoc} semi-empirical model by linearly combining EPIC 2122B's best-fitting day- and night-side model spectra with a tunable day-side filling factor. The color-magnitude trend of the model was derived and shown as a gray solid line in Figure~\ref{fig:CMD}. As expected, as EPIC~2122 B rotates from its day- to night-side, its brightness diminishes and its color reddens.

  \subsection{The Formation and Evolution of WD~0137B and EPIC~2122B}

  \begin{figure}[h]
    \centering
    \includegraphics[width=\columnwidth]{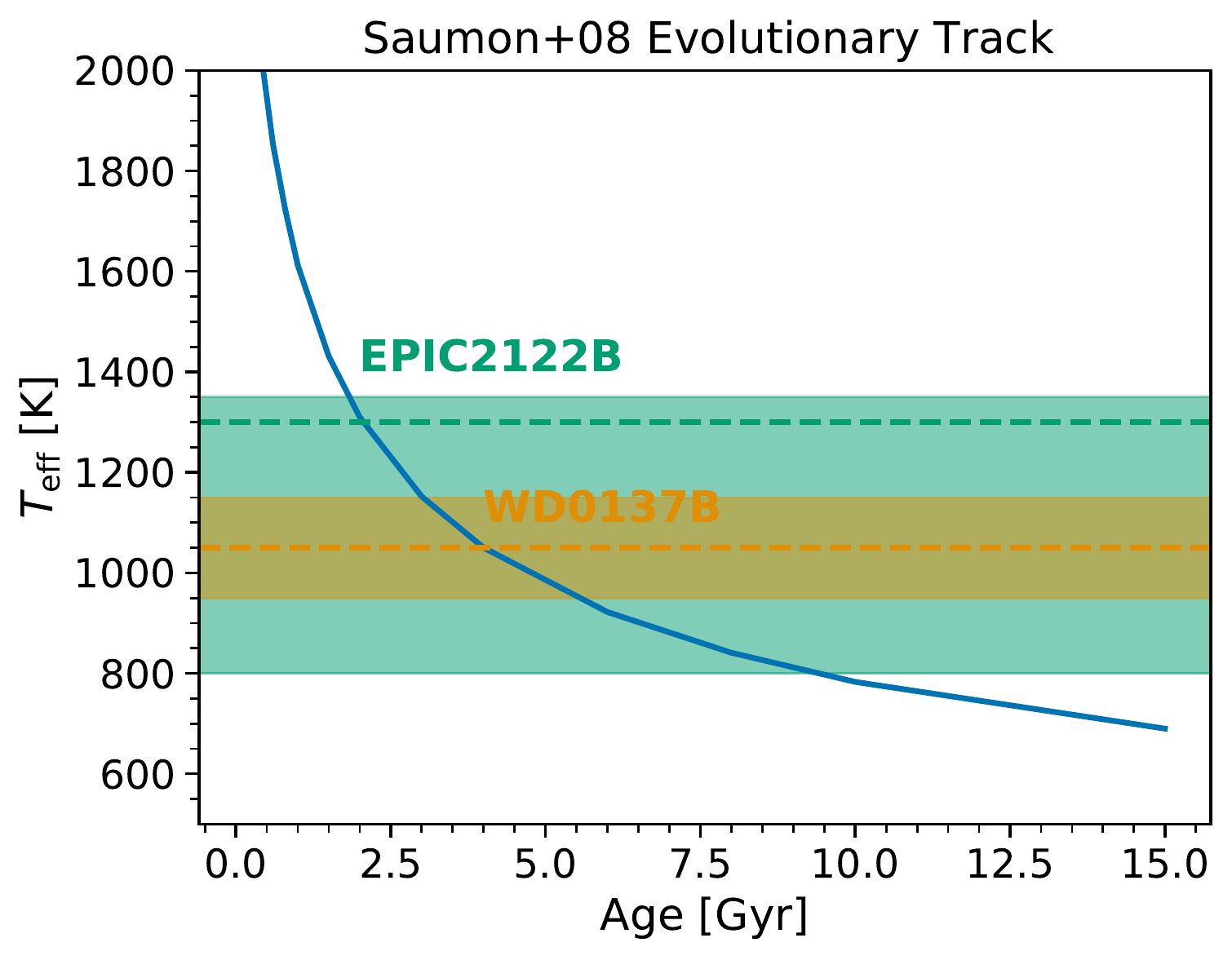}
    \caption{A comprison of the night-side temperatures of the irradiated brown dwarfs and the substellar evolutionary track. The blue curve is the $M=55\,M_{\mathrm{Jup}}$ substellar cooling track in the \citet{Saumon} model. The yellow and green dashed lines mark the night-side temperatures of WD~0137B and EPIC~2122B, respectively, and the shades show the uncertainties (note the asymmetrical error in $T_{\mathrm{eff\,night-side}}$ of EPIC~2122B). The night-side temperature parallels intercept the cooling curve at 4.0~Gyr and 2.1~Gyr.}
    \label{fig:evolution}
  \end{figure}

  The night-side temperatures of WD~0137B and EPIC~2122B can  probe their interior heat flux. We define a non-irradiated temperature ($T_{\mathrm{non-irr}}$) to represent the effective temperature of a hypothesized isolated WD~0137B or EPIC~2122B. The night-side temperatures are upper limits of $T_{\mathrm{non-irr}}$.  We can then compare their night-side temperatures to substellar evolutionary models \citep[e.g.,][]{Saumon} to probe their cooling stages. In Figure~\ref{fig:evolution}, we compare the best-fit night-side temperatures to a $M=55\,M_{\mathrm{Jup}}$ theoretical cooling track in the \citet{Saumon} model, and find the cooling ages of WD~0137B and EPIC~2122B to be $4.0^{+1.6}_{-1.0}$~\si{Gyr} and $2.1^{+7.3}_{-0.3}$~\si{Gyr}, respectively. Considering that these temperatures are upper limits for $T_{\mathrm{non-irr}}$ and irradiation may slows down the cooling processes, these results are the lower limits of the ages of the brown dwarfs.

  These results are significantly longer than the white dwarf cooling ages of $0.25\pm0.8$~Gyr and $0.014$ to $0.018$~Gyr for WD~0137 and EPIC~2122, respectively \citep{Maxted2006, Burleigh2006, Casewell2018}. Therefore, our results support a formation and evolution scenario has been advocated for in previous studies \citep{Maxted2006, Burleigh2006, Casewell2018}: in these systems the brown dwarfs used to be the companions of the white dwarf progenitors and experienced the common envelope evolution phase. Such an evolutionary path is further supported by the low masses of the white dwarfs in these systems. 

  \subsection{Comparing WD~0137B and EPIC~2122B}

  WD~0137B and EPIC~2122B demonstrate many similarities. The phase curves of both objects require multi-order Fourier series to fit and are consistent with the shapes of theoretical phase curves predicted by the GCM simulations. This favors a surface map including a hot spot that is located at the  substellar point. The phase-resolved spectra of both objects can be reduced to a two-dimensional linear space spanned by their respective day- and night-side spectra. The day-side spectra of both objects are very well explained by the irradiated brown dwarf models \citep{Lothringer2020} scaled by a filling factor of ${\sim}50\%$. The $T_{\mathrm{non-irr}}$ of WD~0137B and EPIC~2122B favor cooling ages much older than those of their host white dwarfs, supporting the formation scenario where the brown dwarfs have formed before the post-main-sequence evolution and undergone the entire common envelope phases.

  The two brown dwarfs also have distinctive differences. While EPIC~2122B's spectra are featureless at all phases, WD~0137B has a significant water absorption at every phase except the day-side. This is likely due to the higher level of  UV irradiation that EPIC~2122B receives, which efficiently heats the upper atmospheres, create temperature inversion, and cause water molecules to dissociate \citep{Lothringer2018}. EPIC~2122B has greater day/night temperature contrasts than WD~0137B, perhaps due to its shorter radiative cooling timescale and faster rotation rate slowing down the day/night heat transfer. The high-order Fourier components are also more significant in EPIC~2122B's phase curves, suggesting a more confined hot spot and less efficient heat transfer. However, the difference in phase curve shapes can also be attributed to differences in viewing geometry or surface structures.

  \section{Conclusions}
  \label{sec:conclusions}
The key results of our study are as follows:\\
1. We present HST/WFC3/G141 phase-resolved spectra of white dwarf-brown dwarf binaries WD~0137 and EPIC~2122. Both binaries demonstrate strong photometric and spectroscopic  modulations over their respective orbital periods. These modulations probe the day-night cycles of the tidally locked brown dwarfs.

2. We find that multi-order Fourier series models fit well to the observed phase curves. The best-fitting model to the \SIrange{1.12}{1.65}{\micro\meter} broadband phase curve of WD~0137 contains $k=1$ and $2$ waves with semi-amplitudes of $5.28\pm0.02$\% and $0.45\pm0.02$\%, respectively. For EPIC~2122, the best-fitting model consists of $k=1, 2,\,\mbox{and}\,4$ waves with semi-amplitudes of $29.1\pm0.1$\%, $3.7\pm0.1$\%, and $0.5\pm0.1$\%, respectively. The high-order components and the base order do not have significant phase offsets.

3. The shapes of the observed phase curves agree well with the theoretical predictions from GCM simulations, as well as models assuming radiative-convective equilibrium. The good fit of the RCE models implies inefficient day/night heat transfer, which results from the strong Coriolis force counter-balancing the day/night pressure gradients. In these models, hot spots form at the substellar points and can introduce high-order harmonics in the hemispherically integrated phase curves.

4. The modulations show significant wavelength dependence. In both binaries, because the flux contributions of the brown dwarfs increase towards longer wavelength, the modulation amplitudes rise from \SIrange{1.1}{1.7}{\micro\meter}. For WD~0137, the modulation amplitude is further enhanced in the \SI{1.4}{\micro\meter} water absorption band, revealing strong phase-dependent variations of water vapor absorption.

4. By removing the white dwarf components with model spectra, we obtained phase-resolved brown dwarf spectra. The day/night spectral variations are primarily driven by temperature differences. The spectra of  WD~0137B show a significant water absorption feature in the night-side and gradually become nearly featureless from night to day. In comparison, the spectra of EPIC~2122B are featureless at all phases.

5. For both brown dwarfs, the phase-resolved spectra can be accurately represented as a two dimensional linear space spanned by the day- and night-side spectra. For WD~0137B, its night-side spectrum agrees well with a ${\sim}1050\pm100$\,K non-irradiated, cloudless brown dwarf model and the day-side is best-fit by an irradiated model with day-side only heat redistribution and a \teff of \SI{2430}{\kelvin}. For EPIC~2122B, the night-side emission is only marginally detected and best-fit by a \SI{1300}{K} blackbody. Its day-side is also best-fit by an irradiated brown dwarf model with day-side only heat redistribution and a \teff of ${\sim}$\SI{4050}{\kelvin}. Both day-sides spectra require a filling factor of ${\sim}50\%$ for the model, supporting the notion that the irradiated hemispheres of the brown dwarfs contain a hot spot that dominates the day-side emission.

6. The day- and night-side spectra of WD~0137B and EPIC~2122B demonstrate their flux and brightness temperature contrasts are strongly wavelength dependent. The spectral variations of the day/night contrasts are likely a joint consequence of two properties of these brown dwarfs: 1) different wavelengths trace different pressure levels of the atmosphere; 2) the irradiated and non-irradiated hemispheres differ in their thermal profiles. This wavelength-dependent contrast highlights the complexity in establishing the relationship between temperature contrasts and fundamental properties of irradiated atmospheres. Based on this result, we also advocate for specifying the wavelengths when cross-comparing day/night contrasts of multiple irradiated objects.

7. The day- and night-side of WD~0137B have the color and brightness similar to those of early-L and mid-T types brown dwarfs, respectively. On the substellar CMD, its day-to-night evolution directly connects these two regions by approximately a straight line rather than following the L/T transition turn. This suggests that cloud formation/dissipation, as well as disequilibrium chemistry, which are essential mechanisms that drive the color-magnitude evolution of non-irradiated brown dwarfs, do not play a critical role in the highly-irradiated and tidally-locked atmosphere of WD~0137B.

8. Using the night-side temperatures as proxies for internal thermal temperatures, we constrain the cooling timescales for WD~0137B and EPIC~2122B to be 4.0 and 2.0 Gyr. They are significantly longer than the cooling ages of the respective white dwarfs. This supports the formation scenario where the brown dwarfs form before the white dwarfs and have experienced the common envelope evolution.

Finally, our study exemplifies the remarkable opportunities that the white dwarf-brown dwarf binaries provide for investigating irradiated substellar atmospheres. The phase curve amplitudes are more than two orders of magnitude greater than those of hot Jupiters, resulting in extremely high signal-to-noise ratios. The telescope time request is only a fraction of that required for a complete hot Jupiter phase curve. The irradiated brown dwarfs' properties are in an interesting parameter space, exhibiting fast rotation rates, high UV irradiation, and extreme day/night contrasts. These properties cannot be probed with hot Jupiters or free-floating brown dwarfs. Continuing phase-resolved spectroscopic observations with a larger sample and a broader wavelength coverage will result in excellent data that help deepen the understanding of irradiated atmospheres and strengthen the connection between studies of brown dwarfs and transiting exoplanets.

\vspace{2em}
  We thank the referee for a constructive report that helps improve the quality of this paper. We would like to acknowledge Dr. Adam P. Showman for his contribution to the observing proposal. We thank Dr. Tad Komacek for sharing the code to produce the theoretical relationships in Figure~\ref{fig:Tdaynight}. Y.Z. acknowledges support from the Harlan J. Smith McDonald Observatory Fellowship and the Heising-Simons Foundation 51 Pegasi b Fellowship. This research has made use of the NASA Exoplanet Archive, which is operated by the California Institute of Technology, under contract with the National Aeronautics and Space Administration under the Exoplanet Exploration Program.  The observations and data analysis works were supported by program HST-GO-15947 and the irradiated atmospheric model development was supported by program HST-AR-16142. Supports  for  Program  numbers  HST-GO-15947 and HST-AR-16142 were provided by NASA through a grantfrom the Space Telescope Science Institute, which is operated by the Association of Universities for Research in Astronomy, Incorporated, under NASA contract NAS5-26555.  This  work  is  partly  supported  by  the  international Gemini Observatory, a program of NSF’s NOIRLab, which is managed by the Association of Universities for Research in Astronomy (AURA) under a cooperative agreement with the National Science Foundation,on behalf of the Gemini partnership of Argentina, Brazil, Canada, Chile, the Republic of Korea, and the United States of America.  

\software{Numpy \citep{2020NumPy-Array}, Scipy \citep{2020SciPy-NMeth}, matplotlib \citep{Hunter2007}, Astropy \citep{Robitaille2013}, Pysynphot \citep{2013ascl.soft03023S}, starry \citep{Luger2019}}

\facility{Hubble Space Telescope, Exoplanet Archive}


\appendix
\section{The general circulation model for brown dwarfs around white dwarfs}
\label{sec:GCM}
We briefly describe the GCM used in this study for WD~0137B and EPIC~2122B. The general model structure is similar to that used in \cite{Komacek2017} and the ones without hydrogen dissociation and recombination in \cite{Tan2019}. The  GCM solves the global, three-dimensional hydrostatic primitive equations that govern the large-scale flow in  atmospheres using the dynamical core of the MITgcm \citep{adcroft2004}. An idealized semi-grey, two-stream radiative transfer scheme that includes a broad thermal band and two sub bands in the visible is used to represent the permanent day-side irradiation and nightside cooling.  The opacity in the thermal band is a function of pressure alone and is the same as that used in \cite{Tan2019}.  In the visible band, rather than using only one broad band as previous studies, the stellar irradiation is partitioned into two channels with different opacities:
\begin{equation}
    \begin{split}
        & F_{\rm v1}=(1-A)(1-\beta)\sigma T^4_{\rm irr}\mu_{\rm v}\exp(-\frac{\tau_{\rm v1}}{\mu_{\rm v}}), \\
        & F_{\rm v2}=(1-A)\beta\sigma T^4_{\rm irr}\mu_{\rm v}\exp(-\frac{\tau_{\rm v2}}{\mu_{\rm v}}),
    \end{split}
\end{equation}
where $A$ is the Bond albedo, $\mu_{\rm v}$ is the local zenith angle of the irradiation, $\tau_{\rm v1}=\kappa_{\rm v1}p/g$ and $\tau_{\rm v2}=\kappa_{\rm v2}p/g$ are optical depth of each channel, $p$ is pressure, $g$ is surface gravity, $\kappa_{\rm v1}$ and $\kappa_{\rm v2}$ are opacities which are assumed to be constants, $T_{\rm irr}$ is the irradiated temperature at the substellar point, $\sigma$ is the Stefan-Boltzmann constant and $\beta$ is the partition constant. $A$, $\beta$,  $\kappa_{\rm v1}$ and $\kappa_{\rm v2}$ are free parameters. The total irradiative flux at a location on the day-side is $F_{\rm v}=F_{\rm v1}+F_{\rm v2}$.
This treatment  crudely mimics effects of  strong UV absorption on generating strong thermal inversion in the day-side atmosphere \citep{Lothringer2020}. We consider absorption only and omit scattering. Effects of scattering in energy balance is implicitly included as nonzero bond albedo. We apply a fixed, globally uniform temperature at the bottom pressure to represent the assumption that the thermal structure merge to the same  adiabatic profile in the deep interior. This bottom temperature is a free parameter. We apply two types of frictional drag as in Komacek et al. (2017): a week frictional drag with a spatially independent drag timescale of $\tau_{\rm drag}=10^7$ s; and a basal drag that is applied only to pressure larger than 200 bars with a bottom drag timescale of $10^4$ s. 

For EPIC~2122B, we apply $A=0.2$, $\beta=0.25$, $\kappa_{\rm v1}=0.01~{\rm m^2kg^{-1}}$ and $\kappa_{\rm v2}=0.05~{\rm m^2kg^{-1}}$. These parameters result in a reasonable global-mean PT profile compared to that using comprehensive models in \cite{Lothringer2020} and has been used regularly in the hot Jupiter studies \citep[e.g.,][]{Line2013,Parmentier2014}. The  pressure domain of this model is between 0.01 and 500 bars, and temperature at the bottom pressure is assumed to be 4500 K. 

For WD~0137B, we apply $A=0.3$, $\beta=0.3$, $\kappa_{\rm v1}=0.01~{\rm m^2kg^{-1}}$ and $\kappa_{\rm v2}=0.02~{\rm m^2kg^{-1}}$ to generate a  global-mean PT profile similar to that in \cite{Lothringer2020}. The  pressure domain of this model is between $2\times10^{-3}$ and 200 bars, and temperature at the bottom pressure is assumed to be 3500 K. 

In all models, we assumed a radius of $7\times10^7$ m, similar to that of Jupiter. The models have a horizontal resolution  equivalent to $512\times256$ in longitude and latitude, and have  36 vertical layers evenly discretized in log-pressure space. A 4th-order Shapior filter is applied to maintain numerical stability.

\end{CJK*}
\end{document}